\documentclass[12pt]{article}
\usepackage{a4wide}
\usepackage{amssymb}
\usepackage{amsmath}
\usepackage{graphicx}
\usepackage{graphicx,caption}
\usepackage{soul}

\usepackage{tensor}

\usepackage{comment}

\usepackage[dvipsnames]{xcolor}

\begin{document}
{\renewcommand{\thefootnote}{\fnsymbol{footnote}}
\begin{center}
{\LARGE Black holes in effective loop quantum gravity:\\[1.5 mm]
Covariant holonomy modifications}\\
\vspace{1.5em}
Idrus Husin Belfaqih,$^1$\footnote{e-mail address: {\tt i.h.belfaqih@sms.ed.ac.uk}}
Martin Bojowald,$^2$\footnote{e-mail address: {\tt bojowald@psu.edu}}
Suddhasattwa Brahma$^1$\footnote{e-mail address: {\tt suddhasattwa.brahma@gmail.com}}
and Erick I.\ Duque$^2$\footnote{e-mail address: {\tt eqd5272@psu.edu}}
\\
\vspace{0.5em}
$^1$Higgs Centre for Theoretical Physics, School of Physics \& Astronomy,\\
University of Edinburgh, Edinburgh EH9 3FD, Scotland, UK\\
\vspace{0.5em}
$^2$ Institute for Gravitation and the Cosmos,\\
The Pennsylvania State
University,\\
104 Davey Lab, University Park, PA 16802, USA\\
\vspace{1.5em}
\end{center}
}

\setcounter{footnote}{0}

\begin{abstract}
  Emergent modified gravity provides a covariant, effective framework for
  obtaining spherically symmetric black hole solutions in models of loop
  quantum gravity with scale-dependent holonomy modifications.
  Exact solutions for vacuum black holes in the presence of a cosmological
  constant are derived here and analyzed in four different gauges, explicitly
  related to one another by standard coordinate transformations.
  The global structure is obtained by gluing space-time regions corresponding
  to the gauge choices, reconstructing a non-singular wormhole space-time for
  an arbitrary scale-dependent holonomy parameter. This outcome demonstrates
  the robustness of black-hole models with covariant holonomy modifications
  under quantization ambiguities.
  Compared with previous constructions, full covariance of the resulting
  space-time models as derived here implies subtle new effects and leads to a
  novel understanding of the parameters in holonomy modifications,
  distinguishing a constant holonomy length from a possibly scale-dependent
  function that may change coefficients of holonomy terms. New physical
  results are obtained for instance in the context of a non-trivial zero-mass
  limit of holonomy-modified space-times. The existence of a consistent effective space-time structure implies various novel aspects of a net
  gravitational stress-energy
  and related thermodynamical properties.
\end{abstract}

\tableofcontents

\section{Introduction}
Black holes are now everyday objects encountered in astronomical observations indicating that much of their exterior is well described by General Relativity (GR). Yet, the region beyond the event horizon remains inaccessible to observations and we must rely on theoretical extrapolations to understand the interior physics. In particular, GR robustly predicts a singularity at the center of black holes \cite{PhysRevLett.14.57}, which has long been interpreted as the result of the breakdown of the theory. Hence, it has been argued that a quantum theory of gravity would be able to resolve such singularities in analogy to how the quantum theory of light resolves the ultraviolet catastrophe. One candidate background-independent and non-perturbative theory is loop quantum gravity (LQG) \cite{rovelli2004quantum,thiemann2008modern}. We shall explore the fate of stationary, spherically symmetric (vacuum) solutions due to quantum corrections from LQG in an effective framework, making sure that full covariance is realized in any effective line element used to describe the physics of black holes.

The canonical quantization in LQG is based on the holonomy-flux algebra. This is so because the regularization scheme followed in LQG does not allow for quantum operators corresponding to the connection, but rather its parallel transport along one-dimensional curves. In effective black-hole models, one focuses on the spherically symmetric reduced theory and incorporates quantum corrections corresponding to this feature by modifying the Hamiltonian constraint from a simple polynomial dependence on the connection to a trigonometric one resulting from the real-valued combinations of holonomies. This procedure is often done by hand, rather than derived from first principles, and covariance of the resulting models is seldom checked or even acknowledged as a problem. 

General covariance is harder to grasp in the canonical formulation than in its more common geometric or Lagrangian cousins.
A modified Hamiltonian constraint will in general modify its Poisson brackets.
One must therefore ensure that the brackets remain `first class' for the equations of motion to be gauge covariant.
Even if this condition is fulfilled, there often remains a modified structure function in the Poisson bracket of two Hamiltonian constraints, whose proper interpretation is a modified kinematical dependence of the space-time metric on the phase-space variables \cite{bojowald2018effective,EMG}.
Contrary to what is commonly believed, preserving the first-class nature of the constraint brackets does not imply that the space-time metric tensor is covariant, constructed in the usual way from phase-space variables, the structure function, as well as a lapse function and shift vector field: In a canonically modified theory, this object is not guaranteed to be subject to gauge transformations that are on-shell equivalent to the tensor transformation law.
Rather, additional covariance conditions must be imposed on the structure function, which further restrict compatible modifications of the Hamiltonian constraint \cite{EMGCov}.
In order to implement these conditions, we will use Emergent Modified Gravity (EMG) \cite{EMG,EMGCov} as the underlying framework for effective LQG.
In particular, the spherically symmetric model of EMG can be solved for the most general, covariant, modified constraint in (3+1)-dimensions with a dependence on the spatial derivatives of the phase space up to the second order, which contains terms that can be interpreted as holonomy modifications.

An earlier, special case of EMG in spherical symmetry was studied in \cite{alonso2022nonsingular,BBVeffBH} where the holonomy modifications were shown to lead to a nonsingular black-hole solution in vacuum.
The global structure of such a solution is an interuniversal wormhole joining a black hole to a white hole through their interiors.
The holonomy parameter $\mu$ in these works was assumed constant, $\mu=\mu_0$, a framework referred to as the $\mu_0$ scheme in some of the LQG literature.
This holonomy parameter is associated with the coordinate length of the links in the triangulation of space as required by LQG.
More precisely, in the spherically symmetric effective framework where the space-time topology may be fixed to $\mathcal{M} = \mathbb{R}^2 \times S^2$, one only considers the angular discretization of the spheres (the submanifold $S^2\subset \mathcal{M}$) and hence $\mu_0$ corresponds to the angular coordinate length of these links. A constant $\mu_0$ is required if the constraint operator is constructed from basic holonomy operators. However, holonomies depend on a canonical variable such as a spatial connection component or extrinsic curvature of the spheres, which unlike space-time curvature scalars are not guaranteed to be small in low-curvature regimes. Depending on specific properties of dynamical solutions, a constant $\mu_0$ could therefore imply large holonomy modifications in regions that should remain close to the classical limit.

It is therefore necessary to introduce scale-dependent holonomy parameters $\mu$ that become smaller in semiclassical regions in order to compensate for potentially growing extrinsic curvature. Heuristically, such a behavior can be motivated as including lattice-refinement effects of an underlying discrete spatial state. As spherical areas in a semiclassical region grow larger, a discrete lattice state would have to be refined more and more in order to avoid discreteness being magnified to macroscopic sizes. The coordinate length of lattice links correspondingly decreases, implying a scale-dependent $\mu$ as a function of the spherical area. The scale dependence is not unique and determined largely by phenomenological reasoning, making sure that the modified theory is consistent with classical behavior in the right regimes. In homogeneous cosmological models, there is only one relevant semiclassical regime given by large volumes. In spherical symmetry and the related black-hole models, however, there are different semiclassical regimes, including large radii as well as near-horizon geometries, which should also be semiclassical for astrophysical black holes. In the latter case, however, the conditions on a suitable scale dependence of $\mu$ are different from what is required for large radii because they refer to different components of the spatial metric. Moreover, these conditions are dependent on the space-time gauge or slicing used to express the near-horizon geometry. This observation re-inforces the condition that holonomy modifications must be made compatible with general covariance of an effective line element, and that a large set of scale-dependent $\mu$ must be compatible with a given class of modified theories.

A simple ad-hoc version of scale-dependent holonomy parameters is often
referred to as a $\bar{\mu}$-scheme \cite{APSII}, but it is not uniquely defined. The original motivation was to replace the coordinate length of the links of a triangulation with a more tangible length such as the Planck length $\ell_{\rm Pl}$, characteristic of quantum gravity. Since the holonomy around a closed loop of links is related to the curvature of the gauge connection, one may use the Planck area, or the smallest non-zero eigenvalue $\Delta$ in the area spectrum of loop quantum gravity, as an ingredient of the holonomy parameter $\mu$. If the exponent of $\Delta$ in the holonomy parameter is chosen, which might be proportional to $\Delta$ itself as an area or $\sqrt{\Delta}$ as a side length, or perhaps some other power law depending on one's intuition, dimensional arguments  require that this quantity be divided by a function of metric components with suitable length units. For this purpose, one can employ the component $q_{\vartheta\vartheta}$ that determines the areas of spheres. This procedure may then result in $\mu\propto q_{\vartheta\vartheta}^{-1}$ or $\mu\propto q_{\vartheta\vartheta}^{-1/2}$ or with some other exponent. (The square-root version is closest to the $\bar{\mu}$-scheme in spatially flat cosmological models, where the relevant connection component is proportional to $\dot{a}$ in terms of the scale factor. This component is then multiplied by $\sqrt{\Delta}$ for dimensional reasons, and divided by $a$ in order make the expression independent of spatial rescalings. Here, we have a simple covariance argument, but, compared with spherical symmetry, only referring to a small subset of possible coordinate transformations.)
Such a construction does not lead to a unique form of a scale-dependent $\mu$
because the final result depends on the exponent chosen for $\Delta$, and the
radial metric component $q_{xx}$ could also be used in such an ad-hoc
argument \cite{SchwarzN}. 

A more systematic approach is therefore needed, which we construct here based
on the condition of covariance. As we will see, this condition, while still
not leading to a unique analog of $\mu$, implies additional restrictions on possible
phase-space dependencies of this function, which turn out to include a dependence on
$q_{\vartheta\vartheta}$ but not on $q_{xx}$.  Such a behavior is well-defined
even when crossing the horizon because the physical length of the links
remains spacelike everywhere and discretizes the submanifold
$S^2 \subset \mathcal{M}$.  In Schwarzschild coordinates, the spatial slices
have the topology $\Sigma = \mathbb{R} \times S^2$, such that the exterior is
foliated by spheres with squared areal radius $q_{\vartheta\vartheta}$, while
the interior geometry is that of hypercylinders with squared radius
$q_{\vartheta\vartheta}$.  In the heuristic lattice picture, both cases can be
interpreted as discrete submanifold $S^2$ with plaquettes of constant size if
$\mu\propto q_{\vartheta\vartheta}^{-1/2}$ is used.

The purpose of this paper is to generalize the results of
\cite{alonso2022nonsingular,BBVeffBH} to an arbitrary holonomy parameter as a
function of the radius, $\mu (q_{\vartheta\vartheta})$, and hence obtaining a
general triangulation scheme in its heuristic interpretation. We will note
crucial conceptual differences in our new scheme as well, in particular its
general covariance and its role in light of a complete discussion of
canonical transformations that may be used to change a functional dependence
such as $\mu(q_{\vartheta\vartheta})K_{\varphi}$ where $K_{\varphi}$ is one of
the momenta, classically related to extrinsic curvature. While some properties
of $\mu_0$ and $\bar{\mu}$-type schemes can still be recognized in specific
choices of canonical variables, our covariant construction (both in terms of
space-time and phase-space transformations) provides a novel and unified
picture of holonomy modifications. Since the relevant equations are obtained
in the framework of emergent modified gravity, we follow the latter's notation and
refer to the new holonomy parameter as $\lambda(q_{\vartheta \vartheta})$,
highlighting its new origin and features.

The new conceptual understanding of holonomy modifications leads to novel physical insights. In particular, we extend previous constructions for black holes in LQG to asymptotically de Sitter space-times with a well-defined asymptotic infinity, rather than ending at a maximal radius. For this result, our covariant treatment of scale-dependent holonomy parameters is crucial. Covariance also allows us to apply standard space-time analysis in an unambiguous way, including the derivation of timelike and likelight geodesics, effective stress-energy tensors, and thermodynamics.

After a general discussion of holonomy modifications, we apply our results to
specific versions, including one that in its dynamical implications
corresponds to the traditional choice of
$\mu(q_{\vartheta\vartheta})\propto q_{\vartheta\vartheta}^{-1/2}$.  To this
end we begin Section~\ref{sec:Spherically symmetric model} with a discussion
of emergent modified gravity and describe how some modification functions can
be interpreted as holonomy corrections and hence used as a covariant effective
model for LQG.  The black-hole solution with a general holonomy parameter is
obtained in Section~\ref{Sec:Black hole solution}. The procedure follows the
solution of the equations of motion in four different gauges which we show are
all related by standard coordinate transformations covering distinct but
overlapping regions of the spacetime. These four patches are then sewed
together to recover a global structure with the properties of a wormhole and
present the maximal extension in null coordinates in Section~\ref{sec:Global structure}. We further show that the physical global structure is non-unique, owing to ambiguities in the slicing and gluing of the maximal extension.  Section~\ref{sec:Interior physics} is devoted to an analysis of the interior structure of space-time with the black-hole horizon. We highlight the robust feature that the classical singularity is replaced by a non-singular hypersurface of reflection symmetry, at which the initially collapsing interior can be glued to an expanding version. This outcome is largely independent of the specific scale dependence of holonomy parameters. Geodesic analysis can be used in order to relate this non-singular behavior to heuristic effects such as repulsive modified gravity. In Section~\ref{sec:Exterior physics},
we study several physical effects of holonomy modifications in the
black-hole exterior including corrections to the precession frequency of
nearly circular orbits, to the deflection angle of null rays, and to black
hole thermodynamics. 
In Section~\ref{sec:mu0 scheme}, we revisit asymptotic
problems in holonomy modifications that correspond to the traditional
$\mu_0$-type scheme of scale-independent $\mu$, including asymptotic corrections to
the deflection angle, to the Brown-York quasilocal energy, and to the Bekenstein-Hawking entropy, as well as the
existence of a maximum radius of the universe.  In Section~\ref{sec:mubar scheme} we study the effects of a specific version of scale-dependent $\mu$
and confirm that it removes unexpected and potentially problematic features of the $\mu_0$-type scheme in asymptotic regimes.


\section{Emergent modified gravity in spherical symmetry}
\label{sec:Spherically symmetric model}

The framework of emergent modified gravity plays an important role in discussions of general covariance in a canonical setting and is therefore reviewed in this section. Its main feature is that the {\it space-time metric is emergent rather than effective}: It is not a corrected version of a pre-existing classical metric tensor. Rather, it is the only metric tensor admissible in a modified canonical theory because the classical expression as a phase-space function fails to obey the tensor transformation law when gauge transformations are modified. The entire emergent metric is derived from covariance conditions, in contrast to what is usally referred to as an effective metric if perturbative correction terms are derived for the classical metric.

\subsection{Spherically symmetric canonical gravity}

Canonical gravity, originally derived by Dirac, is conveniently based on the Arnowitt--Deser--Misner (ADM) decomposition where the line element in spherical symmetry takes the form \cite{ADM,arnowitt2008republication}
\begin{equation}
    {\rm d} s^2 = - N^2 {\rm d} t^2 + q_{x x} ( {\rm d} x + N^x {\rm d} t )^2 + q_{\vartheta \vartheta} {\rm d} \Omega^2
    \ ,
\label{eq:ADM line element - spherical}
\end{equation}
where $q_{a b}$ is the three-dimensional metric induced on the spatial hypersurfaces foliating the manifold, while $N$ and $N^x$ are the lapse function and shift vector field associated to the observer frame with time coordinate $t$.
The Hamiltonian is composed of the Hamiltonian constraint $H$ and the diffeomorphism constraint $H_x$.
As phase-space functions, depending on the spatial metric and its momenta, these expressions vanish on physical solutions, $H=0$ and $H_x=0$, bringing the system on-shell.
The Hamiltonian that generates evolution with respect to $t$ is given by $H[N,N^x] =H[N] + H_x[N^x]$ where the square brackets indicate that these terms are smeared (or spatially integrated) by the functions $N$ and $N^x$. The constraints then generate time evolution via Poisson brackets, $\dot{\cal O} = \{ {\cal O} , H[N,N^x] \}$ for any phase-space function ${\cal O}$.

The same constraints also generate gauge transformations by use of different smearing functions $\epsilon^0$ and $\epsilon^x$ instead of the fixed $N$ and $N^x$ for a given frame, and hence the generator of a general gauge transformation is $H[\epsilon^0,\epsilon^x]$.
The vanishing of the constraints must be consistent in different gauges, which means that the Poisson brackets of the constraints with themselves must vanish on-shell.
The specific brackets found in canonical gravity are known as the hypersurface-deformation brackets, given by
\begin{subequations}
\begin{eqnarray}
    \{ H_x [N^x] , H_x[\epsilon^x] \} &=& - H_x [\epsilon^x (N^x)' - N^x (\epsilon^x)']
    \,, \\
    \{ H [N] , H_x [\epsilon^x] \} &=& - H[\epsilon^x N'] 
    \,,
    \label{eq:H,H_x bracket - spherical}
    \\
    \{ H [N] , H[\epsilon^0] \} &=& - H_x \left[ q^{x x} \left( N' \epsilon^0 - N (\epsilon^0)' \right)\right]
    \label{eq:H,H bracket - spherical}
    \,,
\end{eqnarray}
\end{subequations}
where the structure-function $q^{xx}$ is the inverse of the radial metric component, $q^{xx}=1/q_{xx}$.
These brackets indeed vanish on-shell when the constraints equal zero.
Furthermore, if the system is fully on-shell, also imposing equations of motion generated by the constraints, the gauge freedom of canonical gravity is equivalent to the freedom of choosing a space-time slicing or a coordinate system.

Therefore, the constraints are an essential ingredient of encoding general covariance. However, starting with gauge transformations of phase-space functions, two additional steps are needed for a full realization of general covariance, understood as gauge transformations being equivalent to coordinate changes, or as the existence of a tensor-transformation law for the space-time metric: First, we need gauge transformations not only for the spatial part $q_{xx}$, which is a basic phase-space degree of freedom, but also for the time-time and the mixed space-time components, depending on $N$ and $N^x$. While the latter space-time functions do not have momenta and therefore are not included in the phase space (unless an extended version is used), their gauge transformations are uniquely determined by the constraints and their Poisson brackets through consistency conditions between evolution and gauge equations. Secondly, coordinate changes and the tensor transformation law require explicit time derivatives, rather than momentum components as they appear in the constraints and the gauge transformation they generate. Time derivatives are related to momenta by some of the canonical equations of motion generated by the constraints. No additional ingredient is therefore required in order to obtain full expressions of coordinate transformations, but this can be achieved only if the theory is taken fully on-shell by imposing not only the constraint equations but also some of the equations of motion. 

We will include these ingredients in our present review in the more general setting of modified canonical gravity.

\subsection{Modified canonical gravity}

If one considers a modified Hamiltonian constraint $\tilde{H}$, as implied for instance by the regularization procedure of LQG, then we must first ensure that the constraint brackets remain first class for the constraints to vanish consistently in all gauges.
More specifically, we require that the general form of hypersurface-deformation brackets be preserved, 
\begin{subequations}
\label{eq:Hypersurface deformation algebra - spherical - EMG}
\begin{eqnarray}
    \{ H_x [N^x] , H_x[\epsilon^x] \} &=& - H_x [\epsilon^x (N^x)' - N^x (\epsilon^x)']
    \,, \\
    \{ \tilde{H} [N] , H_x [\epsilon^x] \} &=& - \tilde{H}[\epsilon^x N'] 
    \,,
    \label{eq:H,H_x bracket - spherical - EMG}
    \\
    \{ \tilde{H} [N] , \tilde{H}[\epsilon^0] \} &=& - H_x \left[ \tilde{q}^{x x} \left( N' \epsilon^0 - N (\epsilon^0)' \right)\right]
    \label{eq:H,H bracket - spherical - EMG}
    \,,
\end{eqnarray}
\end{subequations}
where the structure-function $\tilde{q}^{x x}$ is determined by demanding that no additional terms appear on the right-hand sides. However, the specific structure function $\tilde{q}^{xx}$ need not equal the classical one, given by a specific function of one of the basic phase-space variables, $q_{xx}$, on which the modified $\tilde{H}$ depends (or an equivalent version in triad variables).

This condition is stronger than anomaly freedom, which would be compatible with, say, an additional $H_x$-term appearing in the Poisson bracket of two $\tilde{H}$. Both conditions, anomaly-freedom and the specific form of hypersurface-deformation brackets, restrict permissible modifications of the constraint.
A comparison with the classical geometrical structure of space-time, together with an analysis of the transformation properties of lapse and shift to be discussed in the next section, suggests that the inverse of the modified structure function $\tilde{q}^{xx}$ plays the role of the radial component of an effective line element (in the stricter sense of emergence) replacing $q_{xx}$ in (\ref{eq:ADM line element - spherical}).
However, anomaly freedom or brackets of hypersurface-deformation form do not imply that this effective line element is invariant. Rather, a further condition must be imposed on $\tilde{H}$, derived from the tensor transformation law of $(1/\tilde{q}^{xx},N,N^x)$, such that the gauge transformations of the effective metric it generates correspond to infinitesimal coordinate transformations \cite{EMGCov}. (The angular component $\tilde{q}_{\vartheta\vartheta}$ also plays a role in the covariance condition, but it is less direct because this component does not appear in the structure functions of spherically symmetric gravity.)

According to the classic analysis of \cite{hojman1976geometrodynamics,kuchar1974geometrodynamics}, if one considers the classical spatial metric as a configuration variable of the phase space, then the Hamiltonian constraint is uniquely determined by the hypersurface-deformation brackets up to the choice of Newton's and the cosmological
constant.
These results therefore imply that $H$  must be the classical constraint of GR, and generally covariant modifications are ruled out unless one includes higher-derivative terms that in a canonical formulation enlarge the classical phase space.
However, in the recent formulation of emergent modified gravity \cite{EMG,EMGCov} the assumption that the spatial metric $q_{ab}$ is a configuration variable of the phase-space is not necessary in order to obtain a consistent field-theory description of space-time.
Instead, emergent modified gravity assumes that the phase space is composed of certain fundamental fields different from the metric, and the metric is an emergent object to be regained by the imposition of anomaly freedom and covariance.
In fact, the most general modified Hamiltonian constraint depending on spatial derivatives of the phase space up to second order and satisfying anomaly freedom and covariance in vacuum spherical symmetry has been obtained exactly in \cite{EMGCov}.
If LQG is to be treated in a covariant framework of effective line elements, without,  as usually assumed in this context, an extension of the classical phase space in order to account for perturbative higher-curvature corrections or higher time derivatives, its effective Hamiltonian constraint must be included in the family of constraints allowed by emergent modified gravity, and its effective line element must be given by the corresponding emergent space-time.

\subsection{Covariance conditions}

The hypersurface-deformation brackets (\ref{eq:Hypersurface deformation algebra - spherical - EMG}) together with general properties of a Poisson bracket, such as the Jacobi identity, imply that off-shell gauge transformations for the lapse function and shift vector are given by \cite{pons1997gauge,salisbury1983realization,bojowald2018effective}
\begin{subequations}\label{eq:Off-shell gauge transformations for lapse and shift - spherical}
\begin{eqnarray}
    \delta_\epsilon N &=& \dot{\epsilon}^0 + \epsilon^x N' - N^x (\epsilon^0)' ,\\
    \delta_\epsilon N^x &=& \dot{\epsilon}^x + \epsilon^x (N^x)' - N^x (\epsilon^x)' + \tilde{q}^{x x} \left(\epsilon^0 N' - N (\epsilon^0)' \right)
    \,.
\end{eqnarray}
\end{subequations}
These equations play a central role in general covariance as explained below.

Building on \cite{bojowald2018effective}, the brackets (\ref{eq:Hypersurface deformation algebra - spherical - EMG}) and the transformations (\ref{eq:Off-shell gauge transformations for lapse and shift - spherical}) suggest the existence of the space-time metric
\begin{equation}
    {\rm d} s^2 = - N^2 {\rm d} t^2 + \tilde{q}_{x x} ( {\rm d} x + N^x {\rm d} t )^2 + \tilde{q}_{\vartheta \vartheta} {\rm d} \Omega^2
    \ ,
\label{eq:ADM line element - spherical - modified}
\end{equation}
where $\tilde{q}_{x x}$ is the inverse of the (possibly non-classical) structure-function, assuming for now that $\tilde{q}_{xx}>0$, while $\tilde{q}_{\vartheta \vartheta}$ is not given by structure functions and must hence be chosen on different grounds. (This feature is a consequence of the symmetry-reduced theory under consideration and its space-time structure.  Since the component is subject to canonical transformations, it is possible to prescribe a specific form of $\tilde{q}_{\vartheta\vartheta}$ as a phase-space function in order to reduce the freedom of performing canonical transformations.)
Because the spatial metric is not directly determined by kinematical properties of the phase space but rather depends, at least in part, on the inverse of the structure-function, we refer to it as an emergent metric. This nomenclature will make more sense when dealing with non-classical structure functions which arise from the requirement of anomaly-freedom of the modified constraints incorporating potential non-perturbative quantum corrections.

Anomaly-freedom of the brackets (\ref{eq:Hypersurface deformation algebra - spherical - EMG}), even if they are of hypersurface-deformation form, does not ensure covariance of the  resulting space-time (\ref{eq:ADM line element - spherical - modified}), contrary to what is commonly stated.
Instead, we have to {\it impose the stronger covariance condition}
\begin{align}
    \delta_\epsilon \tilde{g}_{\mu \nu} \big|_{\text{O.S.}} =&
    \mathcal{L}_\xi \tilde{g}_{\mu \nu} \big|_{\text{O.S.}}
    \ ,
    \label{eq:Covariance condition}
\end{align}
which ensures that canonical gauge transformations generate infinitesimal diffeomorphisms. This requirement contains \eqref{eq:Off-shell gauge transformations for lapse and shift - spherical} as a necessary condition, but it is also non-trivial when applied to the spatial part of the metric and imposes yet another set of restrictions on allowed modifications of the classical constraints.

\subsection{Covariant modified constraints}
Following \cite{bojowald2000symmetry,bojowald2004spherically} we consider the vacuum spherically symmetric theory with canonical gravitational variables $\left\{K_\varphi\left(x\right),E^\varphi\left(y\right)\right\}=\delta\left(x-y\right)$ and $\left\{K_x\left(x\right),E^x\left(y\right)\right\}=\delta\left(x-y\right)$.  (We use units such that Newton's constant equals one.)
In the classical theory, the momenta $E^x$ and $E^\varphi$ are components of the densitized triad, while the configuration variables are directly related to the extrinsic-curvature components ${\cal K}_\varphi =  K_\varphi$ and ${\cal K}_x = 2 K_x$.
Leaving the diffeomorphism constraint \begin{align}
    H_x =& E^\varphi K_\varphi'
    - K_x (E^x)'
    \label{eq:Diffeomorphism constraint - spherical symmetry}
\end{align}
unmodified
and building on \cite{alonso2021anomaly}, we consider the most general ansatz for the Hamiltonian constraint up to second-order derivatives and quadratic first-order derivative terms \cite{EMGCov}
\begin{align}
    \tilde{H} =& a_0
    + ((E^x)')^2 a_{x x}
    + ((E^\varphi)')^2 a_{\varphi \varphi}
    + (E^x)' (E^\varphi)' a_{x \varphi}
    + (E^x)'' a_2
    \notag\\
    &+ (K_\varphi')^2 b_{\varphi \varphi}
    + (K_\varphi)'' b_2
    + (E^x)' K_\varphi' c_{x \varphi}
    + (E^\varphi)' K_\varphi' c_{\varphi \varphi}
    + (E^\varphi)'' c_2
    \,.
    \label{eq:Hamiltonian constraint ansatz - Generalized vacuum - Extended}
\end{align}
Here, $a_0$, $a_{xx}$, $a_{\varphi \varphi}$, $a_{x \varphi}$, $a_2$, $b_{\varphi \varphi}$, $b_2$, $c_2$, $c_{\varphi \varphi}$, $c_{x \varphi}$ are all functions of the phase-space variables, but not of their derivatives.
We have not included terms quadratic in the radial derivatives of $K_x$ in (\ref{eq:Hamiltonian constraint ansatz - Generalized vacuum - Extended}) because it can be readily shown that the covariance condition \eqref{eq:Covariance condition} does not allow them if $\tilde{q}_{\vartheta \vartheta} = \tilde{q}_{\vartheta \vartheta} (E^x)$ which is our case of interest.

Demanding that this ansatz satisfies the anomaly-free brackets of hypersurface deformation form  and imposing the additional covariance conditions encoded in \eqref{eq:Covariance condition}, the general Hamiltonian constraint reduces to
\begin{eqnarray}
    \tilde{H}
    &=& - \sqrt{E^x}\; \frac{g}{2} \bigg[ E^\varphi \left( f_0 + \frac{K_x}{E^\varphi} f_1 \right)
    + \frac{((E^x)')^2}{E^\varphi} \left( f_2
    - \frac{K_x}{2E^\varphi}  \left( \frac{\partial \ln g}{\partial K_\varphi}
    + C_{x \varphi} \right) \right)
    \notag\\
    && \hspace{2cm}
    + \frac{(E^x)' (E^\varphi)'}{(E^\varphi)^2}
    - \frac{(E^x)''}{E^\varphi}
    + \frac{(E^x)' K_\varphi'}{E^\varphi} C_{x \varphi}
    \bigg]
    \,.
    \label{eq:Hamiltonian constraint - Anomaly-free - covariant - spherical - vacuum - general - extended}
\end{eqnarray}
Now, $g$, $f_0$, $f_1$, $f_2$, and $C_{x \varphi}$ are functions of $E^x$ and $K_\varphi$, and are related to one another by four differential equations.
Consequently, the structure function $\tilde{q}^{x x}$ is composed of these functions plus an additional direct dependence on the phase-space variables. More explicitly, it is given by
\begin{eqnarray}
    \tilde{q}^{x x} =
    \left(
    \frac{\partial f_1}{\partial K_\varphi}
    - f_1 C_{x \varphi}
    - \frac{1}{2}
    \left(
    \frac{\partial^2 \ln g}{\partial K_\varphi^2}
    + C_{x \varphi}^2
    + \frac{\partial C_{x \varphi}}{\partial K_\varphi}
    + \frac{\partial \ln g}{\partial K_\varphi} C_{x \varphi}
    \right)
    \left( \frac{(E^x)'}{E^\varphi} \right)^2
    \right)
    \frac{g^2}{4}
    \frac{E^x}{(E^\varphi)^2}
    \ .
    \label{eq:Structure function - Anomaly-free - covariant - spherical - vacuum - general - extended}
\end{eqnarray}

The constraint brackets \eqref{eq:Hypersurface deformation algebra - spherical - EMG} and the covariance condition \eqref{eq:Covariance condition}, which can be formulated in terms of Poisson brackets as well, are both invariant under canonical transformations. We can therefore apply canonical transformations in order to further simplify the constraint and eliminate some of the free functions.
An interesting subset of canonical transformations is given by those that preserve the diffeomorphism constraint \eqref{eq:Diffeomorphism constraint - spherical symmetry}:
\begin{align}
    K_\varphi =& f_c (E^x , \tilde{K}_\varphi)
    \ ,
    \hspace{5cm}
    E^\varphi = \tilde{E}^\varphi \left( \frac{\partial f_c}{\partial \tilde{K}_\varphi} \right)^{-1}
    \ ,
    \notag\\
    K_x =& \frac{\partial (\alpha_c^2 E^x)}{\partial E^x} \tilde{K}_x + \tilde{E}^\varphi \frac{\partial f_c}{\partial E^x} \left( \frac{\partial f_c}{\partial \tilde{K}_\varphi} \right)^{-1}
    \ ,
    \hspace{1cm}
    \tilde{E}^x = \alpha_c^2 (E^x) E^x
    \,,
    \label{eq:Diffeomorphism-constraint-preserving canonical transformations - Spherical - general}
\end{align}
where the new variables are written with a tilde.
It can be shown that there is a canonical transformation with a function $f_c = f_c (K_\varphi)$ and $\alpha_c = 1$, such that the transformed $C_{x \varphi}$ vanishes, at least locally in the phase space.

This canonical transformation allows us to include $C_{x \varphi} = 0$ among  the conditions for covariance, resulting in five equations for the five functions in \eqref{eq:Hamiltonian constraint - Anomaly-free - covariant - spherical - vacuum - general - extended}. The equations, which include partial differential ones, can be solved exactly, yielding 
\begin{subequations}
    \label{eq:Covariant Hamiltonian constraint terms}
\begin{equation}
    g = \chi \cos^2 \left( \lambda \left( K_\varphi + \lambda_\varphi \right) \right)
    \ ,
    \label{eq:Covariant global factor - spherical - generalized vacuum}
\end{equation}
\begin{equation}
    g f_1 = 4 \chi \left(c_f \frac{\sin (2 \lambda (K_\varphi + \lambda_\varphi))}{2 \lambda}
    + q \cos(2 \lambda (K_\varphi + \lambda_\varphi))\right)
    \ ,
\end{equation}
\begin{equation}
    f_2 =
    - \frac{\alpha_2}{4 E^x}
    + \frac{\sin \left(2 \lambda ( K_{\varphi} + \lambda_\varphi) \right)}{2 \lambda \cos^2\left(\lambda (K_{\varphi}+ \lambda_\varphi) \right)} \left( \lambda \frac{\partial (\lambda \lambda_\varphi)}{\partial E^x}
    + \lambda K_{\varphi} \frac{\partial \lambda}{\partial E^x}\right)
    \ ,
\end{equation}
\begin{eqnarray}
    g f_0 &=&
    \chi \Bigg(
    \frac{\alpha_0}{E^x}
    + 2 \frac{\sin^2 \left(\lambda  \left(K_{\varphi }+\lambda_\varphi \right)\right)}{\lambda^2}\frac{\partial c_{f}}{\partial E^x}
    + 4 \frac{\sin \left(2 \lambda  \left(K_{\varphi }+\lambda_\varphi \right)\right)}{2 \lambda} \frac{\partial q}{\partial E^x}
    \notag\\
    &&+ 4 c_f \left( \frac{1}{\lambda} \frac{\partial (\lambda \lambda_\varphi)}{\partial E^x} \frac{\sin \left(2 \lambda  \left(K_{\varphi }+\lambda_\varphi \right)\right)}{2 \lambda}
    + \left(\frac{\alpha_2}{4 E^x} - \frac{\partial \ln \lambda}{\partial E^x}\right) \frac{\sin^2 \left(\lambda  \left(K_{\varphi }+\lambda_\varphi \right)\right)}{\lambda^2}\right)
    \notag\\
    &&
    + 8 q \left( - \lambda \frac{\partial (\lambda \lambda_\varphi)}{\partial E^x} \frac{\sin^2 \left( \lambda \left(K_{\varphi }+\lambda_\varphi \right)\right)}{\lambda^2}
    + \left(\frac{\alpha_2}{4 E^x} - \frac{1}{2} \frac{\partial \ln \lambda}{\partial E^x}\right) \frac{\sin \left(2 \lambda  \left(K_{\varphi }+\lambda_\varphi \right)\right)}{2 \lambda} \right)
    \notag\\
    &&+ 4 K_{\varphi} \frac{\partial \ln \lambda}{\partial E^x} \left( c_f \frac{\sin \left(2 \lambda  \left(K_{\varphi }+ \lambda_{\varphi}\right)\right)}{2 \lambda}
    + q \cos \left(2 \lambda  \left(K_{\varphi }+\lambda_{\varphi }\right)\right) \right)
    \Bigg)
    \ ,
\end{eqnarray}
\end{subequations}
where $\chi , c_{f} , \alpha_0, \alpha_2, \lambda, q, \lambda_\varphi$ are undetermined functions of $E^x$. 
The classical constraint is recovered in the limit $\chi , c_{f} , \alpha_0, \alpha_2 \to 1$ and $\lambda, q \to 0$. (The cosmological constant can be recovered by instead setting $\alpha_{0} \to 1 - \Lambda E^x$, with $\Lambda>0$ corresponding to asymptotically de Sitter space in the classical limit.)

Noting that the set of canonical transformations has not been exhausted, we can use the residual canonical transformation with $f_c = f_x K_\varphi - \tilde{\lambda}_\varphi$ and $\alpha_c = 1$ (where $f_x$ and $\tilde{\lambda}_\varphi$ are allowed to depend on $E^x$) to further simplify the constraint.
A convenient choice for such a simplification is given by $\tilde{\lambda}_\varphi = \lambda_\varphi$:
\begin{subequations}
\begin{eqnarray}
    g
    &=&
    \chi f_x \cos^2 \left( \lambda f_x K_\varphi \right)
    \ ,
\end{eqnarray}
\begin{eqnarray}
    g f_0
    &=& \frac{\chi}{f_x} \Bigg( \frac{\alpha_0}{E^x}
    + 2 \frac{\sin^2 \left(\lambda f_x K_\varphi\right)}{\lambda^2}\frac{\partial c_{f}}{\partial E^x}
    + 4 \frac{\sin \left(2 \lambda f_x K_\varphi\right)}{2 \lambda} \frac{1}{\lambda} \frac{\partial \left(\lambda q\right)}{\partial E^x}
    \nonumber\\
    &&+ \left(\frac{\alpha_2}{E^x} - 2 \frac{\partial \ln \lambda^2}{\partial E^x}\right) \left( c_f \frac{\sin^2 \left(\lambda f_x K_\varphi\right)}{\lambda^2}
    + 2 q \frac{\sin \left(2 \lambda f_x K_\varphi\right)}{2 \lambda} \right)
    \Bigg)
    \nonumber\\
    &&+ \frac{\partial \ln (\lambda f_x )}{\partial E^x} 4 \chi_0 K_\varphi \left(c_f \frac{\sin (2 \lambda f_x K_\varphi)}{2 \lambda}
    + q \cos(2 \lambda f_x K_\varphi)\right)
    \ ,
\end{eqnarray}
\begin{eqnarray}
    g f_1 &=&
    4 \chi \left(c_f \frac{\sin (2 \lambda f_x K_\varphi)}{2 \lambda}
    + q \cos(2 \lambda f_x K_\varphi)\right)
    \ ,
\end{eqnarray}
\begin{eqnarray}
    g f_2
    &=& \chi f_x \cos^2 \left( \lambda f_x K_\varphi \right) \Bigg(
    - \frac{\alpha_2}{4 E^x}
    - \frac{\partial \ln f_x}{\partial E^x}
    \nonumber\\
    &&
    + \frac{\sin \left(2 \lambda f_x K_\varphi \right)}{2 \lambda \cos^2\left(\lambda f_x K_\varphi \right)} \lambda^2 f_x K_\varphi \frac{\partial \ln (\lambda f_x)}{\partial E^x}
    \Bigg)
    \ .
\end{eqnarray}
    \label{eq:Covariant Hamiltonian constraint terms - CT moded up to f_x}
\end{subequations}
The associated structure function \eqref{eq:Structure function - Anomaly-free - covariant - spherical - vacuum - general - extended} becomes
\begin{eqnarray}
    \tilde{q}^{x x} &=&
    \left(
    \frac{\partial f_1}{\partial K_\varphi}
    - \frac{1}{2}
    \frac{\partial^2 \ln g}{\partial K_\varphi^2}
    \left( \frac{(E^x)'}{E^\varphi} \right)^2
    \right)
    \frac{g^2}{4}
    \frac{E^x}{(E^\varphi)^2}
    \\
    &=&
    \left(
    \left( c_f
    + \lambda^2 f_x^2 \left( \frac{(E^x)'}{2 E^\varphi} \right)^2
    \right)
    \cos^2 \left( \lambda f_x K_\varphi \right)
    - 2 q \lambda^2 \frac{\sin (2 \lambda f_x K_\varphi)}{2 \lambda}
    \right) \chi^2
    f_x^2 \frac{E^x}{(E^\varphi)^2}\nonumber
    \ .
    \label{eq:Structure function - Anomaly-free - covariant - spherical - vacuum - general - extended - CT moded up to f_x}
\end{eqnarray}

If we choose $f_x=1$, then the resulting set of Hamiltonian constraints is given by
\begin{eqnarray}
    \tilde{H}
    &=& - \sqrt{E^x} \frac{\chi}{2} \bigg[ E^\varphi \bigg( \frac{\alpha_0}{E^x}
    + 2 \frac{\sin^2 \left(\lambda K_\varphi\right)}{\lambda^2}\frac{\partial c_{f}}{\partial E^x}
    + 4 \frac{\sin \left(2 \lambda K_\varphi\right)}{2 \lambda} \frac{1}{\lambda} \frac{\partial \left(\lambda q\right)}{\partial E^x}
    \nonumber\\
    &&+ \left(\frac{\alpha_2}{E^x} - 2 \frac{\partial \ln \lambda^2}{\partial E^x}\right) \left( c_f \frac{\sin^2 \left(\lambda K_\varphi\right)}{\lambda^2}
    + 2 q \frac{\sin \left(2 \lambda K_\varphi\right)}{2 \lambda} \right)
    \nonumber\\
    &&
    + 4 \left(\frac{K_x}{E^\varphi} + \frac{K_\varphi}{2} \frac{\partial \ln \lambda^2}{\partial E^x} \right) \left(c_f \frac{\sin (2 \lambda K_\varphi)}{2 \lambda}
    + q \cos(2 \lambda K_\varphi)\right)
    \bigg)
    \nonumber\\
    &&
    + \frac{((E^x)')^2}{E^\varphi} \bigg(
    - \frac{\alpha_2}{4 E^x} \cos^2 \left( \lambda K_\varphi \right)
    + \left( \frac{K_x}{E^\varphi} + \frac{K_\varphi}{2} \frac{\partial \ln \lambda^2}{\partial E^x} \right) \lambda^2 \frac{\sin \left(2 \lambda K_\varphi \right)}{2 \lambda} \bigg)
    \nonumber\\
    &&
    + \left(\frac{(E^x)' (E^\varphi)'}{(E^\varphi)^2}
    - \frac{(E^x)''}{E^\varphi}\right) \cos^2 \left( \lambda K_\varphi \right)
    \bigg]
    \ ,
    \label{eq:Hamiltonian constraint - modified - non-periodic}
\end{eqnarray}
with the associated structure function 
\begin{eqnarray}
    \tilde{q}^{x x} &=&
    \left(
    \left( c_f
    + \lambda^2 \left( \frac{(E^x)'}{2 E^\varphi} \right)^2
    \right)
    \cos^2 \left( \lambda K_\varphi \right)
    - 2 q \lambda^2 \frac{\sin (2 \lambda K_\varphi)}{2 \lambda}
    \right) \chi^2
    \frac{E^x}{(E^\varphi)^2}
    \ .
    \label{eq:Structure function - modified - non-periodic}
\end{eqnarray}

If we instead choose $f_x = \tilde{\lambda} / \lambda$, where $\tilde{\lambda}$ is a constant reference value of $\lambda(E^x)$, the resulting set of Hamiltonian constraints is given by 
\begin{eqnarray}
    \tilde{H}
    &=& - \frac{\tilde{\lambda}}{\lambda} \chi \frac{\sqrt{E^x}}{2} \Bigg[ E^\varphi \Bigg(
    \frac{\lambda^2}{\tilde{\lambda}^2} \frac{\alpha_0}{E^x}
    + 2 \frac{\sin^2 \left(\tilde{\lambda} K_\varphi\right)}{\tilde{\lambda}^2}\frac{\partial c_{f}}{\partial E^x}
    + 4 \frac{\sin \left(2 \tilde{\lambda} K_\varphi\right)}{2 \tilde{\lambda}} \frac{\partial}{\partial E^x} \left(\frac{\lambda}{\tilde{\lambda}} q\right)
    \nonumber\\
    &&+ \left( \frac{\alpha_2}{E^x} - 2 \frac{\partial \ln \lambda^2}{\partial E^x}\right) \left( c_f \frac{\sin^2 \left(\tilde{\lambda} K_\varphi\right)}{\tilde{\lambda}^2}
    + 2 \frac{\lambda}{\tilde{\lambda}} q \frac{\sin \left(2 \tilde{\lambda} K_\varphi\right)}{2 \tilde{\lambda}} \right)
    \Bigg)
    \nonumber\\
    &&
    + 4 K_x \left(c_f \frac{\sin (2 \tilde{\lambda} K_\varphi)}{2 \tilde{\lambda}}
    + \frac{\lambda}{\tilde{\lambda}} q \cos(2 \tilde{\lambda} K_\varphi)\right)
    \nonumber\\
    && - \frac{((E^x)')^2}{E^\varphi} \left(
    \left( \frac{\alpha_2}{4 E^x} - \frac{1}{2} \frac{\partial \ln \lambda^2}{\partial E^x}\right) \cos^2 \left( \tilde{\lambda} K_\varphi \right)
    - \frac{K_x}{E^\varphi} \tilde{\lambda}^2 \frac{\sin \left(2 \tilde{\lambda} K_\varphi \right)}{2 \tilde{\lambda}}
    \right)
    \nonumber\\
    &&+ \left( \frac{(E^x)' (E^\varphi)'}{(E^\varphi)^2}
    - \frac{(E^x)''}{E^\varphi} \right) \cos^2 \left( \tilde{\lambda} K_\varphi \right)
    \Bigg]
    \ ,
    \label{eq:Hamiltonian constraint - modified - periodic}
\end{eqnarray}
with the structure-function
\begin{eqnarray}
    \tilde{q}^{x x}
    &=&
    \left(
    \left( c_{f}
    + \left(\frac{(E^x)' \tilde{\lambda}}{2 E^\varphi} \right)^2 \right) \cos^2 \left(\tilde{\lambda} K_\varphi\right)
    - 2 \frac{\lambda}{\tilde{\lambda}} q \tilde{\lambda}^2 \frac{\sin \left(2 \tilde{\lambda} K_\varphi\right)}{2 \tilde{\lambda}}\right)
    \frac{\tilde{\lambda}^2}{\lambda^2} \chi^2 \frac{E^x}{(E^\varphi)^2}
    \ .
    \label{eq:Structure function - modified - periodic}
\end{eqnarray}
Note that the constraint (\ref{eq:Hamiltonian constraint - modified - periodic}), unlike (\ref{eq:Hamiltonian constraint - modified - non-periodic}), is bounded in $K_\varphi$ and fully periodic with ``frequency'' $\tilde{\lambda}$.
The two constraints, however, are related by a simple canonical transformation rescaling the component $K_\varphi$. Therefore, they describe the same system, just in different phase-space coordinates. In the modified context, the phase-space variable $K_{\varphi}$ has lost its direct classical relationship with extrinsic curvature, which is instead determined by the emergent space-time metric. Therefore, there is no canonical or geometrical preference in favor of one of the two versions. (However, the periodic nature of (\ref{eq:Hamiltonian constraint - modified - periodic}) may suggest that this version and the associated canonical variables are more relevant for a fundamental loop quantization in terms of basic holonomy operators.)

The classical constraint is recovered from (\ref{eq:Hamiltonian constraint - modified - non-periodic}) in the limit $\chi , c_{f}, \alpha_2 \to 1$, $\lambda \to 0$, and $\alpha_0 \to 1 - \Lambda E^x$ with cosmological constant $\Lambda$.
The constraint (\ref{eq:Hamiltonian constraint - modified - periodic}) cannot reproduce the classical limit directly  because the canonical transformation $K_\varphi\to (\tilde{\lambda}/\lambda) K_\varphi$ does not exist in the classical limit. Since the canonical transformation has introduced a new constant $\tilde{\lambda}$, its limit behavior must be specified separately. One possibility is the combination of $\lambda \to \tilde{\lambda}$ followed by $\tilde{\lambda}\to0$.

If we redefine
\begin{eqnarray}
    \chi \to \bar{\chi} \frac{\lambda}{\tilde{\lambda}}
    \ , \
    q \to \bar{q} \frac{\tilde{\lambda}}{\lambda}
    \ , \
    \alpha_0 \to \frac{\tilde{\lambda}^2}{\lambda^2} \bar{\alpha}_0
    \ , \
    \alpha_2 \to \bar{\alpha}_2 + 4 E^x \frac{\partial \ln \lambda}{\partial E^x}
    \ ,
    \label{eq:Redefinitions of lambda}
\end{eqnarray}
the free function $\lambda$ disappears from the constraint (\ref{eq:Hamiltonian constraint - modified - periodic}) by being absorbed by the other parameters, and is completely replaced by the constant $\tilde{\lambda}$.
A possible interpretation of $\tilde{\lambda}$ is given in the next subsection as the length of a reference curve along which the holonomy is computed.
Replacing the function $\lambda$ with a constant $\tilde{\lambda}$ simplifies the constraint, but it obscures the effects of non-constant $\lambda$ by combining it with the other modification functions according to (\ref{eq:Redefinitions of lambda}). Since we are specifically interested in non-constant $\lambda$ for the purpose of this paper, we will not make much use of this redefinition.

A modified angular metric component $\tilde{q}_{\vartheta \vartheta} = \tilde{q}_{\vartheta \vartheta} (E^x)$ may also be considered.
However, through a residual canonical transformation (\ref{eq:Diffeomorphism-constraint-preserving canonical transformations - Spherical - general}) with $f_c = K_\varphi$ and the appropriate $\alpha_c$, the angular metric component can always be rendered into its classical form $\tilde{q}_{\vartheta \vartheta} = E^x$. The appearance of $\alpha_c$ in the Hamiltonian constraint can thereby be absorbed in the undetermined functions of $E^x$, obtaining the same result (\ref{eq:Hamiltonian constraint - modified - periodic}).
Any covariant Hamiltonian constraint of second order  in spatial derivatives in the spherically symmetric sector is therefore contained in (\ref{eq:Hamiltonian constraint - modified - periodic}).

It is important to note that canonical transformations cannot have an effect on the physical system because they merely represent a change of phase-space coordinates.
The canonical transformations used here, and discussed in detail in \cite{EMGCov}, serve the purpose of simplifying the covariance conditions such that they can be solved exactly.
Factoring out canonical transformations is also important because a different choice of phase-space coordinates may lead one to misclassify a modified Hamiltonian  constraint different from (\ref{eq:Hamiltonian constraint - modified - periodic}) as a new theory, even if it is in fact equivalent to a well-known one.
For instance, \cite{Gambini2022} argued that the non-invertibility of a canonical transformation of the form $K_\varphi \to \sin(\lambda K_\varphi)/\lambda$ applied to the classical system could model holonomy modifications. This assertion is incorrect since the construction merely amounts to using the classical theory with phase-space coordinates that are valid in a bounded range. It is therefore not surprising that the results on critical collapse of scalar matter, reported in \cite{Gambini2022}, did not show any deviation from GR.
Similarly, \cite{alonso2022nonsingular} partially attributed modified gravitational effects to non-invertible canonical transformations, misidentifying the correct reason for singularity removal. The relevant modifications instead originate in a linear combination of constraints with phase-space dependent coefficients which result in additional terms for the new Hamiltonian constraint, as explained in \cite{EMGCov}.

\subsection{Effective loop quantum gravity}\label{sec:Effective loop quantum gravity}

The covariant Hamiltonian defined in (\ref{eq:Hamiltonian constraint -  modified - periodic}) contains various independent modification functions in
several of its terms. How does it relate to effective loop quantum gravity,
which is usually considered with a smaller number of modifications?  Phenomenologically,
the parameter $\lambda$ is special since it is responsible for the
non-singular black-hole solution obtained  for constant $\lambda=\tilde{\lambda}$
in \cite{alonso2022nonsingular}.
In the functional behavior of the constraint, this parameter has an appearance
related to the holonomy length $\mu$ in LQG.

\subsubsection{Traditional ingredients}

The starting point of any loop quantization consists in replacing a direct
quantization of the classical phase space by operators in the holonomy-flux
representation.
In spherically symmetric LQG \cite{bojowald2000symmetry,bojowald2004spherically}, the basic holonomies are given by
\begin{eqnarray}
    h^x_e [K_x] &=& \exp \int_e {\rm d} x\ i K_x
    \,, \label{eq:Radial holonomy}
    \\
    h^\varphi_{v, \mu} [K_\varphi] &=& \exp \int_\mu {\rm d} \vartheta\ i K_\varphi = \exp \left(i \mu K_\varphi (v)\right)
    \,, \label{eq:Angular holonomy}
\end{eqnarray}
where $e$ stands for an arbitrary radial curve of finite coordinate length, while $v$ stands for an arbitrary point in the radial line, and $\mu$ is the coordinate length of an arbitrary longitudinal curve on the 2-sphere intersecting the point $v$.
The explicit integration in the angular holonomy (\ref{eq:Angular holonomy}) is possible due to spherical symmetry, but the radial holonomy integration must remain formal.
Similarly, the fluxes are given by direct integration of the densitized triads $E^x$ and $E^\varphi$ over finite, 2-dimensional surfaces with normals in the radial and angular direction, respectively. Fluxes do not require exponentiation and are therefore closer to the original phase-space variables.
For the following argument, we need only the expressions of holonomies.

Because loop quantization is based on the holonomy-flux variables, the Hamiltonian constraint must be rewritten as a function of holonomies rather than the bare (extrinsic) curvatures $K_x$ and $K_\varphi$, leading to a modified constraint in the spirit of emergent modified gravity, as outlined in the previous section.
However, the radial holonomy (\ref{eq:Radial holonomy}) is essentially non-local, and cannot be studied within this framework, which has  thus far only been formulated locally and up to second order in derivatives.
On the other hand, the angular holonomies (\ref{eq:Angular holonomy}) are indeed local when restricting the state space to spherically symmetric states.
Therefore, it is natural to expect that quantum effects due to the angular holonomy components can be captured by the local equations of emergent modified gravity.
Furthermore, since the angular holonomies can be integrated, they become simple complex exponentials of the angular extrinsic curvature (\ref{eq:Angular holonomy}).
Because the Hamiltonian constraint must be Hermitian as an operator in the quantum theory, or simply real-valued prior to quantization, the holonomy modifications in the Hamiltonian constraint will appear as trigonometric functions of $K_\varphi$ with `frequency' $\mu$. 
This is indeed the case of the Hamiltonian constraint (\ref{eq:Hamiltonian constraint - modified - periodic}) derived in the previous section.
Therefore, the parameter $\lambda$ in the notation of emergent modified gravity can be given the interpretation of an angular holonomy length within LQG,
and hence acquires the status of a quantum parameter under this
interpretation. However, it also inherits new features from the enhanced symmetry
properties of emergent modified gravity compared with traditional models of
LQG.

In LQG, different triangulations of space are allowed, and a specific choice would determine the form of the quantum parameter $\mu (E^x)$ appearing in the holonomy modification function, using the phase-space variable $E^x$ as the squared radius of our geometry.
For example, we can choose a fine lattice such that the spheres are triangulated by small regular polygons with $n$ sides of angular length $\mu$. (It is common to choose tetrahedra for the three-dimensional lattice and hence triangles for the two-dimensional lattice of the spheres, setting $n=3$ for this purpose.)
Each polygon, also referred to as a plaquette, at radius $\sqrt{E^x}$ then covers a geometrical area of $n E^x \mu^2 / (4 \tan (\pi/n))$.
Choosing the plaquettes at different radii to preserve the same geometrical area, we obtain the formula
\begin{equation} \label{mur}
    \mu = \frac{\hat{r}}{\sqrt{E^x}} \hat{\mu}
    \ ,
\end{equation}
where $\hat{r}$ is a constant reference radius at which $\mu = \hat{\mu}$.
This function satisfies $\mu \to 0$ as $E^x \to \infty$, as desired for recovering an asymptotically smooth geometry without triangulation effects at large radii. 

\subsubsection{Shortcomings}

Several assumptions are used in arriving at this form. An additional, previously unrecognized issue is that the classical component $E^x$ of a space-time metric is used in the geometrical construction, which in general may deviate from the correct geometrical meaning provided by the emergent space-time metric. As we have seen, in spherically symmetric models it is always possible to choose phase-space variables such that the angular component of the metric is still given by $E^x$. The construction of holonomy length may therefore be applied as described here. However, strict proposals such as the $\bar{\mu}$-scheme would require an application of these arguments to any surface, even if it extends into the radial direction which may still be homogeneous in the black-hole interior. In this case, an emergent space-time metric may have to be taken into account, but this is available only after the modified theory has been analyzed. In general, the $\bar{\mu}$-scheme therefore does not provide a self-contained recipe for the construction of holonomy modifications.

One ingredient of a $\bar{\mu}$-type scheme that may be applied here is a relationship between holonomy parameters and areas of interest, such as the smallest non-zero eigenvalue of an area operator in a loop quantization. Also this step is heuristic and not fully justified because holonomies of a gauge connection do not depend on a metric, and therefore need not have any relationship with geometrical areas.
Nevertheless, such a relationship may be used in order to fix free parameters. (In practice, using area eigenvalues merely determines numerical factors multiplying suitable powers of the Planck length, whose form could easily be obtained from dimensional arguments.)
To be specific, by choosing the plaquette size derived above to equal a certain distinguished area, $n \hat{r}^2 \hat{\mu}^2 / (4 \tan (\pi/n)) = \Delta$, we arrive at the relation
\begin{equation}
    \mu (E^x)^2 = \tilde{\Delta}/E^x
 \label{eq:mu-scheme holonomy function}
\end{equation}
with $\tilde{\Delta}=4n^{-1} \tan (\pi/n) \Delta$.
The square of the quantum parameter (\ref{eq:mu-scheme holonomy function}) is proportional to the solid angle of a single plaquette on a sphere of radius $\sqrt{E^x}$. Since this expression equals $\tilde{\Delta}/E^x$,
the number of plaquettes triangulating the sphere of radius $\sqrt{E^x}$ is given by
\begin{equation}
    \mathcal{N}_{\Delta} (E^x) = 4 \pi E^x/\tilde{\Delta}
    \ .
    \label{eq:Number of plaquettes on sphere - mu-scheme}
\end{equation}

Loop quantization requires the Hamiltonian constraint to be explicitly periodic in $K_\varphi$. Having $\mu$ depend on $E^x$ preserves periodicity in $K_{\varphi}$ at any fixed value of $E^x$, but it turns the holonomy $\exp (i \mu (\hat{E}^x) \hat{K}_\varphi)$ into a holonomy-flux hybrid operator upon quantization. Such expressions do not obey the basic holonomy-flux algebra, and they do not even provide a closed set of commutators because an expression such as
\begin{eqnarray}
    \left\{ h^x_e [K_x] , h^\varphi_{v, \mu} [K_\varphi] \right\}
    = - h^x_{e_1} [K_x] \frac{\partial \mu (\nu)}{\partial E^x (\nu)} K_\varphi (v) h^\varphi_{v, \mu} [K_\varphi] h^x_{e_2} [K_x]
    \,, \label{hh}
\end{eqnarray}
where $e = e_1 \cup e_2$, such that $e_1$ and $e_2$ join at $x=\nu$, violates the periodicity condition.
In the basic holonmy-flux algebra,  the right-hand side of the Poisson bracket between two holonomy operators should be zero since the connection components Poisson commute.  The factor of $K_{\varphi}(v)$ in (\ref{hh}) shows that trying to implement an $E^x$-dependent holonomy length is not compatible with a closed algebra of basic operators.
Hence, the periodic version (\ref{eq:Hamiltonian constraint - modified - periodic}) of the Hamiltonian constraint is preferred over (\ref{eq:Hamiltonian constraint - modified - non-periodic}) when carrying out the loop quantization because it resolves two important problems: It depends on pure holonomies of the form $\exp (i \tilde{\lambda} \hat{K}_\varphi)$ with constant $\tilde{\lambda}$, and can therefore be expressed in terms of the generators of a closed basic algebra. At the same time, it allows for scale-dependent $\lambda$-effects because a non-trivial $\lambda(E^x)$ appears in the coefficients of (\ref{eq:Hamiltonian constraint - modified - periodic}) after it has been removed as a factor of $K_{\varphi}$ in holonomies by a canonical transformation.
It is then possible to choose the holonomy length that appears in trigonometric functions as being equal to the $E^x$-independent reference value $\tilde{\lambda}=\hat{\mu}$. At the same time, dynamical implications of non-constant holonomy parameters are realized through modified coefficients of holonomy terms, as we demonstrate in detail below.

A common motivation for an $E^x$-dependent $\mu$ in models of loop quantum
gravity is not based on properties of basic holonomy operators, but rather on
phenomenological questions because a connection or extrinsic-curvature
component such as $K_{\varphi}$ may be large in classical regimes, depending
on the space-time slicing. A holonomy length $\mu(E^x)$ might then be used to
counteract growing holonomy modifications if it decreases at a suitable rate
as $E^x$ grows, implying large spherical areas. It is often argued that this
reason requires an explicit appearance of a non-constant $\mu(E^x)$ in
holonomies. This reasoning seems to favor (\ref{eq:Hamiltonian constraint - modified -
  non-periodic}) with a corresponding $\lambda(E^x)$ in holonomy-like terms, even though covariance
prevents this constraint from being strictly periodic in $K_{\varphi}$. Moreover, a
quantum constraint operator for this version cannot be based on basic
holonomies in a closed holonomy-flux algebra.

\subsubsection{New features}

Our new discussion of canonical transformations overturns this established understanding because it is possible to map any constraint of the form (\ref{eq:Hamiltonian constraint - modified - non-periodic}) into an expression (\ref{eq:Hamiltonian constraint - modified - periodic}) that is strictly periodic in $K_{\varphi}$. The canonical transformation implies that both versions generate exactly the same dynamics and physical observables. Effects of a non-constant holonomy length in (\ref{eq:Hamiltonian constraint - modified - non-periodic}) are realized in (\ref{eq:Hamiltonian constraint - modified - periodic}) by a modified $E^x$ dependence of coefficients of holonomy terms, indicated by the non-constant ratios $\tilde{\lambda}/\lambda$ in (\ref{eq:Hamiltonian constraint - modified - periodic}), in particular in the $K_x$-term of the modified Hamiltonian constraint. This term is relevant for Hamilton's equation 
\begin{equation} \label{Exdot}
    \dot{E}^x=\frac{\tilde{\lambda}}{\lambda} \chi \frac{\sqrt{E^x}}{2} \left(4c_f+\tilde{\lambda}^2 \frac{((E^x)')^2}{(E^{\varphi})^2}\right)  \frac{\sin(2\tilde{\lambda}K_{\varphi})}{2\tilde{\lambda}}
  \end{equation}
for the spherical area $E^x$  (assuming vanishing $N^x$ for the sake of simplicity).
As an equation of motion generated by the periodic constraint, this result implies that the classical relationship between the canonical variable $K_{\varphi}$ and the time derivative $\dot{E}^x$
receives an additional factor of $\lambda/\tilde{\lambda}$, making $K_{\varphi}$ decrease as $\sqrt{E^x}$ grows according to the modified dynamics. (For small $\tilde{\lambda}$ and $\lambda/\tilde{\lambda}$ independent of $\tilde{\lambda}$, we have $K_{\varphi}\approx (\lambda/\tilde{\lambda} )\dot{\sqrt{E^x}}/(\chi c_f)$, where $c_f,\chi \rightarrow 1$ in the classical limit.) This behavior has the same effect of suppressing holonomy modifications in classical regimes as an explicitly $E^x$-dependent holonomy length would imply. It is only harder to see this outcome immediately because it requires an analysis of equations of motion. 

By dismissing this option, the traditional arguments for holonomy modifications suggest an unjustified preference for one of the possible versions of canonically equivalent constraints, based on  flawed heuristics that implicitly relate the canonical variable $K_{\varphi}$ in a modified theory to a classical component of extrinsic curvature: Classically, $K_{\varphi}$ is not necessarily small enough  for $\sin(\tilde{\lambda} K_{\varphi})/\tilde{\lambda}$ to be close to $K_{\varphi}$ in  regimes of small space-time curvature. 
It is therefore important to distinguish between small-$\lambda$ regimes (as a formal limit of the constraint) and small-$\lambda K_\varphi$ regimes (as a dynamical feature of solutions for $K_{\varphi}$). In a modified theory, one cannot use classical heuristics in order to estimate the magnitude of $K_{\varphi}$, interpreted as a component of extrinsic curvature, in order to determine whether $\lambda K_{\varphi}$ is sufficiently small in a given semiclassical regime. Since $\lambda K_{\varphi}$ appears in a modified term of the Hamiltonian constraint when holonomies are used and since the emergent space-time metric is not the same as the classical metric, there is no kinematical relationship between the magnitude of $K_{\varphi}$ in a modified theory and the magnitude of extrinsic curvature in a corresponding classical slicing. An evaluation of suitable holonomy modifications can therefore be made only if equations of motion as well as the emergent metric are considered. However, traditional arguments in favor of certain holonomy modifications are held entirely at a kinematical level.

The distinction between small-$\lambda$ and small-$\lambda K_\varphi$ is relevant in these considerations even if one is interested only in leading-order corrections to the classical behavior. The leading order of a small-$\lambda$ expansion of the theory is simply the limit of $\lambda \to 0$, resulting in the classical dependence of the constraint on $K_{\varphi}$ and an emergent metric identical with the classical metric. (The overall theory may still be modified in this limit if the remaining modification functions are non-classical.) Holonomy terms can also be expanded on a subspace of the solution space where $\lambda K_{\varphi}$ is small, even if $\lambda$ is not taken to zero in a limit. This expansion is relevant for physical properties of the semiclassical limit. 
In the low-curvature regime of small $\lambda K_{\varphi}$,  the quadratic order in this expansion determines near-classical physics. The non-periodic version of the Hamiltonian constraint, (\ref{eq:Hamiltonian constraint - modified - non-periodic}), then becomes
\begin{eqnarray}
    \tilde{H}
    &\approx& - \sqrt{E^x} \frac{\chi}{2} \bigg[ E^\varphi \bigg( \frac{\alpha_0}{E^x}
    + 2 K_\varphi^2\frac{\partial c_{f}}{\partial E^x}
    + 4 K_\varphi \frac{1}{\lambda} \frac{\partial \left(\lambda q\right)}{\partial E^x}
    \nonumber\\
    &&+ \left(\frac{\alpha_2}{E^x} - 2 \frac{\partial \ln \lambda^2}{\partial E^x}\right) \left( c_f K_\varphi^2
    + 2 q K_\varphi \right)
    \nonumber\\
    &&
    + 4 \left(\frac{K_x}{E^\varphi} + \frac{K_\varphi}{2} \frac{\partial \ln \lambda^2}{\partial E^x} \right) \left(c_f K_\varphi
    + q \left(1-2 \lambda^2 K_\varphi^2\right)\right)
    \bigg)
    \nonumber\\
    &&
    + \frac{((E^x)')^2}{E^\varphi} \bigg(
    - \frac{\alpha_2}{4 E^x} \left( 1-\lambda^2 K_\varphi^2 \right)
    + \left( \frac{K_x}{E^\varphi} + \frac{K_\varphi}{2} \frac{\partial \ln \lambda^2}{\partial E^x} \right) \lambda^2 K_\varphi \bigg)
    \nonumber\\
    &&
    + \left(\frac{(E^x)' (E^\varphi)'}{(E^\varphi)^2}
    - \frac{(E^x)''}{E^\varphi}\right) \left( 1-\lambda^2 K_\varphi^2 \right)
    \bigg]
    \,,
    \label{eq:Hamiltonian constraint - modified - non-periodic - low curvature}
\end{eqnarray}
with structure function 
\begin{eqnarray}
    \tilde{q}^{x x} &\approx&
    \left(
    \left( c_f
    + \lambda^2 \left( \frac{(E^x)'}{2 E^\varphi} \right)^2
    \right)
    \left( 1-\lambda^2 K_\varphi^2 \right)
    - 2 q \lambda^2 K_\varphi
    \right) \chi^2
    \frac{E^x}{(E^\varphi)^2}
    \,.
    \label{eq:Structure function - modified - non-periodic - low curvature}
\end{eqnarray}
(A similar result follows for the periodic version.)
Some holonomy effects therefore survive in the low-curvature regime and may lead to unexpected dynamical implications, unless the strict limit of $\lambda\to0$ is taken in this regime too.
This approximation of the Hamiltonian constraint is valid only in slices where $\lambda K_\varphi$ is sufficiently small. Physical implications can only be analyzed if equations of motion are studied, revealing through $\tilde{q}^{xx}$ the geometrical meaning and dynamical behavior of $K_{\varphi}$ in suitable semiclassical regimes.  

An important additional flaw in several discussions of holonomy modifications is the lack of covariance. It is then possible for $\sin(\tilde{\lambda} K_{\varphi})/\tilde{\lambda}$ to be close to $K_{\varphi}$ in some space-time slicings but deviate strongly in others, even in classical regimes. This problem is solved by our constructions, which explicitly realize slicing independence together with the correct geometrical meaning of $K_{\varphi}$. As already emphasized, covariance is also the main reason why non-periodic terms are required in a Hamiltonian constraint with $E^x$-dependent holonomy length. Our constructions not only implement full covariance, they also relate different versions of holonomy modifications by carefully considering canonical transformations.  For effective equations and their solutions, we may use either of the two, or perhaps others related by further canonical transformations. One may simply prefer an option that generates the simplest equations of motion for a given purpose which, however, will in general be gauge dependent.

Covariant holonomy modifications therefore cannot be expressed in the
straitjacket of $\mu_0$ and $\bar{\mu}$-like schemes that have only a
heuristic basis motivated by expectation about the holonomy length combined
with a classical understanding of canonical variables in a modified
theory. The appearance of holonomy terms, classified in this way, depends on
the canonical variables used, and a $\mu_0$-like scheme can be transformed
into an equivalent $\bar{\mu}$-like scheme if only the arguments of
trigonometric functions are used in the definition of holonomy
modifications. As an important implication of our new $\lambda$-scheme,
dynamical properties of holonomy modifications and their suppression in
classical regimes can be realized with strictly periodic holonomy terms,
provided their coefficients are amended by suitable modification
functions. (In traditional models of loop quantum gravity, such modifications
are usually understood as inverse-triad corrections because these coefficients
classically depend only on the densitized triad and its inverse but not on the
momenta.)

\subsubsection{Self-consistent treatment of holonomy effects}

Taking these lessons into account, we arrive at a novel understanding of
holonomy modifications in models of loop quantum gravity. The freedom of
performing canonical transformations can be fixed by requiring that all terms
in the Hamiltonian constraint depending on $K_{\varphi}$ are strictly periodic
in this variable. According to emergent modified gravity, covariance then
implies that only the combination $\tilde{\lambda}K_{\varphi}$ with a constant
$\tilde{\lambda}$ can appear in these periodic functions. At this point, the
choice of canonical variables is unique if it is required to preserve the
classical limit for $\tilde{\lambda}\to0$, such that $\tilde{\lambda}K_{\varphi}$
cannot be used as a new phase-space variable conjugate to
$E^{\varphi}/\tilde{\lambda}$.  There is therefore an unambiguous interpretation
of $K_{\varphi}$-dependent functions as an effective description of angular
holonomies. Accordingly, we refer to the constant $\tilde{\lambda}$ as the
holonomy length of the theory. By definition, this length is always constant
and does not depend on $E^x$.

Different schemes are implemented by choosing specific functions for
$\lambda(E^x)$, which does not appear in holonomy terms but rather in
their coefficients in the Hamiltonian constraint. The choice of $\lambda(E^x)$
changes the geometrical meaning and the dynamical behavior of $K_{\varphi}$ in
holonomy terms according to (\ref{Exdot}). As already mentioned, this
modification makes it difficult to assess the significance of holonomy
modifications in different regimes because this task would require an estimate
of the values taken by $K_{\varphi}$, which depends on the dynamics and the
space-time gauge. In fact, the equation of motion for $E^x$ in a modified
theory does not directly determine $K_{\varphi}$ but rather the holonomy
function
\begin{equation} \label{Kphi}
 \frac{\sin(2\tilde{\lambda}K_{\varphi})}{2\tilde{\lambda}}=
 \frac{\lambda}{\tilde{\lambda}}\frac{2\dot{E}^x}{\chi\sqrt{E^x}}
\frac{1}{4c_f+\tilde{\lambda}^2 ((E^x)')^2/(E^{\varphi})^2}  
\end{equation}
obtained by inverting (\ref{Exdot}) and still asuming vanishing $N^x$ for the
sake of convenience. For $\chi$ and $c_f$ close to their classical values, the
ratio $\lambda/\tilde{\lambda}$ determines the strength of holonomy
modifications, defined as the deviation of
$(2\tilde{\lambda})^{-1}\sin(2\tilde{\lambda}K_{\varphi})$ from the classical
expression $2\dot{E}^x/\sqrt{E^x}$ of $K_{\varphi}$. For the classical limit
to be realized at $\tilde{\lambda}\to0$, we need
$\lambda(E^x)=\tilde{\lambda}h(E^x)$ with a function $h(E^x)$ independent of
$\tilde{\lambda}$ (or finite and non-zero for $\tilde{\lambda}\to0$). Since
this new function $h(E^x)=\lambda(E^x)/\tilde{\lambda}$ describes the relative
strength of holonomy modifications on different scales, we call it the
holonomy function. Its appearance shows that it is independent of the holonomy
length and instead describes the dynamical behavior of holonomy modifications
or the variable $K_{\varphi}$.

Some choices of the holonomy function $h(E^x)$ resemble properties of the
traditional $\mu_0$ or $\bar{\mu}$-type schemes, but in some cases the usual
heuristics turns out to be misleading. For instance, a constant $h(E^x)=1$, or
$\lambda=\tilde{\lambda}$, could be related to a $\mu_0$-type scheme of
constant holonomy parameter $\mu$ in the traditional formulation. A common
argument against using such a constant is that the interpretation of
$\tilde{\lambda}$ as a holonomy length derived from coordinates would imply a
growing physical length $\tilde{\lambda}\sqrt{E^x}$ as measured by the relevant spatial geometry. The growth in $E^x$ of this function then suggests that holonomy
modifications increase in asymptotic regimes of large $E^x$, which are often
expected to behave in a classical manner. However, equation~(\ref{Kphi}) shows
that holonomy modifications do not grow in this case but merely track the
classical behavior of $K_{\varphi}$: The second term in the denominator on the right-hand side of \eqref{Kphi} is sub-dominant in this regime, and thus the expression $(2\tilde{\lambda})^{-1}\sin(2\tilde{\lambda}K_{\varphi})$ on the left is identical with the classical behavior $2\dot{E}^x/\sqrt{E^x}$ of $K_{\varphi}$ for $\tilde{\lambda} = \lambda$. Modifications then appear only because one has to rewrite $\sin(2\tilde{\lambda}K_{\varphi})$ in terms of $\sin(\tilde{\lambda}K_{\varphi})$ for some terms of the constraint, using trigonometric identities. Holonomy modifications therefore do not grow in an
unbounded manner in the case of a constant holonomy length, but they may not
decrease quickly enough so as to become subdominant in asymptotic
regimes. This behavior can be changed by a holonomy function $h(E^x)$ that
decreases at a sufficient rate for growing $E^x$. We will see examples in our
specific cases to be discussed later.

\subsection{Gravitational observable}

Consider a scalar function on the phase space $\mathcal{M}$.
This function is a weak observable if
$\delta_\epsilon \mathcal{M} = \mathcal{D}_H H + \mathcal{D}_x H_x$, where
$\mathcal{D}_H$ and $\mathcal{D}_x$ are functions on the phase space and
$\epsilon$ is a gauge function. On-shell, the function $\mathcal{M}$ then
remains invariant under time evolution or arbitrary gauge transformations. In
this section, we will keep the most general form of our modification functions
in the Hamiltonian constraint, before restricting them to take values as
required in effective LQG in the next section.
If one considers the dependence $\mathcal{M} = \mathcal{M} (E^x , K_\varphi ,
(E^x)'/E^\varphi)$, using the periodic constraint (\ref{eq:Hamiltonian
  constraint - modified - periodic}), the weak observability condition
$\delta_\epsilon \mathcal{M} = \mathcal{D}_H H + \mathcal{D}_x H_x$ determines
the functional dependence \cite{EMGscalar}
\begin{eqnarray}\label{eq:General weak observable}
    \mathcal{M}
    &=&
    d_0
    + \frac{d_2}{2} \left(\exp \int {\rm d} E^x \ \frac{\bar{\alpha}_2}{2 E^x}\right)
    \left(
    c_f \frac{\sin^2\left(\tilde{\lambda} K_{\varphi}\right)}{\tilde{\lambda}^2}
    + 2 \bar{q} \frac{\sin \left(2 \tilde{\lambda}  K_{\varphi}\right)}{2 \tilde{\lambda}}
    - \cos^2 (\tilde{\lambda} K_\varphi) \left(\frac{(E^x)'}{2 E^\varphi}\right)^2
    \right)
    \notag\\
    &&
    + \frac{d_2}{4} \int {\rm d} E^x \ \left( \frac{\bar{\alpha}_0}{E^x} \exp \int {\rm d} E^x \ \frac{\bar{\alpha}_2}{2 E^x}\right)
    \ ,
\end{eqnarray}
which is unique up to the constants $d_0$ and $d_2$ with classical limits
$d_0\to0$ and $d_2\to1$. (There exists an anomaly-free and
covariant constraint  more general than (\ref{eq:Hamiltonian constraint - modified -
  periodic}), but after imposing the existence of a weak observable, the extra
terms are restricted to vanish and we recover (\ref{eq:Hamiltonian constraint
  - modified - periodic}) with (\ref{eq:General weak observable}) as the
unique weak observable \cite{EMGscalar}.)  Here we use the barred functions
as defined by (\ref{eq:Redefinitions of lambda}) for the sake of conciseness.

If we insert the classical values $\alpha_2=1$, and $\alpha_0 = 1 -\Lambda E^x$, and $d_0 = 0$, $d_2=1$ the weak observable simplifies to
\begin{eqnarray}
    \mathcal{M} =
    \frac{\sqrt{E^x}}{2}
    \left(
    1 - \frac{\Lambda}{3} E^x
    + \frac{\tilde{\lambda}^2}{\lambda^2} \left( c_f \frac{\sin^2\left(\tilde{\lambda} K_{\varphi}\right)}{\tilde{\lambda}^2}
    + 2 \frac{\lambda}{\tilde{\lambda}} q \frac{\sin \left(2 \tilde{\lambda}  K_{\varphi}\right)}{2 \tilde{\lambda}}
    - \cos^2 (\tilde{\lambda} K_\varphi) \left(\frac{(E^x)'}{2 E^\varphi}\right)^2 \right)
    \right)
    \label{eq:Weak observable in simple case}
\end{eqnarray}
where we have chosen the integration constant $\tilde{\lambda}^2$ for the exponential integrals.
As already seen for the constraint, we must be careful in taking the classical limit.
If we first invert the canonical transformation to the non-periodic variables, the classical limit is given by $\lambda\to0$. If this inversion is not performed, we must first take the limit $\lambda\to \tilde{\lambda}$, followed by $\tilde{\lambda}\to0$.
The classical limit of this observable is then the classical mass observable.
The structure function (\ref{eq:Structure function - modified - periodic}) can be written in terms of this observable as
\begin{eqnarray}
    \tilde{q}^{xx}
    =
    \left(c_f + \lambda^2 \left(1 - \frac{2 \mathcal{M}}{\sqrt{E^x}}
    - \frac{\Lambda}{3} E^x \right) \right) \frac{\tilde{\lambda}^2}{\lambda^2} \chi^2 \frac{E^x}{(E^\varphi)^2}
    \ .
    \label{eq:Structure function - modified - in terms of m}
\end{eqnarray}

In the non-periodic version, the mass observable takes the form
\begin{eqnarray}
    \mathcal{M} =
    \frac{\sqrt{E^x}}{2}
    \left(
    1 - \frac{\Lambda}{3} E^x
    + c_f \frac{\sin^2\left(\lambda K_{\varphi}\right)}{\lambda^2}
    + 2 q \frac{\sin \left(2 \lambda  K_{\varphi}\right)}{2 \lambda}
    - \cos^2 (\lambda K_\varphi) \left(\frac{(E^x)'}{2 E^\varphi}\right)^2 \right)
    \,,
    \label{eq:Weak observable in simple case - nonperiodic}
\end{eqnarray}
and the structure-function is given by
\begin{eqnarray}
    \tilde{q}^{xx}
    =
    \left(c_f + \lambda^2 \left(1 - \frac{2 \mathcal{M}}{\sqrt{E^x}}
    - \frac{\Lambda}{3} E^x \right) \right) \chi^2 \frac{E^x}{(E^\varphi)^2}
    \ .
    \label{eq:Structure function - modified - in terms of m - nonperiodic}
\end{eqnarray}
Note that the mass observable is periodic in $K_{\varphi}$ even in the
non-periodic version of the constraint.

\subsection{Reflection symmetry of the transition surface}

The Hamiltonian constraint (\ref{eq:Hamiltonian constraint - modified -
  periodic}), besides being periodic in $K_\varphi$ has another interesting
symmetry in the special case of $q=0$.
In this case, the structure function is symmetric around the surface
$K_\varphi = \pi / (2 \tilde{\lambda})$ in the sense that $\tilde{q}^{x x}$ is
invariant under the reflection $\delta \to - \delta$ where $\delta$ is defined
by $K_\varphi = \pm \pi / (2 \tilde{\lambda}) + \delta$. The Hamiltonian
constraint is invariant under the combined operation $\delta \to - \delta$,
$K_x \to - K_x$.
Thus, the two conditions $K_\varphi = \pm \pi / (2 \tilde{\lambda})$ and $K_x
= 0$ define a surface of reflection symmetry in the phase-space. This property
can manifest itself in dynamical solutions as a reflection surface in space-time \cite{EMGPF}.
If $q\neq 0$, this reflection-symmetry is broken unless $q$, like $K_x$, changes sign in the dynamical solution, but this is unlikely to be the case in general.
Physical implications of breaking the reflection-symmetry breaking effects of $q\neq0$ will be studied elsewhere
\cite{Axion}.
In the following we set $q=0$, such that reflection symmetry is realized.

It will be useful in the following sections to evaluate the mass observable
(\ref{eq:Weak observable in simple case}) at the reflection symmetry surface
$K_\varphi=-\pi/(2 \tilde{\lambda})$, or equivalently its incarnation in the
non-periodic variables (\ref{eq:Weak observable in simple case - nonperiodic})
at $K_\varphi=-\pi/(2 \lambda)$. After some simplifications, we find that $E^x$, for a given value of the observable ${\cal M}$ obeys the equation 
\begin{equation}
    c_f + \lambda^2 \left( 1 - \frac{2 \mathcal{M}}{\sqrt{E^x}}
    - \frac{\Lambda}{3} E^x \right)
    = 0
    \label{eq:Weak observable at reflection}
\end{equation}
in any region of space-time where $K_{\varphi}=-\pi/(2\lambda)$.
Since $\lambda$ is an arbitrary function of $E^x$, this equation may have
multiple solutions according to the chosen modification function. Multiple
reflection surfaces may therefore exist in a space-time solution. As will be clear from dynamical solutions
derived later on, this reflection symmetry happens around the hypersurface in
space-time where the areal radius $\sqrt{E^x}$ reaches an extremum. For the specific
solution $x_{\lambda}^{(-)}$, which we will use to denote the minimal value of
$\sqrt{E^x}$ according to the dynamics defined by a given $\lambda$, this will
correspond to a spacelike transition surface connecting a black hole to a white hole
through a wormhole interior. In later sections, we will discuss the physical interpretation of
these extremal values, including the case where there exists a maximal value
at $x_\lambda^{(+)}$.

\section{Black-hole solution for general triangulation schemes}
\label{Sec:Black hole solution}

We continue to work with a general functional dependence of
$\lambda = \lambda(E^x)$ and now demonstrate that, qualitatively,
singularity resolution of black holes in models of loop quantum gravity based on holonomy
modifications is a triangulation-independent feature. In order to isolate
effects of the holonomy modification $\lambda$, we set the functions
$c_{f}$, $\alpha_0$, $\alpha_2$, and $q$ to their classical values, while keeping a
nonclassical, nontrivial $\chi(E^x)$ for rescaling freedom.  For now, we keep
a cosmological constant $\Lambda$ and only set it to zero when we analyze
suitable conditions for asymptotic flatness. As will become clear later on, it
is necessary to scale the emergent metric using a suitable $\chi$ if we
are interested in asymptotic flatness, depending on the chosen $\lambda$. For instance, a
constant $\lambda=\tilde{\lambda}$ requires $\chi(\infty)=1/\sqrt{1+\tilde{\lambda}^2}$,
while $\lambda=\sqrt{\Delta/E^x}$ requires $\chi(\infty)=1$.  The scaling of
the spatial metric by a constant $\chi$ is not a simple coordinate
transformation of the radial coordinate because the latter would change other
components non-trivially if there is space-time curvature.

Given the classical choice for most of the modification functions, the
non-periodic constraint (\ref{eq:Hamiltonian constraint - modified -
  non-periodic}) simplifies to
\begin{eqnarray}
    \tilde{H} &=&
    - \chi \frac{\sqrt{E^x}}{2} \Bigg[
    E^\varphi \bigg( \frac{1}{E^x} - \Lambda + \frac{1}{E^x} \frac{\sin^2 ( \lambda K_\varphi)}{\lambda^2}
    + 4 \left( K_\varphi \frac{\sin (2 \lambda K_\varphi)}{2 \lambda} - \frac{\sin^2 ( \lambda K_\varphi)}{\lambda^2} \right) \frac{\partial \ln \lambda}{\partial E^x} \bigg)
    \notag\\
    &&
    + 4 K_x \frac{\sin (2 \lambda K_\varphi)}{2 \lambda}
    - \frac{((E^x)')^2}{4 E^\varphi} \left( \frac{1}{E^x} \cos^2 (\lambda K_\varphi)
    - 4 \lambda^2 \left( \frac{K_x}{E^\varphi} + K_\varphi \frac{\partial \ln \lambda}{\partial E^x} \right) \frac{\sin (2 \lambda K_\varphi)}{2 \lambda} \right)
    \notag\\
    &&
    + \cos^2 (\lambda K_\varphi) \left( \frac{(E^x)' (E^\varphi)'}{(E^\varphi)^2} - \frac{(E^x)''}{E^\varphi} \right)
    \Bigg]
    \ .
    \label{eq:Modified constraint - non-periodic version - simple}
\end{eqnarray}
with structure-function
\begin{eqnarray}
    \tilde{q}^{x x} &=&
    \left( 1
    + \lambda^2 \left( \frac{(E^x)'}{2 E^\varphi} \right)^2
    \right)
    \cos^2 \left( \lambda K_\varphi \right) \chi^2 \frac{E^x}{(E^\varphi)^2}
    \nonumber\\
    &=&
    \left(1 + \lambda^2 \left(1 - \frac{2 \mathcal{M}}{\sqrt{E^x}}
    - \frac{\Lambda}{3} E^x \right) \right) \chi^2 \frac{E^x}{(E^\varphi)^2}
    \ .
    \label{eq:Modified structure function - non-periodic version - simple}
\end{eqnarray}
For comparison, the periodic version (\ref{eq:Hamiltonian constraint - modified - periodic}) becomes
\begin{eqnarray}
    \tilde{H}
    &=& - \chi \frac{\tilde{\lambda}}{\lambda} \frac{\sqrt{E^x}}{2} \bigg[ E^\varphi \left( \frac{\lambda^2}{\tilde{\lambda}^2} \left( \frac{1}{E^x} - \Lambda \right)
    + 4 \left(\frac{1}{4 E^x} - \frac{\partial \ln \lambda}{\partial E^x}\right) \frac{\sin^2 \left(\tilde{\lambda} K_\varphi\right)}{\tilde{\lambda}^2}
    \right)
    + 4 K_x \frac{\sin (2 \tilde{\lambda} K_\varphi)}{2 \tilde{\lambda}}
    \nonumber\\
    &&
    + \frac{((E^x)')^2}{E^\varphi} \left( \cos^2 \left( \tilde{\lambda} K_\varphi \right) \left(
    - \frac{1}{4 E^x}
    + \frac{\partial \ln \lambda}{\partial E^x}
    \right)
    + \tilde{\lambda}^2 \frac{K_x}{E^\varphi} \frac{\sin \left( 2 \tilde{\lambda} K_\varphi \right))}{2 \tilde{\lambda}} \right)
    \notag\\
    &&
    + \left(\frac{(E^x)' (E^\varphi)'}{(E^\varphi)^2}
    - \frac{(E^x)''}{E^\varphi} \right) \cos^2 \left( \tilde{\lambda} K_\varphi \right)
    \bigg]
    \ ,
    \label{eq:Modified constraint - periodic version - simple}
\end{eqnarray}
with structure-function
\begin{eqnarray}
    \tilde{q}^{x x} &=&
    \left( 1
    + \tilde{\lambda}^2 \left( \frac{(E^x)'}{2 E^\varphi} \right)^2
    \right)
    \cos^2 \left( \tilde{\lambda} K_\varphi \right) \chi^2
    \frac{\tilde{\lambda}^2}{\lambda^2} \frac{E^x}{(E^\varphi)^2}
    \nonumber\\
    &=&
    \left(1 + \lambda^2 \left(1 - \frac{2 \mathcal{M}}{\sqrt{E^x}}
    - \frac{\Lambda}{3} E^x \right) \right) \chi^2 \frac{\tilde{\lambda}^2}{\lambda^2} \frac{E^x}{(E^\varphi)^2}
    \ .
    \label{eq:Modified structure function - periodic version - simple}
\end{eqnarray}

Since the effective line element can be written as
\begin{eqnarray}
    {\rm d} s^2 = - N^2 {\rm d} t^2 + \tilde{q}_{x x} ( {\rm d} x + N^x {\rm d} t )^2 + E^x {\rm d} \Omega^2
    \,,
\label{eq:line element - spherical - modified}
\end{eqnarray}
we shall now solve for the phase-space variables, imposing different gauges, in order 
to explicitly obtain the resulting space-time geometry.

\subsection{Non-periodic phase-space coordinates}

We will use the non-periodic constraint (\ref{eq:Modified constraint - non-periodic version - simple}) in this subsection to compute regions of a Schwarzschild-type static exterior and homogeneous regions in two different
gauges.  We show that a third gauge results in a modified
Gullstrand-Painlev\'e space-time and is related to the two
previous gauges by a standard coordinate transformation in regions of overlap.

These three gauges possess a second coordinate singularity beyond the
horizons, but they all show regular Ricci and Kretschmann scalars at those
points.  In Section~\ref{sec:Periodic phase space coordinates} we will use
the periodic version of the constraint and obtain another gauge that is free
of such a coordinate singularity and related to the Schwarzschild-type homogeneous
region by a standard coordinate transformation.

Because the overall factor
$\chi$ always multiplies the lapse function in the equations of motion, it
will be useful to define
\begin{equation}
    \bar{N} = \chi N
    \,.
\end{equation}
We encourage readers to consult Appendix~\ref{Appendix A} for a detailed
exposition of the equations of motion utilized in this section. 

\subsubsection{Schwarzschild gauge: Static region}

We define the Schwarzschild gauge by
\begin{eqnarray}
    N^x = 0 \ , \hspace{1cm} E^x = x^2
    \ .
\end{eqnarray}
The consistency condition $\dot{E}^x = 0$ implies the equation
\begin{eqnarray}
    \left( 1
    + \lambda^2 \left(\frac{(E^x)'}{2 E^\varphi}\right)^2 \right) \frac{\sin
  (2 \lambda K_\varphi)}{2 \lambda} 
    = 0
\end{eqnarray}
which, assuming consistency with the classical limit, is solved by imposing $K_\varphi = 0$.
This result implies a second consistency equation $\dot{K}_\varphi = 0$, which
we will address later on by solving it for the lapse function.

The vanishing of the diffeomorphism constraint further implies that $K_x = 0$. Finally, the vanishing of the Hamiltonian constraint implies the equation
\begin{eqnarray}
    0 &=& \left( 1 - \Lambda x^2 \right) (E^\varphi)^2 - 3 x^2 + x^3 (\ln (E^\varphi)^2)'
    \ ,
\end{eqnarray}
which is solved by
\begin{eqnarray}
    E^\varphi = \frac{x}{\sqrt{1 - c_\varphi/x - \Lambda x^2/3}}
    \,,
\end{eqnarray}
with constant $c_\varphi$.
Substituting all these results into the mass observable (\ref{eq:Weak
  observable in simple case}), and rewriting it as $\mathcal{M} = M$
(interpreted as the mass parameter of a line element rather than a phase-space
function), determines
\begin{eqnarray}
    c_\varphi &=& 2 M
    \ .
\end{eqnarray}

Lastly, $\dot{K}_\varphi=0$ implies that $\bar{N}$ acquires the expression of
the classical lapse function, up to a scaling constant $\alpha$, since $K_\varphi = 0$
eliminates all $\lambda$ dependence in the constraint and the equations of
motion it generates. Therefore,
\begin{eqnarray}
    N = \frac{\sqrt{1 - J(x)}}{\alpha \chi}
    \,,
\end{eqnarray}
where we have defined
\begin{equation}
    J(x)= \frac{2 M}{x} + \frac{\Lambda x^2}{3}
\end{equation}
for brevity.

The structure function is then given by
\begin{equation}
    \tilde{q}^{x x} = \chi^2
    \left( 1 + \lambda^2 \left( 1 - J(x)\right)
    \right) \left(1 - J(x)\right)
    \ ,
    \label{eq:mu-scheme structure function - non-periodic version - Schwarzschild}
\end{equation}
where $\lambda$ is an arbitrary function of $x$ through $E^x$. Space-time solutions
therefore depend on $\lambda$ through the emergent space-time metric, while
traditional models of loop quantum gravity always imply $\lambda$-independent
solutions  in the static gauge. This property is one example of covariance
problems in this setting because an unmodified static gauge makes it difficult
to construct an equivalent non-static gauge for the same model.
From (\ref{eq:line element - spherical - modified}), the space-time line element is given by
\begin{equation}
    {\rm d} s^2 =
    - \left(1 - J(x)\right) \frac{{\rm d} t^2}{\alpha^2 \chi^2}
    + \frac{1}{\left( 1 + \lambda^2 \left( 1 - J(x) \right)
    \right) \left(1 - J(x)\right)} \frac{{\rm d} x^2}{\chi^2}
    + x^2 {\rm d} \Omega^2
    \,,
    \label{eq:Spacetime metric - modified - Schwarzschild}
\end{equation}
for arbitrary $\lambda(x)$ and $\chi(x)$, and a constant $\alpha$. For static
solutions as constructed here, $\alpha$ may always be absorbed by a
redefinition of the time coordinate.

The locations of the black-hole and cosmological horizons are still given by their classical expressions since they solve the same classical equation,
\begin{equation}\label{eq:Horizons eq}
    1 - \frac{2 M}{x} - \frac{\Lambda x^2}{3} = 0
  \end{equation}
  implied by factors of $(1-J(x))^{\pm 1}$ in (\ref{eq:Spacetime metric - modified - Schwarzschild}).
Therefore, using a small positive cosmological constant so as to match with observations for our universe, the black-hole horizon appears at
\begin{equation}
    x_{\rm H} \approx 2 M
    \,,
\end{equation}
and the cosmological one at
\begin{equation}
    x_\Lambda \approx \sqrt{\frac{3}{\Lambda}}
    \,.
\end{equation}
Just as for a classical black hole in a de Sitter background, the static
metric is valid only in the region $x_{\rm H}<x < x_\Lambda$, beyond which the
metric turns spatially homogeneous in an extension of the same gauge, just
flipping the roles of $t$ and $x$ as time and space coordinates.

However, now we have further coordinate singularities at the values of $x$ that solve the equation
\begin{eqnarray}
    1+\lambda^2 \left(1 - \frac{2 M}{x} - \frac{\Lambda x^2}{3}\right) = 0
     \label{eq:Minimum radius equation}
\end{eqnarray}
implied by the new factor of ${\rm d}x^2$ in (\ref{eq:Spacetime metric - modified - Schwarzschild}).
This equation can have multiple solutions depending on the chosen
triangulation scheme $\lambda=\lambda(x)$, but we can still formulate several
general results. These new coordinate singularities are closely related to non-classical
hypersurfaces of reflection symmetry implied by holonomy effects.

In particular, note that the term multiplying $\lambda^2$ is the expression on
the left-hand side of the horizons equation (\ref{eq:Horizons eq}) and hence it vanishes at $x_{\rm H}$ and $x_\Lambda$.
Since $\lambda^2>0$, this term must turn negative for (\ref{eq:Minimum radius
  equation}) to hold, and hence a coordinate singularity of this form cannot appear in the static region where this term is always positive.
Thus, the solutions $x=x_\lambda^{(i)}$ of this equation, with $i$ labeling the different solutions, must be either below the Schwarzschild horizon, $x_\lambda^{(-)}<x_{\rm H}$ denoting the largest solution less than the Schwarzschild radius, or above the cosmological horizon, $x_\lambda^{(+)}>x_\Lambda$ denoting the smallest solution greater than the cosmological horizon. Therefore, solutions to (\ref{eq:Minimum radius
  equation}) do not lie in the static region, but rather in the homogenous ones.

Finally, we note that this equation corresponds precisely to  (\ref{eq:Weak
  observable at reflection}) for the reflection-symmetry surface.
Therefore, the solutions $x_\lambda^{(-)}$ and $x_\lambda^{(+)}$ correspond to extrema of the radius $\sqrt{E^x}$, around which the spacetime is symmetric, and will have particularly interesting consequences. 
Solutions of these two types are not guaranteed to exist for any
triangulation scheme $\lambda = \lambda(E^x)$, and there could
be more than two solutions in some cases. Nevertheless, we will show that
there exists an $x_\lambda^{(-)}$, corresponding to the minimum of the areal
radius, for any well-behaved holonomy function, introducing  the quantum
hypersurface where the geometry transitions from a black hole to a white hole through a wormhole interior.
In later sections we will come back to specific solutions corresponding to the traditional $\mu_0$ and $\bar{\mu}$ schemes.
First, however, we must ensure that the homogeneous regions exist as solutions
to the equations of motion in their corresponding gauges, and identify the
coordinate transformations relating the different gauge choices in regions of overlap.

\subsubsection{Schwarzschild gauge: Homogeneous regions}
\label{s:HomS}

We now restrict the gauge by setting all spatial derivatives equal to zero.
We will use the label $t_{\rm h}$ for the time coordinate and $x_{\rm h}$ for the
radial coordinate in a homogeneous patch. 
The homogeneity condition implies the partial gauge fixing
\begin{eqnarray}
    N^x = 0 \ , \hspace{1cm} N' = 0
    \,.
\end{eqnarray}
The on-shell condition $\tilde{H}=0$ can be solved to obtain
\begin{eqnarray}
    K_x &=&
    - \frac{E^\varphi}{4 E^x} \frac{2 \lambda}{\sin (2 \lambda K_\varphi)} \left( 1 - \Lambda E^x + \frac{\sin^2 ( \lambda K_\varphi)}{\lambda^2}\right.\\
&&\qquad\qquad\qquad\qquad\left.    + 2 \left( K_\varphi \frac{\sin (2 \lambda K_\varphi)}{2 \lambda} - \frac{\sin^2 ( \lambda K_\varphi)}{\lambda^2} \right) \frac{\partial \ln \lambda^2}{\partial \ln E^x} \right)
    \,.\nonumber 
\end{eqnarray}
Equations of motion for the remaining phase-space variables are
\begin{eqnarray}
    \dot{E}^x &=&
    2 \bar{N} \sqrt{E^x}\; \frac{\sin (2 \lambda K_\varphi)}{2 \lambda}
    \ ,
\end{eqnarray}
as well as
\begin{eqnarray}
    \frac{\dot{E}^\varphi}{E^\varphi}
    &=&
    \frac{\bar{N}}{2 \sqrt{E^x}} \frac{2 \lambda}{\sin (2 \lambda K_\varphi)} \left(
    \frac{\sin^2 ( \lambda K_\varphi)}{\lambda^2}
    - \left(1-\Lambda E^x\right) \cos (2 \lambda K_\varphi)\right.\\
&&\qquad\qquad\qquad\qquad\left.    - 2 \lambda^2 \frac{\sin^4 (\lambda K_\varphi)}{\lambda^4} \frac{\partial \ln \lambda^2}{\partial \ln E^x}
    \right)
    \ .\nonumber
\end{eqnarray}
and
\begin{eqnarray}
    \dot{K}_\varphi
    &=&
    - \frac{\bar{N}}{2 \sqrt{E^x}} \left( 1 - \Lambda E^x + \frac{\sin^2 ( \lambda K_\varphi)}{\lambda^2}
    + 2 \left( K_\varphi \frac{\sin (2 \lambda K_\varphi)}{2 \lambda} - \frac{\sin^2 ( \lambda K_\varphi)}{\lambda^2} \right) \frac{\partial \ln \lambda^2}{\partial \ln E^x} \right)\,.
    \nonumber \\ 
\end{eqnarray}

Combining these three time derivatives, we can pick $E^x$ as an evolution parameter by using the chain rule, ${\rm d} K_\varphi/{\rm d} E^x = \dot{K}_\varphi/\dot{E}^x$, obtaining the equation
\begin{eqnarray}
    \frac{{\rm d}}{{\rm d} \ln E^x} \left(\frac{\sin^2 (\lambda K_\varphi)}{\lambda^2}\right)
    =
    - \frac{1}{2} \left( 1 - \Lambda E^x + \frac{\sin^2 (\lambda K_\varphi)}{\lambda^2} \right)
    \ ,
\end{eqnarray}
which has the solution
\begin{eqnarray}
    \frac{\sin^2 ( \lambda K_\varphi)}{\lambda^2} &=& \frac{c_x}{\sqrt{E^x}} + \frac{\Lambda E^x}{3} - 1
\end{eqnarray}
with constant $c_x$.
Direct substitution into the mass observable (\ref{eq:Weak observable in simple case - nonperiodic}) and relabeling ${\cal M}=M$ determines $c_x = 2 M$.
The equation of motion for the structure function is given by
\begin{eqnarray}
    \left(\ln \frac{\tilde{q}^{x x}}{\chi^2}\right)^\bullet
    &=&
    \left(\ln \left( \cos^2 (\lambda K_\varphi) \frac{E^x}{(E^\varphi)^2} \right)\right)^\bullet
    \nonumber\\
    &=&
    2 \left( - \lambda^2 \frac{\tan (\lambda K_\varphi)}{\lambda} \left( \dot{K}_\varphi + \frac{1}{2} K_\varphi \frac{\partial \ln \lambda^2}{\partial \ln E^x} \frac{\dot{E}^x}{E^x} \right)
    - \frac{\dot{E}^\varphi}{E^\varphi} \right)
    + \frac{\dot{E}^x}{E^x}
    \nonumber\\
    &=&
    \frac{\bar{N}}{\sqrt{E^x}} \frac{\lambda}{\tan (\lambda K_\varphi)} \left( \frac{2 M}{\sqrt{E^x}} + \frac{\Lambda}{3} E^x - \frac{2 \lambda E^x}{3} \right)
    \ .
\end{eqnarray}

We now complete the gauge fixing by setting $E^x = t_{\rm h}^2$, yielding
\begin{eqnarray}
    \frac{\sin^2 ( \lambda K_\varphi)}{\lambda^2} &=& \frac{2 M}{t_{\rm h}} + \frac{\Lambda}{3} t_{\rm h}^2 - 1
    \ .
\end{eqnarray}
The consistency equation $\dot{E}^x = 2 t_{\rm h}$ can be used to solve for
the lapse function as
\begin{eqnarray}
    \bar{N} &=&
    \frac{2 \lambda}{\sin (2 \lambda K_\varphi)}
    = \left( J(t_{\rm h}) - 1\right)^{-1/2} \left(1-\lambda^2 \left(J(t_{\rm h}) - 1\right)\right)^{-1/2}
\end{eqnarray}
with our previous definition of the function $J$,
and the equation of motion for the structure function is solved by
\begin{eqnarray}
    \tilde{q}^{x x} = \alpha^2 \chi^2 \left(\frac{2 M}{t_{\rm h}} + \frac{\Lambda}{3} t_{\rm h}^2 - 1\right)^{-1}
\end{eqnarray}
with an integration constant $\alpha$.
We therefore obtain the space-time line element
\begin{eqnarray}
    {\rm d} s^2 =
    - \frac{1}{\left( 1 - \lambda^2 \left(J(t_{\rm h}) -1\right)
    \right) \left(J(t_{\rm h})-1\right)} \frac{{\rm d} t_{\rm h}^2}{\chi^2}
    + \left(J(t_{\rm h})-1\right) \frac{{\rm d} x_{\rm h}^2}{\alpha^2 \chi^2}
    + t_{\rm h}^2 {\rm d} \Omega^2
    \ ,
    \label{eq:Spacetime metric homogeneous - modified - Schwarzschild}
\end{eqnarray}
for arbitrary $\lambda$ and $\chi$ as functions of $t_{\rm h}$, which matches
the static Schwarzschild metric (\ref{eq:Spacetime metric - modified -
  Schwarzschild}) up to the label swap $x \to t_{\rm h}$, $t \to x_{\rm h}$.
This effective metric has a new (quantum) coordinate singularity at the time coordinates solving the equation
\begin{eqnarray}
    1-\lambda^2 \left(\frac{2 M}{t_{\rm h}} + \frac{\Lambda t_{\rm h}^2}{3} - 1\right) = 0
    \,, \label{eq:Minimum radius equation - homogeneous}
\end{eqnarray}
which is the same equation as the one obtained in the static region
(\ref{eq:Minimum radius equation}) up to the label change $x\to t_{\rm h}$. As discussed in this context, solutions $t_{\rm h}$ represent hypersurfaces of
reflection symmetry. (The name quantum coordinate
singularity is motivated by the observation that this new singular
hypersurface appears only in the presence of holonomy modifications. As
mentioned earlier, on adopting suitable coordinates, we shall demonstrate
later that one of the solutions of (\ref{eq:Minimum radius equation -
  homogeneous}) corresponds to the transition surface where $\sqrt{E^x}$
acquires a minimum value.)

Therefore, homogeneous Schwarzschild-type coordinates are valid only in the
regions $x_\lambda^{(-)}<t_{\rm h}<x_{\rm H}$ where $x_{\lambda}^{(-)}$ is the
largest solution of (\ref{eq:Minimum radius equation - homogeneous}) smaller
than $x_{\rm H}$, and $x_\Lambda<t_{\rm h}<x_\lambda^{(+)}$ where
$x_{\lambda}^{(+)}$ is the smallest solution of (\ref{eq:Minimum radius
  equation - homogeneous}) greater than $x_{\Lambda}$. If there is no
$x_\lambda^{(-)}$ obeying the condition that defines it, we set
$x_\lambda^{(-)}=0$.  If there
is no $x_\lambda^{(+)}$ obeying the condition that defines it, we set
$x_\lambda^{(+)}=\infty$. If (\ref{eq:Minimum radius equation - homogeneous})
has more than two distinct solutions, additional regions can be
introduced. However, we will show in Section~\ref{sec:Extremality} that the solutions
  $x_{\lambda}^{(\pm)}$ closest to $x_{\rm H}$ and $x_{\Lambda}$,
  respectively, represent extremal radii, such
  that $x_{\lambda}^{(-)}$ is a minimum radius, or a ``bounce,'' and
  $x_{\lambda}^{(+)}$ is a maximum radius of a recollapsing homogeneous
  region. Solutions $x_{\lambda}^{(i)}$ less than $x_{\lambda}^{(-)}$ or
  greater than $x_{\lambda}^{(+)}$ are therefore never reached by dynamical
  solutions if we start in the static Schwarzschild region and extend it
  across coordinate singularities. Nevertheless, these alternative solutions
  might be used to define independent space-time models that could perhaps be
  connected with the static region by tunneling processes of a more complete
  quantum theory.

As already indicated, the homogenous solutions can formally be obtained from
the static one by a simple flip of coordinates. However, the corresponding
space-time regions are disjoint and separated by horizons.  For a more precise
connection between these regions, we should construct a gauge that describes a
space-time region overlapping with a portion of the static solution and a
portion of a homogeneous one, crossing the horizon.  Classically, an example
for such a gauge choice is given by the Gullstrand-Painlev\'e coordinates,
which we now construct for the modified equations.

\subsubsection{From static Schwarzschild to Gullstrand--Painlev\'e}

The metric (\ref{eq:Spacetime metric - modified - Schwarzschild}) is static
and thus has a timelike Killing vector $\xi=\partial_t$.  The quantity
$\varepsilon=-\xi\cdot {\bf u}=-u_t = -g_{t \nu}\, u^\nu$ is therefore
conserved along geodesics, using the 4-velocity ${\bf u}$. Compared with the
classical Gullstrand--Painlev\'e system, we are interested in geodesics for
free fall starting at rest at a finite radius, $x=x_0$, because the asymptotic
regime may be non-static if there is a cosmological horizon or non-classical
asymptotic behavior implied by certain types of holonomy modifications. The
4-velocity of such an object has a time component
$u^t |_{x= x_0}=g^{tt} u_t |_{x=x_0}=-g^{tt}|_{x=x_0}\varepsilon$, using the
  diagonal nature of the static metric in the second step, while all spatial
  components vanish.  Timelike normalization,
  $-1=g_{\mu\nu} u^\mu u^\nu|_{x=x_0}=g_{tt}(u^t)^2|_{x=x_0}$, then implies
\begin{equation} \label{GP_energy}
 \varepsilon^2= -\frac{1}{g^{tt}|_{x=x_0}}= \frac{1 - J(x_0)}{\alpha^2 \chi(x_0)^2}\,.
\end{equation}

For $x\not=x_0$, $u^x$ is non-zero and is related to $u^t$ by the
normalization condition. Since $u_t=-\varepsilon$ is conserved, it is convenient to
evaluate normalization in the form $g^{\mu\nu}u_{\mu}u_{\nu}=-1$, such that
\begin{eqnarray}
  u_x&=&s \sqrt{g_{xx}(x)(-g^{tt}(x)\epsilon^2-1)}\nonumber\\
  &=&s
  \sqrt{\frac{1}{(1-J(x))(1+\lambda(x)^2(1-J(x)))\chi(x)^2}\left(
      \frac{\chi(x)^2}{\chi(x_0)^2} \frac{1-J(x_0)}{1-J(x)}-1\right)}
      \nonumber\\
  &=& \frac{s}{\chi(x)\chi(x_0)}
      \frac{\sqrt{\chi(x)^2(1-J(x_0))-\chi(x_0)^2(1-J(x))}}{|1-J(x)|
      \sqrt{1+\lambda(x)^2(1-J(x))}} \label{ux}
\end{eqnarray}
with a sign choice $s$ that distinguishes outgoing from ingoing GP
coordinates.

The normalization condition also implies that we can obtain the proper-time
interval along our geodesics from the co-velocity components:
\begin{equation}
  {\rm d}\tau = -\frac{g^{\mu\nu}{\rm d}x_{\mu}{\rm d}x_{\nu}}{{\rm d}\tau}=
  - u_{\mu} g^{\mu\nu} {\rm d}x_{\nu}=-u_{\mu}{\rm d}x^{\nu}\,.
\end{equation}
Using proper time as the new GP coordinate, we obtain
\begin{eqnarray}
    {\rm d} t_{\rm GP} &=& - u_\mu d x^\mu
    \nonumber\\
    &=&
    \varepsilon {\rm d} t
        - \frac{s}{\chi(x)\chi(x_0)}
      \frac{\sqrt{\chi(x)^2(1-J(x_0))-\chi(x_0)^2(1-J(x))}}{|1-J(x)|
      \sqrt{1+\lambda(x)^2(1-J(x))}}
        {\rm d} x
    \,. \label{tGP}
\end{eqnarray}
The coordinate transformation $t_{\rm GP} (t , x)$ is defined by an explicit
integration of the 1-form (\ref{tGP}), which depends on the choice of
$\lambda(x)$ and $\chi(x)$ and therefore cannot be performed at a general
level. However, since $\varepsilon$ is constant and the coefficient of ${\rm d}x$
is independent of $t$, (\ref{tGP}) defines a closed 1-form, such that an
integral $t_{\rm GP}(t,x)$ exists as a local coordinate for any choice of
$\lambda(x)$ and $\chi(x)$. 

Substituting ${\rm d} t = \varepsilon^{-1} {\rm d} t_{\rm GP} + \varepsilon^{-1} u_x {\rm d} x$ in (\ref{eq:Spacetime metric - modified - Schwarzschild}), we obtain
\begin{eqnarray}
    {\rm d} s^2_{\rm GP}
    &=&
    - N^2 \varepsilon^{-2} ( {\rm d}t_{\rm GP}^2 + 2 u_x {\rm d} t_{\rm GP} {\rm d} x + u_x^2 {\rm d} x^2)
    + \tilde{q}_{x x} {\rm d} x^2
    + x^2 {\rm d} \Omega^2
    \nonumber\\
    &=&
    - \frac{N^2}{\varepsilon^2} {\rm d} t_{\rm GP}^2
    - 2 u_x \frac{N^2}{\varepsilon^{2}} {\rm d} t_{\rm GP} {\rm d} x
    + \left(\tilde{q}_{x x} - \frac{N^2}{\varepsilon^{2}} u_x^2\right) {\rm d} x^2
    + x^2 {\rm d} \Omega^2
    \nonumber\\
    &=&
    - {\rm d} t_{\rm GP}^2
    + \frac{\tilde{q}_{x x} N^2}{\varepsilon^{2}} \left( {\rm d} x
    - u_x \tilde{q}^{x x} {\rm d} t_{\rm GP} \right)^2
    + x^2 {\rm d} \Omega^2
    \nonumber\\
    &=&
    - {\rm d} t_{\rm GP}^2
    + \frac{\chi^{-4}}{\alpha^2 \varepsilon^{2}} \left( 1 + \lambda^2 \left( 1 - J(x) \right)
        \right)^{-1}\nonumber\\
  &&\times\left( {\rm d} x 
    - s \chi \sqrt{J(x)-1 -\frac{\chi^2}{\chi(x_0)^2}( J(x_0)-1)} \sqrt{1 +
        \lambda^2 \left( 1 - J(x) \right)} {\rm d} t_{\rm GP} \right)^2 
    \nonumber\\
    &&
    + x^2 {\rm d} \Omega^2
    \label{eq:Spacetime metric - mu-scheme - GP}
\end{eqnarray}
where we used $u_x^2=\tilde{q}_{xx}(\varepsilon^2/N^2-1)$ twice in the third
line, and then inserted the explicit $u_x$ from (\ref{ux}).
The metric remains regular at the horizon coordinates, but still diverges at the new (quantum) singular coordinates $x=x_\lambda^{(i)}$.

In the case of a vanishing cosmological constant, $\Lambda=0$, it is useful to take the limit $x_0 \to \infty$, in which case $\alpha \varepsilon \to 1/\lim_{x\to\infty}\chi(x)=:1/\chi_{\infty}$ from (\ref{GP_energy}) and the metric simplifies to 
\begin{eqnarray}
    {\rm d} s^2_{\rm GP}
    &=&
    - {\rm d} t_{\rm GP}^2
    + \frac{\chi_\infty^2}{\chi^{4}} \left( 1 + \lambda^2 \left( 1 - \frac{2M}{x} \right)
        \right)^{-1}\nonumber\\
  &&\times\left( {\rm d} x 
    + s \chi \sqrt{\frac{2M}{x}+\frac{\chi^2}{\chi_{\infty}^2}-1} \sqrt{1 + \lambda^2 \left( 1 - \frac{2M}{x} \right)} {\rm d} t_{\rm GP} \right)^2\nonumber\\
    && \,\;\;\;\;\;\;\;    + x^2 {\rm d} \Omega^2
    \,.
    \label{eq:Spacetime metric - modified - GP simplified}
\end{eqnarray}

\subsubsection{From homogeneous Schwarzschild to Gullstrand-Painlev\'e}

We can now start with the GP metric (\ref{eq:Spacetime metric - mu-scheme -
  GP}) we just derived and extend it across the horizons.  Because we expect
the causal meaning of time and space coordinates to change in the interior of
the black hole, let us define $T_{\rm GP} = x$ and
$X_{\rm GP} = t_{\rm GP}$ and rewrite the GP metric (\ref{eq:Spacetime metric
  - mu-scheme - GP}) as
\begin{eqnarray}
    {\rm d} s^2_{\rm GP}&=&
    - {\rm d} X_{\rm GP}^2
    + \frac{\chi^{-4}}{\alpha^2 \varepsilon^{2}} \left( 1 + \lambda^2 \left( 1 - J(T_{\rm GP}) \right)
    \right)^{-1} \bigg( {\rm d} T_{\rm GP}
    \nonumber\\
    &&\qquad
    + s \frac{\chi}{\chi(x_0)} \sqrt{\chi^2 (1-J(x_0))-\chi(x_0)^2 (1-J(T_{\rm GP}))} \sqrt{1 + \lambda^2 \left( 1 - J(T_{\rm GP}) \right)} {\rm d} X_{\rm GP} \bigg)^2
    \nonumber\\
    &&
    + T_{\rm GP}^2 {\rm d} \Omega^2
    \nonumber\\
    &=&
    - \left( 1 - \lambda^2 \left(J(T_{\rm GP}) - 1\right)
    \right)^{-1}
    \left( J(T_{\rm GP}) - 1 \right)^{-1} \frac{{\rm d} T_{\rm GP}^2}{\chi^2}
    \nonumber\\
    &&
    + \frac{J(T_{\rm GP}) - 1}{\alpha^2 \chi^2 \varepsilon^{2}} \Bigg( {\rm d} X_{\rm GP} 
    + \frac{s}{\chi} \sqrt{\frac{J(T_{\rm GP})-1+(\chi^2/\chi(x_0)^2)(1-
       J(x_0))}{ 1 - \lambda^2 (J(T_{\rm GP}) - 1)}} \frac{{\rm d}T_{\rm GP}}{ J(T_{\rm GP}) - 1}
    \Bigg)^2
    \nonumber\\
    &&
    + T_{\rm GP}^2 {\rm d} \Omega^2
    \,.
    \label{eq:Spacetime metric - mu-scheme - GP - interior labels}
\end{eqnarray}
A comparison of (\ref{eq:Spacetime metric - mu-scheme - GP - interior labels}) with (\ref{eq:Spacetime metric homogeneous - modified - Schwarzschild}) shows that the two are related by the coordinate transformation $t_{\rm h} = T_{\rm GP}$ and
\begin{eqnarray}
{\rm d} x_{\rm h}
    &=&
    \frac{{\rm d} X_{\rm GP}}{\varepsilon}
    \\
    &&
    - \frac{s}{\varepsilon\chi} \sqrt{\frac{J(T_{\rm GP})-1+(\chi^2/\chi(x_0)^2)\left(1-J(x_0)\right)}{  1 - \lambda^2 \left( J(T_{\rm GP}) - 1\right)}}
      \frac{{\rm d} T_{\rm GP}}{J(T_{\rm GP})-1}
    \nonumber
\end{eqnarray}
for arbitrary $\lambda$ and $\chi$.
The explicit integration can only be completed after specifying the choice of
$\lambda (t_{\rm h})$ and $\chi(t_{\rm h})$, but such an integration always
exists locally because ${\rm d}x_{\rm h}$ is a closed 1-form.

The metric in (\ref{eq:Spacetime metric - mu-scheme - GP -
  interior labels}) remains regular at the horizon coordinates  if we
  choose $\alpha\varepsilon=\sqrt{1-J(x_0)}/\chi(x_0)$, but it diverges
at the reflection surfaces $T_{\rm GP}=x_\lambda^{(i)}$ for any $\alpha$.  Therefore, the GP
chart is valid only in the region $x_\lambda^{(-)}<x<x_\lambda^{(+)}$, where
for more than two solutions to (\ref{eq:Minimum radius equation}) the
$\pm$-versions of $x_{\lambda}^{(i)}$ are as defined at the end of
Section~\ref{s:HomS}.

In the case of a vanishing cosmological constant, we can take $x_0\to \infty$ and $\alpha \varepsilon=1/\lim_{x\to\infty}\chi(x)=:1/\chi_{\infty}$, simplifying the metric to
\begin{eqnarray}
    {\rm d} s^2_{\rm GP}
    &=&
    - \left( 1 + \lambda^2 \left( 1 - J(T_{\rm GP}) \right)
    \right)^{-1}
    \left( J(T_{\rm GP}) - 1 \right)^{-1} \frac{{\rm d} T_{\rm GP}^2}{\chi^2}
    \nonumber\\
    &&
    + \frac{\chi_{\infty}^2}{\chi^2} \left( J(T_{\rm GP}) - 1\right) \left( {\rm d} X_{\rm GP}
       - \frac{s}{\chi}
\sqrt{\frac{J(T_{\rm GP})+\chi^2/\chi_{\infty}^2-1}{ 1 - \lambda^2 (J(T_{\rm GP}) - 1)}} \frac{{\rm d}T_{\rm GP}}{ J(T_{\rm GP}) - 1}
    \right)^2
    \nonumber\\
    &&
    + T_{\rm GP}^2 {\rm d} \Omega^2
    \ .
    \label{eq:Spacetime metric - modified - GP - interior labels}
\end{eqnarray}

\subsubsection{Gullstrand-Painlev\'e gauge}
The GP metric (\ref{eq:Spacetime metric - mu-scheme - GP}) covers both the static and the homogeneous regions in a single coordinate chart.
We now perform a consistency check by showing that the solution obtained above satisfies the canonical equations of motion in an appropriate GP gauge.

Based on the metric (\ref{eq:Spacetime metric - mu-scheme - GP}), we define the GP gauge as
\begin{eqnarray}
    N = 1 \ , \hspace{1cm} E^x = x^2
    \ .
\end{eqnarray}
The consistency equation $\Dot{E}^x = 0$ determines the shift vector
\begin{eqnarray}
    N^x
    &=&
    - \chi \left( 1
    + \lambda^2 \frac{x^2}{(E^\varphi)^2} \right) \frac{\sin (2 \lambda K_\varphi)}{2 \lambda}
    \ .
\end{eqnarray}
The on-shell conditions $H_x=0$ and $\tilde{H}=0$, respectively, determine $K_x
= E^\varphi K_\varphi' / (2 x)$ and 
\begin{eqnarray}
    \frac{\sin^2 (\lambda K_\varphi)}{\lambda^2}
    &=&
    \left( 1 + \lambda^2 \frac{x^2}{(E^\varphi)^2} \right)^{-1} \left( \frac{c_\varphi}{x} + \frac{\Lambda x^2}{3} + \frac{x^2}{(E^\varphi)^2} - 1 \right)
    \ ,
\end{eqnarray}
for arbitrary $\lambda$, where $c_\varphi$ is an integration constant.

Finally, the equation of motion for $E^\varphi$ reads
\begin{eqnarray}\label{EphiGP}
    &&\left(\ln \left(\chi^2 \frac{(E^\varphi)^2}{x^2}\right)\right)^\bullet\\
    &=&
    - \sqrt{\left(\frac{c_\varphi}{x} + \frac{\Lambda x^2}{3} + \frac{x^2}{(E^\varphi)^2} - 1\right)
    \left(1 + \lambda^2 \left(1 - \frac{c_\varphi}{x} - \frac{\Lambda x^2}{3} \right)\right)}
    \left( \ln \left(\chi^2 \frac{(E^\varphi)^2}{x^2}\right) \right)' \nonumber
\end{eqnarray}
and is solved by $E^\varphi = x / (\alpha \chi \varepsilon)$ with constant
$\alpha$ and $\varepsilon$. (We introduce two integration constants for the
sake of comparing with our previous line element, which will relate
$\varepsilon$ to the constant denoted earlier by the same letter. As a partial
differential equation, (\ref{EphiGP}) has additional solutions that depend on
the specific choice of $\lambda(x)$ but will not be required here.)  The
resulting metric can then be derived from \eqref{eq:line element -
  spherical - modified}, which matches exactly with (\ref{eq:Spacetime metric
  - mu-scheme - GP}) upon the identification of $c_\varphi=2 M$ and
$\varepsilon^2 = (1 - J(x_0))/(\alpha^2 \chi(x_0)^2)$.

Therefore, the modified GP metric (\ref{eq:Spacetime metric - mu-scheme - GP -
  interior labels}) is related by standard coordinate transformations to both
the static and homogeneous solutions in regions of overlap.  Furthermore, it
is well-defined at the Schwarzschild horizon $x=x_{\rm H}$ provided
$x_0>x_{\rm H}$, and at the cosmological horizon $x=x_\Lambda$ provided
$x_0>x_\Lambda$. In the latter case, $x_0$ (defined as the reference position
for the inertial observer) can no longer be interpreted as the coordinate
where the observer is at rest, but the coordinate transformation still holds.

\subsection{Internal-time gauge of homogeneous regions}
\label{sec:Periodic phase space coordinates}

As we have seen, the homogeneous Schwarzschild charts terminate at the quantum
hypersurfaces of reflection symmetry at $x_\lambda^{(i)}$ solving (\ref{eq:Minimum
  radius equation}).  As we will show in Section \ref{sec:Interior physics},
the Ricci scalar remains regular at those surfaces, suggesting that another
chart must exist for which the metric remains regular there.  Since the
coordinates $x_\lambda^{(i)}$ always lie in the homogeneous regions, we will
formulate a new homogeneous gauge where the time coordinate is not given by
the radius in an extension of the Schwarzschild exterior, but is instead strategically
chosen to track the angular momentum $K_\varphi$.  To this end, it is useful
to switch to the constraint (\ref{eq:Hamiltonian constraint - modified -
  periodic}) because it is periodic in the variable of interest, $K_\varphi$.

We set all spatial derivatives to zero and
label $t_\varphi$ as the time coordinate in our new gauge as well as $x_{\rm
  h}$ as the spatial coordinate, which agrees with the previous homogeneous coordinate.
The partial gauge-fixing 
\begin{eqnarray}
    N^x = 0 \ , \hspace{1cm} N' = 0
    \ ,
\end{eqnarray}
is used for the homogeneous gauge. The on-shell condition $\tilde{H}=0$ can be solved for $K_x$:
\begin{eqnarray}
    K_x
    =
    - \frac{E^\varphi}{4 E^x} \frac{2 \tilde{\lambda}}{\sin (2 \tilde{\lambda} K_\varphi)} \left( \frac{\lambda^2}{\tilde{\lambda}^2} \left(1- \Lambda E^x\right)
    + \left(1 - 2 \frac{\partial \ln \lambda^2}{\partial \ln E^x}\right) \frac{\sin^2 \left(\tilde{\lambda} K_\varphi\right)}{\tilde{\lambda}^2}
    \right)
    \,.
\end{eqnarray}
The relevant equations of motion are
\begin{eqnarray}
    \frac{\dot{E}^x}{E^x} =
    \frac{\tilde{\lambda}}{\lambda} \frac{2 \bar{N}}{\sqrt{E^x}} \frac{\sin (2 \tilde{\lambda} K_\varphi)}{2 \tilde{\lambda}}
\end{eqnarray}
and
\begin{eqnarray}
    \frac{\dot{E}^\varphi}{E^\varphi}
    &=&
    \frac{\tilde{\lambda}}{\lambda} \frac{\bar{N}}{2 \sqrt{E^x}} \frac{2 \tilde{\lambda}}{\sin (2 \tilde{\lambda} K_\varphi)} \Bigg( \left(1 - 2 \frac{\partial \ln \lambda^2}{\partial \ln E^x}\right) \frac{\sin^2 \left(\tilde{\lambda} K_\varphi\right)}{\tilde{\lambda}^2}
    - \frac{\lambda^2}{\tilde{\lambda}^2} \left(1- \Lambda E^x\right) \cos (2 \tilde{\lambda} K_\varphi) \Bigg)
    \ ,\nonumber\\
\end{eqnarray}
as well as
\begin{eqnarray}
    \dot{K}_\varphi
    &=&
    - \frac{\tilde{\lambda}}{\lambda} \frac{\bar{N}}{2 \sqrt{E^x}} \left( \frac{\lambda^2}{\tilde{\lambda}^2} \left(1- \Lambda E^x\right)
    + \left( 1 - 2 \frac{\partial \ln \lambda^2}{\partial \ln E^x}\right) \frac{\sin^2 \left(\tilde{\lambda} K_\varphi\right)}{\tilde{\lambda}^2}
    \right)
    \ .
\end{eqnarray}
(Recall that $\bar{N}=\chi N$.)

Combining these equations of motion we obtain
\begin{eqnarray}
    \frac{{\rm d}}{{\rm d} \ln E^x} \left( \frac{\tilde{\lambda}^2}{\lambda^2} \frac{\sin^2 \left(\tilde{\lambda} K_\varphi\right)}{\tilde{\lambda}^2} \right)
    &=&
    - \frac{1}{2} \left( 1 - \Lambda E^x
    + \frac{\tilde{\lambda}^2}{\lambda^2} \frac{\sin^2 \left(\tilde{\lambda} K_\varphi\right)}{\tilde{\lambda}^2}
    \right)
    \ ,
\end{eqnarray}
which has a general solution
\begin{eqnarray}
    \frac{\tilde{\lambda}^2}{\lambda^2} \frac{\sin^2 \left(\tilde{\lambda} K_\varphi\right)}{\tilde{\lambda}^2}
    &=&
    \frac{c_x}{\sqrt{E^x}} + \frac{\Lambda E^x}{3} - 1
    \ ,
    \label{eq:Curvature-triad - Homogeneous - modified}
\end{eqnarray}
for arbitrary $\lambda(E^x)$.
Inserting into the mass observable (\ref{eq:Weak observable in simple case})
and relabeling $\mathcal{M}=M$ a constant (rather than phase-space function) determines $c_x = 2 M$.
A different combination of the time derivatives gives us the equation of motion
\begin{eqnarray}
    \left(\ln \frac{\tilde{q}^{xx}}{\chi^2}\right)^\bullet
    %
    &=&
    - 2 \tilde{\lambda} \tan \left(\tilde{\lambda} K_\varphi\right) \dot{K}_\varphi
    - 2 \frac{\dot{E}^\varphi}{E^\varphi}
    + \left(1 - \frac{\partial \ln \lambda^2}{\partial \ln E^x} \right) \frac{\dot{E}^x}{E^x}
    \nonumber\\
    &=&
    \frac{\tilde{\lambda}}{\lambda} \frac{\bar{N}}{\sqrt{E^x}}
    \Bigg[
    \frac{\sin (2 \tilde{\lambda} K_\varphi)}{2 \tilde{\lambda}}
    + \frac{\lambda^2}{\tilde{\lambda}^2} \left( \tilde{\lambda}^2 \frac{\tan \left(\tilde{\lambda} K_\varphi\right)}{\tilde{\lambda}}
    + \frac{2 \tilde{\lambda}}{\tan (2 \tilde{\lambda} K_\varphi)} \right)
    \Bigg]
    \nonumber\\
    &=&
    \frac{\bar{N}}{\sqrt{E^x}} \frac{\lambda}{\tilde{\lambda}} \frac{\tilde{\lambda}}{\tan \left(\tilde{\lambda} K_\varphi\right)}
    \Bigg[
    1 - \Lambda E^x
    + \frac{\tilde{\lambda}^2}{\lambda^2} \frac{\sin^2 (\tilde{\lambda} K_\varphi)}{\tilde{\lambda}^2}
    \Bigg]
\end{eqnarray}
for the structure-function.

We now complete the gauge fixing by choosing $K_\varphi = - t_{\varphi}$.
The consistency equation $\dot{K}_\varphi = -1$ determines the lapse function
\begin{eqnarray}
    N
    &=&
    \frac{\lambda}{\tilde{\lambda}} \frac{2 \sqrt{E^x}}{\chi} \frac{\tilde{\lambda}^2}{\lambda^2} \left( 1 - \Lambda E^x
    + \left( 1 - 2 \frac{\partial \ln \lambda^2}{\partial \ln E^x}\right) \frac{\tilde{\lambda}^2}{\lambda^2} \frac{\sin^2 \left(\tilde{\lambda} K_\varphi\right)}{\tilde{\lambda}^2}
    \right)^{-1}
    \,,
    \label{eq:Lapse homogeneous - modified - Internal time}
\end{eqnarray}
such that the equation of motion for the structure function becomes
\begin{eqnarray}
    \left(\ln \frac{\tilde{q}^{xx}}{\chi^2}\right)^\bullet
    &=&
    2 \frac{\tilde{\lambda}}{\tan \left(\tilde{\lambda} K_\varphi\right)}
    \left(
    1 - \Lambda E^x
    + \frac{\tilde{\lambda}^2}{\lambda^2} \frac{\sin^2 (\tilde{\lambda} K_\varphi)}{\tilde{\lambda}^2}
    \right)
    \nonumber\\
    &&\qquad\qquad \times
    \left( 1 - \Lambda E^x
    + \left( 1 - 2 \frac{\partial \ln \lambda^2}{\partial \ln E^x}\right) \frac{\tilde{\lambda}^2}{\lambda^2} \frac{\sin^2 \left(\tilde{\lambda} K_\varphi\right)}{\tilde{\lambda}^2}
    \right)^{-1}
    \ .
    \label{eq:Structure function EoM - modified - Internal time}
\end{eqnarray}

This equation is hard to solve directly, but we can perform the associated
coordinate transformation from $t_{\rm h}$ to $t_\varphi$ using
(\ref{eq:Curvature-triad - Homogeneous - modified}) with $E^x = t_{\rm h}^2$ and $K_\varphi = - t_\varphi$:
\begin{equation}
    \frac{\tilde{\lambda}^2}{\lambda(t_{\rm h}^2)^2} \frac{\sin^2 \left(\tilde{\lambda} t_\varphi\right)}{\tilde{\lambda}^2}
    =
    \frac{2 M}{t_{\rm h}} + \frac{\Lambda t_{\rm h}^2}{3} - 1
  \end{equation}
  such that
  \begin{eqnarray}
    \frac{\sin \left(2 \tilde{\lambda} t_\varphi\right)}{2 \tilde{\lambda}} {\rm d} t_\varphi
    &=&
    \frac{1}{2} \left( \frac{\lambda^2}{\tilde{\lambda}^2} \frac{\partial \ln \lambda^2}{\partial t_{\rm h}} \left(\frac{2 M}{t_{\rm h}} + \frac{\Lambda t_{\rm h}^2}{3} - 1\right)
    - \frac{1}{t_{\rm h}} \frac{\lambda^2}{\tilde{\lambda}^2} \left(\frac{2 M}{t_{\rm h}} - \frac{2 \Lambda t_{\rm h}^2}{3}\right) \right) {\rm d} t_{\rm h}
    \nonumber\\
    &=&
    \frac{1}{2 \sqrt{E^x}} \frac{\lambda^2}{\tilde{\lambda}^2} \left( 
    - \frac{2 M}{\sqrt{E^x}}
    + \frac{2 \Lambda E^x}{3}
    + 2 \frac{\partial \ln \lambda^2}{\partial \ln E^x} \frac{\tilde{\lambda}^2}{\lambda^2} \frac{\sin^2 \left(\tilde{\lambda} K_\varphi\right)}{\tilde{\lambda}^2} \right) {\rm d} t_{\rm h}
    \nonumber\\
    &=&
    - \frac{1}{2 \sqrt{E^x}} \frac{\lambda^2}{\tilde{\lambda}^2} \left( 
    1 - \Lambda E^x
    + \left( 1 - 2 \frac{\partial \ln \lambda^2}{\partial \ln E^x} \right) \frac{\tilde{\lambda}^2}{\lambda^2} \frac{\sin^2 \left(\tilde{\lambda} K_\varphi\right)}{\tilde{\lambda}^2} \right) {\rm d} t_{\rm h}
  \end{eqnarray}
  or
  \begin{equation}
    {\rm d} t_{\rm h}
    =
    - 2 \sqrt{E^x} \frac{\tilde{\lambda}^2}{\lambda^2}
    \frac{\sin \left(2 \tilde{\lambda} t_\varphi\right)}{2 \tilde{\lambda}}
    \left( 1 - \Lambda E^x + \left( 1 - 2 \frac{\partial \ln \lambda^2}{\partial \ln E^x} \right) \frac{\tilde{\lambda}^2}{\lambda^2} \frac{\sin^2 \left(\tilde{\lambda} t_\varphi\right)}{\tilde{\lambda}^2} \right)^{-1} {\rm d} t_\varphi
    \label{eq:Schwarzchild-Internal time transformation - Homogeneous - mu-scheme}
\end{equation}

The homogeneous Schwarzschild metric (\ref{eq:Spacetime metric homogeneous -
  modified - Schwarzschild}) then becomes
\begin{eqnarray}
    {\rm d} s^2 &=&
    - 4 E^x \frac{\tilde{\lambda}^2}{\lambda^2}
    \left( 1 - \Lambda E^x + \left( 1 - 2 \frac{\partial \ln \lambda^2}{\partial \ln E^x} \right) \frac{\tilde{\lambda}^2}{\lambda^2} \frac{\sin^2 \left(\tilde{\lambda} t_\varphi\right)}{\tilde{\lambda}^2} \right)^{-2} \frac{{\rm d} t_\varphi^2}{\chi^2}\nonumber\\
&&    + \frac{\tilde{\lambda}^2}{\lambda^2} \frac{\sin^2 \left(\tilde{\lambda} t_\varphi\right)}{\tilde{\lambda}^2} \frac{{\rm d} x_{\rm h}^2}{\alpha^2 \chi^2}
    + E^x {\rm d} \Omega^2
    \,,
    \label{eq:Spacetime metric homogeneous - modified - Internal time}
\end{eqnarray}
for arbitrary functions $\lambda$ and $\chi$, where $E^x$ is implicitly given in terms of $t_\varphi$ by the solution to (\ref{eq:Curvature-triad - Homogeneous - modified}).
Notice that the reference value $\tilde{\lambda}$, which is always constant, can be eliminated completely by scaling the time coordinate to $\tilde{t}_\varphi=\tilde{\lambda} t_{\varphi}$.
One can now check that the lapse function of (\ref{eq:Spacetime metric homogeneous - modified - Internal time}) matches the one obtained canonically, (\ref{eq:Lapse homogeneous - modified - Internal time}).
Finally, we can take the inverse of the radial component of (\ref{eq:Spacetime
  metric homogeneous - modified - Internal time}) as the structure function,
take its time derivative and confirm that it indeed satisfies the canonical
equation of motion (\ref{eq:Structure function EoM - modified - Internal
  time}) since
\begin{eqnarray}
    \partial_{t_\varphi} \ln \left(\frac{1}{\alpha^2 \chi^2} \frac{\tilde{\lambda}^2}{\lambda^2} \frac{\sin^2 \left(\tilde{\lambda} t_\varphi\right)}{\tilde{\lambda}^2}\right)
    &=&
    2 \frac{\tilde{\lambda}}{\tan \left(\tilde{\lambda} t_\varphi\right)} \left( 1 - \Lambda E^x + \frac{\tilde{\lambda}^2}{\lambda^2} \frac{\sin^2 \left(\tilde{\lambda} t_\varphi\right)}{\tilde{\lambda}^2} \right)
    \\
    &&\qquad\quad \times
    \left( 1 - \Lambda E^x
    + \left( 1 - 2 \frac{\partial \ln \lambda^2}{\partial \ln E^x}\right) \frac{\tilde{\lambda}^2}{\lambda^2} \frac{\sin^2 \left(\tilde{\lambda} t_\varphi\right)}{\tilde{\lambda}^2}
    \right)^{-1}
    \nonumber
    \,.
\end{eqnarray}

\subsection{Proof of extremality}
\label{sec:Extremality}

We are now ready to prove our previous claim that any solution of
(\ref{eq:Minimum radius equation}) for the new quantum coordinate
singularities in homogeneous regions of the modified Schwarzchild geometry
(\ref{eq:Spacetime metric - modified - Schwarzschild}) dynamically corresponds
to an extremal radius (with some restrictions in case of higher multiplicities of the roots). Inserting $E^x=x^2$  in equation~(\ref{eq:Minimum
  radius equation}), we obtain the condition that the function defined by
\begin{equation}
  b(E^x)=  1+\lambda^2 \left(1 - \frac{2 M}{x} - \frac{\Lambda x^2}{3}\right) =
  1+\lambda^2 \left(1 - \frac{2 M}{\sqrt{E^x}} - \frac{\Lambda E^x}{3}\right)
\end{equation}
vanishes at $E^x=(x_{\lambda}^{(i)})^2$. With our identification of $c_x=2M$
and equation (\ref{eq:Curvature-triad - Homogeneous
  - modified}), it
then follows that
\begin{equation} \label{Km}
  \sin^2(\tilde{\lambda}K_{\varphi})= 1-b(E^x)
\end{equation}
and $\sin^2(\tilde{\lambda}K_{\varphi})=1$ at any solution of (\ref{eq:Minimum
  radius equation}).

The function $b(E^x)$ is always positive for $E^x$ between $x_{\rm H}^2$ and
$x_{\Lambda}^2$. Outside of this range, its first zeros on the two sides are
defined as $(x_{\lambda}^{(\pm)})^2$. If these solutions have odd
multiplicity, $b(E^x)$ becomes negative for values of $E^x$ less than
$(x_{\lambda}^{(-)})^2$ or greater than $(x_{\lambda}^{(+)})^2$, such that
there is no corresponding $K_{\varphi}$ for which equation (\ref{Km}) could
hold. The only dynamical option is for $E^x$ to start increasing after it
reaches $(x_{\lambda}^{(-)})^2$, or to start decreasing after it reaches
$(x_{\lambda}^{(+)})^2$. These values are therefore dynamical extrema of
$E^x$, as well as of $\sin^2(\tilde{\lambda}K_{\varphi})$ as a measure of
curvature. The extremum of $\sin^2(\tilde{\lambda}K_{\varphi})$ also
characterizes these extrema of $E^x$ as hypersurfaces of reflection symmetry.
If there are solutions of (\ref{eq:Minimum radius equation}) other than
$x_{\lambda}^{(\pm)}$, they will not play a role for dynamical solutions
obtained by extending the static region across coordinate singularities.
Unlike in Schwarzschild or GP coordinates, the metric (\ref{eq:Spacetime
  metric homogeneous - modified - Internal time}) remains regular at any
hypersurface of reflection symmetry, given by
$t_\varphi = \pi / (2\tilde{\lambda})$ and can be used for this purpose.  We
discuss the global structure in Section~\ref{sec:Global structure} and perform
a detailed analysis of singularity resolution in Section~\ref{sec:Interior
  physics}.

For completeness, we mention why the previous conclusions are restricted to
$x_{\lambda}^{(\pm)}$ with odd multiplicity. If $x_{\lambda}^{(-)}$ or
$x_{\lambda}^{(+)}$ has even multiplicity, the function $b(E^x)$ stays
positive in a neighborhood around such a solution, and dynamical values of $\sqrt{E^x}$
beyond this solution are not ruled out by the condition (\ref{Km}). A detailed
dynamical analysis would be required for a complete understanding, which we
will not perform in this paper. Nevertheless, such cases, even though they are
not generic, might be of interest as new types of extremal black holes, defined in the general sense that initially distinct coordinate singularities coincide for some parameter choices. For extremal black holes in classical general relativity, the inner and outer horizon coincide, implying a double root of the defining equation. For new extremal solutions in emergent modified gravity, the black-hole horizon retains its non-extremal form, but the internal structure changes because some of the interior coordinate singularities coincide. Such solutions
could shed light on possible connections, such as tunneling processes, between
space-time regions within $x_{\lambda}^{(\pm)}$ and values beyond these
extrema that lie in disjoint space-time regions in the non-extremal case.

In these discussions, we assumed that the areal gauge is always available in a region around the extremal radius, which generically is the case for physically acceptable solutions: This gauge gauge would not be available in a given region if and only if $E^x$ is completely constant there in space and time. (If $E^x$ is constant only spatially in a given slicing but depends on time, one can always deform the slicing locally such that $(E^x)'\not=0$.) The diffeomorphism constraint then implies that $K_{\varphi}'=0$ in this region because $E^{\varphi}$ is non-zero around these new coordinate singularities. Maintaining $\dot{E}^x=0$ in the region implies that $K_{\varphi}$ cannot change in time according to the $E^x$-equation of motion, and stays at the value required for reflection symmetry on all slices in the region. With these conditions, the equation of motion (\ref{Kphidot}) for $K_{\varphi}$ implies a first-order differential equation for $\lambda$ which, for zero cosmological constant, has the solution $\lambda=(c/\sqrt{E^x}-1)^{-1/2}$ with a constant $c$. This behavior is not of the form usually considered for holonomy modifications, and could only be realized in a finite range of $E^x$.

\section{Global structure}
\label{sec:Global structure}

The global structure of classical space-time solutions may change if singularities are
resolved or if there are large-scale effects from holonomy modifications that
could alter the asymptotic behavior. Different implications are possible,
depending on the form of holonomy modifications as determined by the function $\lambda(E^x)$.

\subsection{Asymptotic flatness and zero-mass limit}

If we set $\Lambda=0$, we may impose asymptotic flatness.  For this purpose,
the metric (\ref{eq:Spacetime metric - modified - Schwarzschild}) for the
modified Schwarzschild exterior is the relevant one, which
takes the asymptotic form
\begin{equation}
    {\rm d} s^2 \approx
    - \frac{{\rm d} t^2}{\alpha^2 \chi(\infty)^2}
    + \left( 1 + \lambda(\infty)^2
    \right)^{-1} \frac{{\rm d} x^2}{\chi(\infty)^2}
    + x^2 {\rm d} \Omega^2 
    \,.
    \label{eq:Spacetime metric - modified - Schwarzschild - asymptotic}
\end{equation}
Therefore, asymptotic flatness partially determines the limiting behavior of the two modification functions by the relations
\begin{eqnarray}\label{eq:Asymptotic flatness condition}
    \alpha = \chi(\infty)^{-1}
    \quad , \quad
    \chi(\infty) = 1 / \sqrt{1+\lambda(\infty)^2}
    \,,
\end{eqnarray}
which require $\chi$, and hence $\lambda$, to be asymptotically constant because $\alpha$ is a constant, implied by one of our integrations.
We will impose this condition in the following.

We may independently ask that flat space-time be recovered in the zero-mass
limit $M\to0$, in which case
the line element takes the form
\begin{equation}
    {\rm d} s^2 \approx
    - \frac{{\rm d} t^2}{\alpha^2 \chi^2}
    + \left( 1 + \lambda^2
    \right)^{-1} \frac{{\rm d} x^2}{\chi^2}
    + x^2 {\rm d} \Omega^2
    \,.
    \label{eq:Spacetime metric - modified - Schwarzschild - zero mass}
\end{equation}
The radial component is rendered flat by the choice
\begin{equation}\label{eq:Zero mass limit condition 1}
    \chi = 1 / \sqrt{1+\lambda^2}
    \,,
\end{equation}
which is stronger than the asymptotic-flatness condition (\ref{eq:Asymptotic flatness condition}), but compatible with it.
On the other hand, the time component becomes flat only if
\begin{equation}\label{eq:Zero mass limit condition 2}
    \chi = {\rm constant} =:\chi_0
    \quad , \quad
    \alpha = \chi_0^{-1}
    \,,
\end{equation}
in which case the metric is asymptotically flat too according to
(\ref{eq:Asymptotic flatness condition}).  Therefore, the radial and time
components can both be flat in the zero-mass limit only if $\lambda$ is
constant, or if $\lambda$ depends not only on $E^x$ but also on the mass (for
instance via renormalization) such that it approaches a constant as
$M\to0$. Here, we will not consider the second possibility.

If one wishes to recover flat space as part of the classical geometry in the zero-mass
limit, one needs a constant radial component. In this case, we impose the
conditions (\ref{eq:Asymptotic flatness condition}) and (\ref{eq:Zero mass
  limit condition 1}) such that the static region is described by
\begin{equation}
    {\rm d} s^2 =
    - \left(1 - \frac{2M}{x}\right) \frac{1+\lambda^2}{1+\lambda_\infty^2} {\rm d} t^2
    + \left( 1 - \frac{\lambda^2}{1+\lambda^2} \frac{2M}{x}
    \right)^{-1} \left(1 - \frac{2M}{x}\right)^{-1} {\rm d} x^2
    + x^2 {\rm d} \Omega^2
    \,,
    \label{eq:Spacetime metric - modified - Schwarzschild - Asymptotic and zero mass - classical space}
\end{equation}
where $\lambda_\infty = \lim_{x\to\infty}\lambda (x)$. The conditon of having flat space in the zero-mass limit could therefore be used to determine one of the modification functions. However, the spatial
quantum geometry realized in LQG on Planckian scales can be expected to leave traces even
in an effective description, such that a non-Euclidean 3-dimensional space
could be acceptable at zero mass.  Because the focus of this paper is on black holes in
effective loop quantum gravity we will adopt this perspective and do not require (\ref{eq:Zero mass
  limit condition 1}).

Therefore, we
impose only the asymptotic-flatness condition (\ref{eq:Asymptotic flatness
  condition}), such that the static region is described by
\begin{equation}
    {\rm d} s^2 =
    - \left(1 - \frac{2M}{x}\right) \frac{{\rm d} t^2}{\chi^2 (1+\lambda_\infty^2)}
    + \left( 1 + \lambda^2 \left( 1 - \frac{2M}{x} \right)
    \right)^{-1} \left(1 - \frac{2M}{x}\right)^{-1} \frac{{\rm d} x^2}{\chi^2}
    + x^2 {\rm d} \Omega^2
\end{equation}
with the overall factor $\chi$ required to have the asymptotic limit
$\chi(\infty)=1/\sqrt{1+\lambda_\infty^2}=:\chi_0$, but otherwise arbitrary.
While several of the following analyses hold for arbitrary $\chi(x)$, we will,
for simplicity of the notation, assume a constant overall factor, $\chi=\chi_0$.
The metric then further simplifies to
\begin{equation}
    {\rm d} s^2 =
    - \left(1 - \frac{2M}{x}\right) {\rm d} t^2
    + \left( 1 + \lambda^2 \left( 1 - \frac{2M}{x} \right)
    \right)^{-1} \left(1 - \frac{2M}{x}\right)^{-1} \frac{{\rm d} x^2}{\chi_0^2}
    + x^2 {\rm d} \Omega^2
    \,,
    \label{eq:Spacetime metric - modified - Schwarzschild - Asymptotic and zero mass}
\end{equation}
This chart, and its homogeneous counterpart, will suffice for most of the following applications.

A second argument to prefer (\ref{eq:Spacetime metric - modified -
  Schwarzschild - Asymptotic and zero mass}) over (\ref{eq:Spacetime metric -
  modified - Schwarzschild - Asymptotic and zero mass - classical space}) is
the realization that the time component dominates in the non-relativistic
regime: The proper-time interval for a timelike object moving along a radial curve with
coordinate velocity $v={\rm d} x / {\rm d} t$ is given by
${\rm d} \tau = \sqrt{-{\rm d} s^2} = \sqrt{-g_{tt} + g_{xx} v^2} \,{\rm d} t =
\left(\sqrt{-g_{tt}} + O (v^2) \right) {\rm d} t$.  Therefore,
(\ref{eq:Spacetime metric - modified - Schwarzschild - Asymptotic and zero
  mass - classical space}) would imply a non-classical gravitational potential even for
slow-moving objects, whereas such effects are suppressed by a factor of $v^2$ if
(\ref{eq:Spacetime metric - modified - Schwarzschild - Asymptotic and zero
  mass}) is used in this regime.

\subsection{The maximal extension}

For any $\lambda(E^x)$ such that (\ref{eq:Minimum radius equation}) has only
two roots $x_{\lambda}^{(\pm)}$ we obtain a similar global structure to that
of \cite{alonso2022nonsingular,alonso2023charged}, where constant $\lambda$
was assumed.
To see this, we start with a static Schwarzschild region with metric
(\ref{eq:Spacetime metric - modified - Schwarzschild}). 
The GP metric (\ref{eq:Spacetime metric - modified - GP simplified}) covers
this region as well as two homogeneous Schwarzschild regions with metric
(\ref{eq:Spacetime metric homogeneous - modified - Schwarzschild}), one for
$x< x_{\rm H}$ and another for $x> x_\Lambda$.
Thus, we can sew all three charts together into a black-hole like space-time region
that ends at the minimum radius $x_{\lambda}^{(-)}$, rather than $x=0$, and at
the maximum radius $x_{\lambda}^{(+)}$, rather than $x\to\infty$ (assuming
that both $x_{\lambda}^{(\pm)}$ exist as finite positive
solutions of (\ref{eq:Minimum radius equation})).
The additional homogeneous gauge based on $K_{\varphi}$ as an internal time,
implying the metric (\ref{eq:Spacetime metric homogeneous - modified -
  Internal time}), covers two homogeneous Schwarzschild regions connected to
each other through $x_{\lambda}^{(+)}$ or $x_{\lambda}^{(-)}$.
This gauge can be used to sew together GP charts with two complete homogeneous
charts at the
extremal radii, forming a single wormhole-like space-time.

A transformation to null coordinates leads to the maximal extension as follows.
Consider two null frames described by the 1-forms ${\rm d} u = v^{(u)}_\mu {\rm d} x^\mu$ and ${\rm d} v = v^{(v)}_\mu {\rm d} x^\mu$.
Replacing the time and the space coordinates with null coordinates $u$ and
$v$ transforms a metric of the general form (\ref{eq:line element - spherical -
  modified}) into a Kruskal-Szekeres form. Following the general results from
\cite{EMGPF}, we obtain the line element
\begin{equation}
    {\rm d} s^2
    =
    - \frac{N^2 - \tilde{q}_{x x}(N^x)^2}{v^{(u)}_t v_t^{(v)}} {\rm d} u {\rm d} v
    \label{eq:Kruskal-Szekeres metric form}
\end{equation} 
(suppressing the angular part of the metric).
Using the null condition $g^{\mu \nu} v^{(i)}_\mu v^{(i)}_\nu = 0$ for $i=u , v$, the components of the null frames can be related to one another by
\begin{equation}\label{eq:Null components relation}
    \frac{v_t^{(i)}}{v_x^{(i)}}
    = s_{(i)} \sqrt{N^2 \tilde{q}^{x x}} + N^x
    \ ,
\end{equation}
where $s_{(u)} = - 1$ and $s_{(v)} = + 1$, such that one is an ingoing family of light rays and the other is outgoing.

Using the exterior metric (\ref{eq:Spacetime metric - modified - Schwarzschild}), which is static, we can choose null geodesics such that the components $v_t^{(i)}$ are constant (being Killing conserved quantities) and can be absorbed into $u$ and $v$.
Using this, the null 1-forms can be written as
\begin{eqnarray}
    {\rm d} u
    &=&
    {\rm d} t + \alpha \left(1 - \frac{2 M}{x} - \frac{\Lambda x^2}{3}\right)^{-1} \left( 1 + \lambda^2 \left( 1 - \frac{2 M}{x} - \frac{\Lambda x^2}{3}\right)
    \right)^{-1/2} {\rm d} x
    \ , \nonumber\\
    {\rm d} v
    &=&
    {\rm d} t - \alpha \left(1 - \frac{2 M}{x} - \frac{\Lambda x^2}{3}\right)^{-1} \left( 1 + \lambda^2 \left( 1 - \frac{2 M}{x} - \frac{\Lambda x^2}{3} \right)
    \right)^{-1/2} {\rm d} x
    \ .
    \label{eq:Null 1-forms - Kruskal-Szekeres}
\end{eqnarray}
The null and Schwarzschild coordinates are then easily related by
\begin{equation}
    u = t - x_*
    \ , \qquad
    v = t + x_*
    \ ,
    \label{eq:Null coordinates - Kruskal-Szekeres}
\end{equation}
where $x_*$ is the direct integration of the $x$-component of the 1-forms (\ref{eq:Null 1-forms - Kruskal-Szekeres}), which can be done once $\lambda$ is specified.

In these null coordinates, the metric is simply
\begin{equation}
    {\rm d} s^2
    =
    - (\alpha\chi)^{-2} \left(1 - \frac{2 M}{x} - \frac{\Lambda x^2}{3}\right) {\rm d} u {\rm d} v
    \ .
    \label{eq:Kruskal-Szekeres metric form - modified}
\end{equation}
In the limit $x \to x_{\rm H}$ the null coordinates take the values $u \to + \infty$ and $v \to - \infty$, while $x_* \to - \infty$; on the other hand, in the limit $x \to x_\Lambda$, they take the values $u \to + \infty$ and $v \to + \infty$, and $x_* \to \infty$. 
We partially overcome these divergences by the usual coordinate transformation
\begin{equation}
    U = - e^{-u / ( 4 M \alpha)}
    \ , \quad
    V = e^{v / ( 4 M \alpha)}
    \ ,
\end{equation}
such that the region $x_{\rm H}<x<x_\Lambda$ corresponds to $U \in ( - \infty , 0)$, $V \in ( 0 , + \infty )$, where $U,V=0$ at the Schwarzschild horizon, and $U=-\infty,V=+\infty$ at the cosmological horizon.
The metric becomes
\begin{equation}
    {\rm d} s^2
    =
    - F_{\text{E}} \left(x\right) {\rm d} U {\rm d} V
\end{equation}
with
\begin{equation}
    F_{\text{E}} (x) =
    \frac{16 M^2}{\chi^2} \left(1 - \frac{2 M}{x} - \frac{\Lambda x^2}{3}\right) e^{ - 2 x_* / ( 4 M \alpha)}
    \ ,
\end{equation}
and $x = x (U,V)$ given implicitly by the solution to
\begin{equation}
    U V = - e^{ 2 x_* / ( 4 M \alpha)}
    \ .
\end{equation}

In the limits $x\to x_{\rm H}$ and $x\to x_\Lambda$ we obtain $F_{\rm E} \to 0$ for finite and differentiable functions $\lambda$ and $\chi$, as expected from null surfaces.
Therefore, we can extend the null coordinates to $x<x_{\rm H}$ by taking
negative values for $U , V$ (which must be possible, since the existence of such a region is implied by the GP metric).
Finally, performing the conformal transformation
\begin{equation}
    \bar{U} = \text{arctan} (U)
    \ , \hspace{1cm}
    \bar{V} = \text{arctan} (V)
    \ ,
\end{equation}
we obtain the metric
\begin{equation}
    {\rm d} s^2
    =
    - \bar{F}_{\text{E}} \left(x\right) {\rm d} \bar{U} {\rm d} \bar{V}
    \,.
    \label{eq:KS vacuum metric - conformal}
\end{equation}
In these new coordinates, the values $U\to-\infty$ and $V\to+\infty$,
corresponding to the horizon $x=x_\Lambda$, are located at $\bar{U}=-\pi/2$
and $\bar{V}=+\pi/2$, respectively.  We can therefore extend the chart past
the cosmological horizon by including values of $\bar{U}<-\pi/2$ and
$\bar{V}>+\pi/2$ in the ranges of the conformal coordinates.

\begin{figure}[h]
    \centering
    \includegraphics[trim=6cm 0cm 6.1cm 0cm,clip=true,width=0.75\columnwidth]{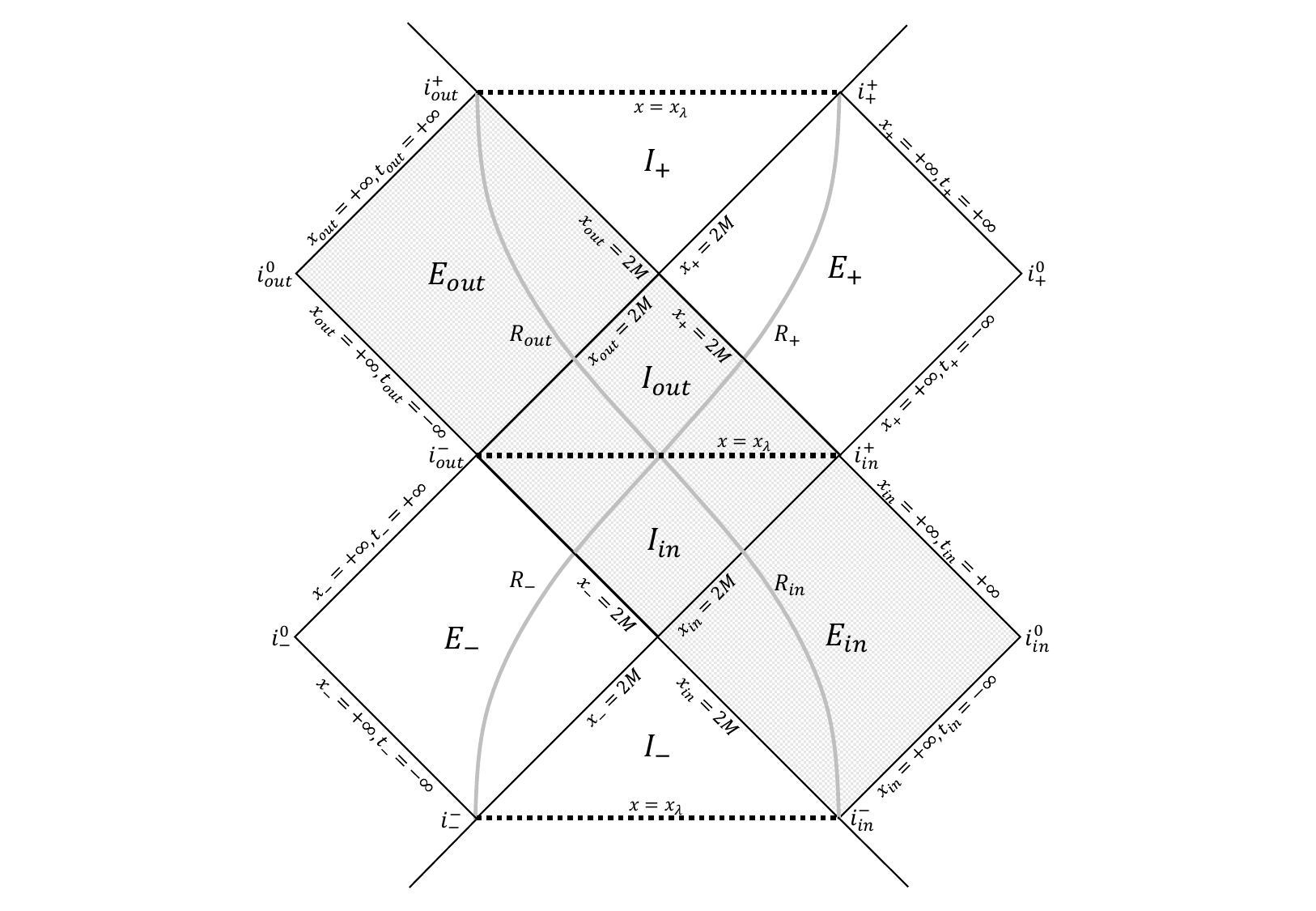}
    \caption{Maximal extension of the wormhole solution in vacuum with vanishing cosmological constant.
    This case has only one solution for the coordinate $x_\lambda$ of the
    hypersurface of reflection symmetry.
    The region $E_{\rm in}\cup I_{\rm in}\cup I_{\rm out}\cup E_{\rm out}$, corresponding to the shaded area, is the wormhole solution obtained by the sewing process.
    The gray lines $R_{i}$ denote geodesics falling from (to) the remote past (far future).}
    \label{fig:Holonomy_KS_Vacuum_Wormhole-Periodic}
  \end{figure}

The usual periodic Penrose diagram for the maximal extension therefore
follows, see Figures
\ref{fig:Holonomy_KS_Vacuum_Wormhole-Periodic}-\ref{fig:Holonomy_KS_Vacuum_Wormhole-Periodic-Cosmo-NO
  MaxRadius}. This global structure was first obtained in
\cite{alonso2022nonsingular,alonso2023charged} for constants
$\lambda=\tilde{\lambda}$, $\chi=\chi_0$, and $\alpha=1$, provided we still have
exactly two finite solutions $x_{\lambda}^{(\pm)}$ for the extremal radii if
$\Lambda\not=0$, and only one solution $x_{\lambda}^{(-)}$ for $\Lambda=0$.
We therefore have a similar global structure with the exception that the
metric is allowed to have arbitrary functions $\lambda (x)$ and $\chi (x)$,
and a free constant $\alpha$.  An important result of this generalization is
that the maximum radius can be avoided depending on the choice of $\lambda$.
In particular, non-constant $\lambda\propto 1/\sqrt{E^x}$, corresponding to
the $\bar{\mu}$-scheme in models of loop quantum gravity but now realized in a
covariant fashion, does not develop such a maximum radius surface as we will
see below.

\begin{figure}[h]
    \centering
    \includegraphics[trim=1.65cm 0cm 2cm 0cm,clip=true,width=0.75\columnwidth]{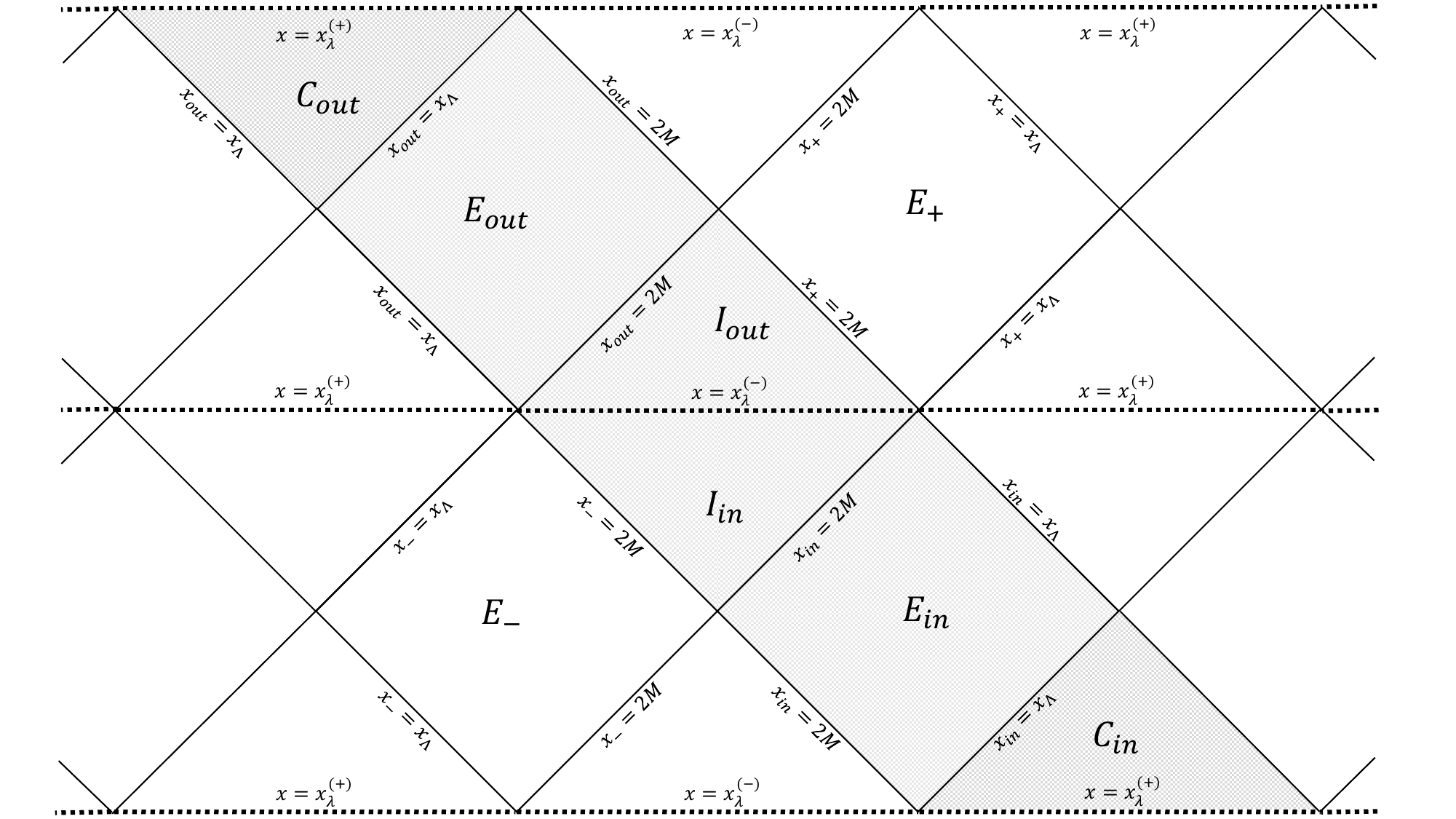}
    \caption{Maximal extension of the wormhole solution in vacuum on a bounded de Sitter background.
    The region $E_{\rm in}\cup I_{\rm in}\cup I_{\rm out}\cup E_{\rm out}$, corresponding to the shaded area, is the wormhole solution obtained by the sewing process.}
    \label{fig:Holonomy_KS_Vacuum_Wormhole-Periodic-Cosmo-MaxRadius}
  \end{figure}
  
\begin{figure}[h]
    \centering
    \includegraphics[trim=1.65cm 0cm 2cm 0cm,clip=true,width=0.75\columnwidth]{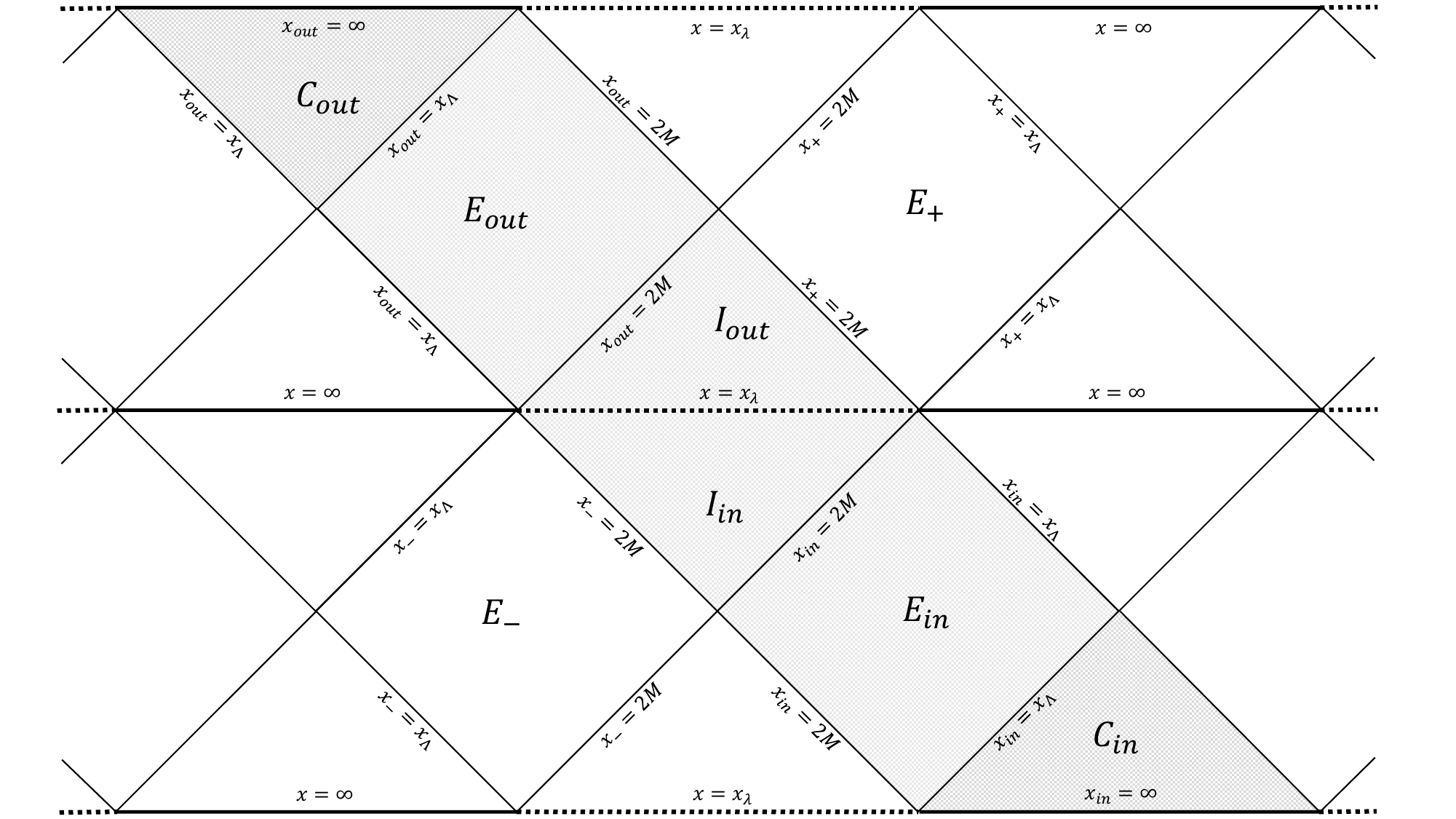}
    \caption{Maximal extension of the wormhole solution in vacuum on an unbounded de Sitter background.
    The region $E_{\rm in}\cup I_{\rm in}\cup I_{\rm out}\cup E_{\rm out}$ is the wormhole solution obtained by the sewing process.}
    \label{fig:Holonomy_KS_Vacuum_Wormhole-Periodic-Cosmo-NO MaxRadius}
\end{figure}

\subsection{Physical global structure}

We emphasize that the global solution is not fully determined by
solving the equations of motion alone.  For simplicity, we will consider the
case of a vanishing cosmological constant given by the diagram in
Fig.~\ref{fig:Holonomy_KS_Vacuum_Wormhole-Periodic}.  This is not necessarily
the physical space-time as we show now with a simple explicit example.

\begin{figure}[h]
    \centering
    \includegraphics[trim=5.5cm 5cm 5.6cm 5.25cm,clip=true,width=0.75\columnwidth]{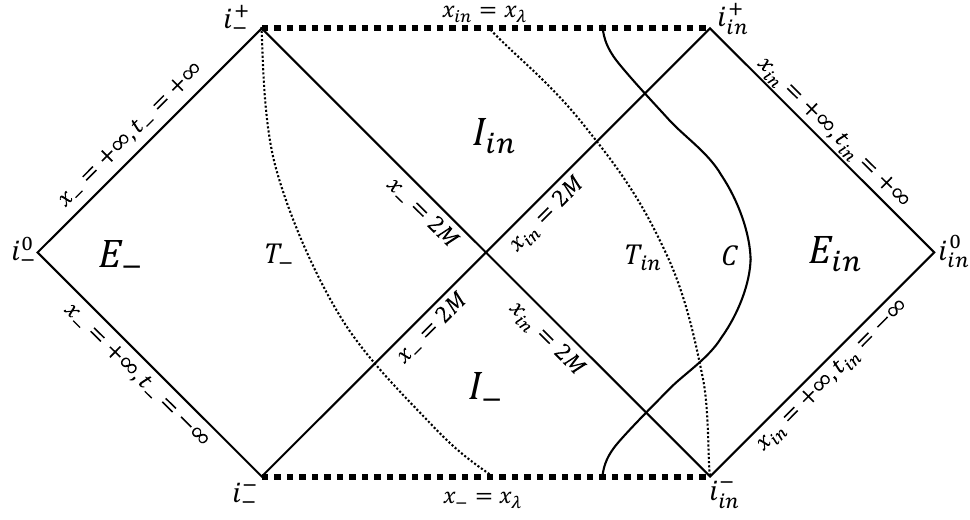}
    \caption{Alternative wormhole solution in vacuum with vanishing
      cosmological constant, periodically identified at the dashed lines of $x_{\lambda}$.
    The solid line labeled by $C$ is an example of a closed timelike curve.
    The dotted line given by $T=T_{\rm in}\cup T_-$ is an example of a
    timelike curve falling from past timelike infinity of $E_{\rm in}$,
    crossing the wormhole, and continuing all the
    way to future timelike infinity of $E_-$.}
    \label{fig:Holonomy_KS_Vacuum_Wormhole-Closed}
\end{figure}

Any sliced portion of the diagram Fig.~\ref{fig:Holonomy_KS_Vacuum_Wormhole-Periodic} is a local vacuum solution. Therefore, if we take multiple slices of this diagram and glue them smoothly we arrive at a new global solution.
As a simple example of such a procedure, consider the region $E_{\rm in} \cup I_{\rm in} \cup E_- \cup I_-$, sliced from the maximal extension at $x_{\rm in}=x_\lambda$ and $x_-=x_\lambda$.
We can then glue smoothly the two boundaries $x_{\rm in}=x_\lambda$ and $x_-=x_\lambda$ because the two regions are locally identical.
The result is shown in Fig.~\ref{fig:Holonomy_KS_Vacuum_Wormhole-Closed}, which has the peculiarity that closed timelike curves exist.

The diagrams  in Fig.~\ref{fig:Holonomy_KS_Vacuum_Wormhole-Periodic} and
Fig.~\ref{fig:Holonomy_KS_Vacuum_Wormhole-Closed} are both valid
representations of the local solutions of the equations of motion, but they
describe two different physical space-times as evidenced by their global
properties such as the existence of closed timelike curves in the latter but
not in the former.  We conclude that the determination of the physical global
solution requires extra input (such as the absence of closed timelike curves)
which may be seen as extended boundary conditions.

\subsection{Presence of matter}

The solutions obtained above are in a vacuum because no matter terms were
considered in the constraints. Nevertheless, the solution contains the parameter $M$ which,
through the classical limit and its equivalence to the
observable of the system, we may interpret as the mass of the black hole that
produces the gravitational field and the curvature of space-time.
In classical black holes, this mass can be interpreted as resting at the
singularity, representing a trace of all the matter that collapsed to form the
black hole.
However, the space-time implied by
Fig.~\ref{fig:Holonomy_KS_Vacuum_Wormhole-Periodic} has no such singularity
and forces us to face an interpretational quandary as evidenced by the
following thought experiment.

An observer may fall into the black hole from $E_{\rm in}$, cross the hypersurface $x=x_\lambda^{(-)}$, and exit to the region $E_{\rm out}$ or even $E_+$ without ever seeing any matter because the system is in vacuum.
So, is there really a mass or not?
The answer could be no, and in that case, the global space-time is akin to an
eternal black hole or a wormhole. 
If the answer is that there must be some matter somewhere that accounts for
the parameter $M$, then we need a model of emergent modified gravity coupled to matter such that the
space-time solution will differ from the vacuum case somewhere near the
maximum-curvature hypersurface.
Figure~\ref{fig:Holonomy_KS_Vacuum_Wormhole-Periodic} then cannot give us the
physical global structure because it cannot result from the gravitational
collapse of matter. In particular, stability of singularity resolution in
the presence of matter is not guaranteed \cite{EMGscalar,EMGPF}.  Without knowing the
non-vacuum space-time solution near the hypersurface of $x=x_\lambda^{(-)}$, we cannot extend the
space-time through it in an obvious way.  However, we may still use the
existing vacuum solution to draw possible, though not definite, conclusions of
what gravitational collapse would look like, proceeding as follows.

\begin{figure}[h]
    \centering
    \includegraphics[trim=13.25cm 1cm 12.5cm 1.25cm,clip=true,width=0.5\columnwidth]{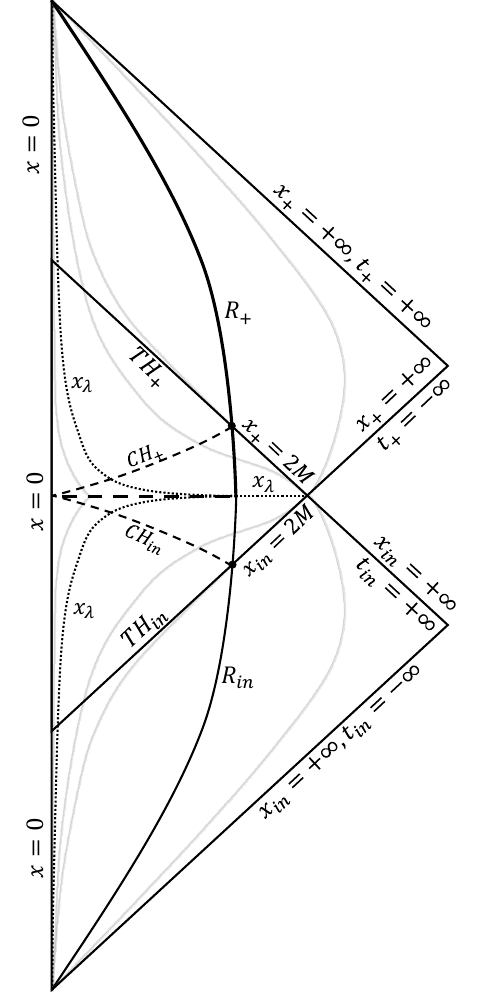}
    \caption{Formation of an interuniversal wormhole.
    The dashed line at the center is the maximum-curvature hypersurface, which
    need not be $x=x_\lambda$ in the interior of a material star.}
    \label{fig:Holonomy_KS_Formation_Wormhole}
\end{figure}

Consider a star with radius $R_0$, such that the region $x>R_0$ is
described by our vacuum solution but the region $x<R_0$ is not. 
Under gravitational collapse, the radius of the star will follow a timelike
worldline $R_{\rm in}$ in Fig.~\ref{fig:Holonomy_KS_Vacuum_Wormhole-Periodic}
starting at $i_{\rm in}^-$, crossing the horizon and arriving at the
reflection-symmetry hypersurface $x=x_\lambda^{(-)}$. 
After this, there are two possibilities: the worldline can either proceed
outwards all the way to $i_{\rm out}^+$, or matter coupling effects may make
it bounce back and proceed to $i_+^+$.
We can now slice the vacuum portions at $R_1= R_{\rm in} \cup R_{\rm out}$ or $R_2= R_{\rm in} \cup R_{+}$, arriving at different space-time solutions for the exterior of the star.
In the latter case, for example, we obtain the diagram in
Fig.~\ref{fig:Holonomy_KS_Formation_Wormhole}, denoting the formation of a
wormhole connecting two universes. We have simply extrapolated the diagram to
the interior of the star whose explicit solution requires specific matter couplings. See \cite{EMGPF} for an exact solution for the collapse of dust.

\begin{figure}[h]
    \centering
    \includegraphics[trim=13cm 1cm 13cm 1.25cm,clip=true,width=0.5\columnwidth]{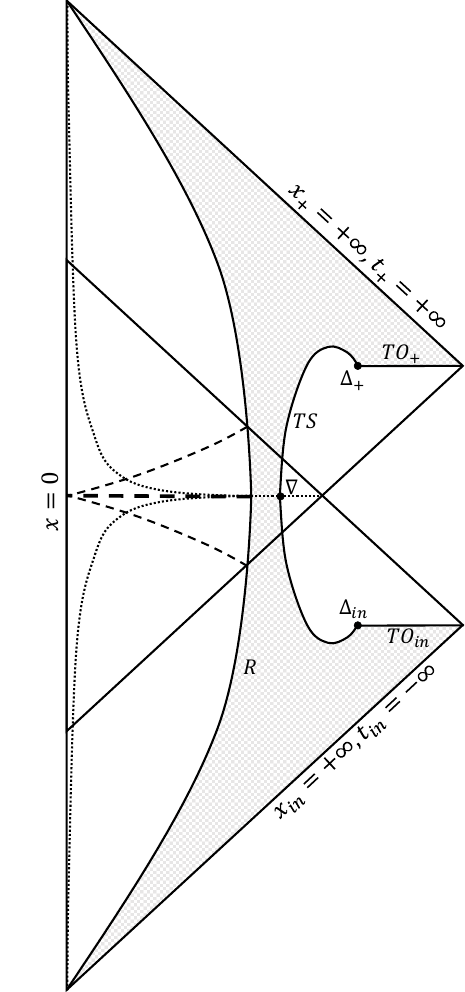}
    \caption{Star-exterior of a black-hole to white-hole transition in a space-time diagram.
    The shaded region is a solution to the modified equations of the star
    exterior, with the two boundaries $R$ and $S = TO_{\rm in} \cup TS \cup TO_{+}$.
    The point where $TS$ intersects the minimum-radius hypersurface is denoted by $\nabla$.}
    \label{fig:Holonomy_KS_Formation_Wormhole_sclicing}
\end{figure}

\begin{figure}[h]
    \centering
    \includegraphics[trim=14cm 2.9cm 13.5cm 3.5cm,clip=true,width=0.5\columnwidth]{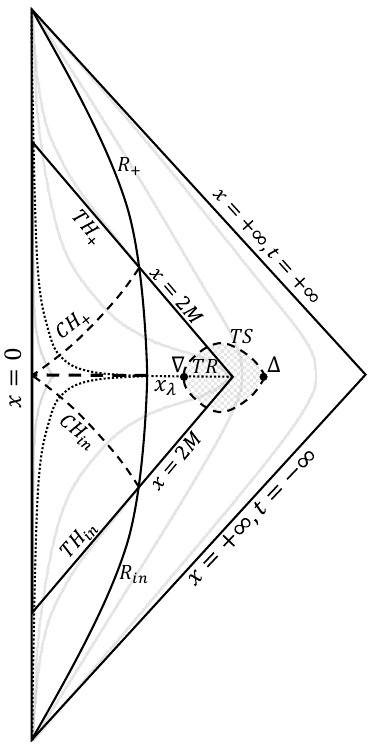}
    \caption{Black-hole to white-hole transition in a conformal diagram.
    The dashed line $(t_{\rm GP} = 0, x<x_\lambda)$ at the center is the
    maximum-curvature hypersurface, enclosing the space-time singularity.
    We refer to the region $TR$ as the transition region, bounded by the surface $TS$.
    The point $\Delta$ is chosen close to the horizon at the coordinates $( t_{\rm GP} = 0 , x = 2 M + \delta)$, with small, positive $\delta$, but otherwise undetermined.
    While the transition region has an undetermined geometry, one can schematically extrapolate the trajectories of constant-$x$ curves.}
    \label{fig:Black-to-White_Hole_Transition_KS}
\end{figure}

The diagram in Fig.~\ref{fig:Holonomy_KS_Formation_Wormhole} can be further sliced as shown in Fig.~\ref{fig:Holonomy_KS_Formation_Wormhole_sclicing}.
The shaded area is a valid solution to the vacuum equations of motion with two boundaries, one given by the star's radius $R$, and the other given by $S$.
The two surfaces $TO_+$ and $TO_{\rm in}$ can then be smoothly joined because the space-time surrounding them are locally identical.
The resulting conformal diagram is shown in Fig.~\ref{fig:Black-to-White_Hole_Transition_KS}. We conclude that a detailed investigation of matter models for gravitational collapse is required before reliable conclusions can be drawn about potential astrophysical implications of non-singular black holes.

\section{Interior physics}
\label{sec:Interior physics}

Unlike the homogeneous Schwarzschild or the GP coordinates, the internal time
coordinates allow us to cross the reflection-symmetry hypersurfaces at
$x_\lambda^{(\pm)}$ since the metric (\ref{eq:Spacetime metric homogeneous -
  modified - Internal time}) is regular at any $x_{\lambda}^{(i)}$ solving (\ref{eq:Minimum radius equation}).
What we find then is that (\ref{eq:Spacetime metric homogeneous - modified -
  Internal time}) connects two different regions on the two sides of the
reflection surface, each with the same geometry (\ref{eq:Spacetime metric homogeneous - modified - Schwarzschild}).
We therefore obtain a quantum extension of the modified Schwarzschild solution,
connected by a reflection-symmetry hypersurface at a minimum radius, in which
case we have two black/white-hole interiors with a range
of $x_\lambda^{(-)}<x<x_{\rm H}$, or at a maximum radius, in which case we have
two de Sitter regions with a range
$x_{\Lambda}<x<x_\lambda^{(+)}$. In this section, we present a detailed
analysis of the interior and the mechanism of singularity resolution implied
by it.

\subsection{Curvature around reflection-symmetry hypersurfaces}

The classical Schwarzschild geometry is a vacuum solution and therefore has
vanishing Ricci curvature. Even though we did not include matter in our new
solutions, they solve modified field equations and there is no guarantee that
the emergent space-time geometry remains Ricci flat. In particular, it could
be possible that these geometries form new physical singularities where even
the Ricci scalar becomes infinite, unlike at the classical Schwarzschild
singularity. We first show that this possible outcome is not realized at a
reflection-symmetry hypersurface, where Ricci curvature is not zero but finite.

The Ricci scalar of the metric (\ref{eq:Spacetime metric - modified - Schwarzschild}) for arbitrary $\lambda$ is given by 
\begin{eqnarray}
    {\cal R} &=& \frac{2}{x^2} \left(
    1 - \chi^2 \left(
    1 + \lambda^2 \left( 
    1 - \frac{3 M^2}{x^2}
    + x \left(1 - \frac{2 M}{x}\right) \left(1 - \frac{3 M}{2 x}\right) \left(\ln \lambda^2\right)' \right)
    \right)
    \right)
    \nonumber\\
    &&
    + 4 \chi^2 \Lambda \left( 1
    + \lambda^2 \left( \frac{3}{2} \left( 1 - \frac{4 M}{3 x} \right)
    + \frac{5 x}{12} \left(1- \frac{9 M}{5 x}\right) (\ln \lambda^2)'\right) \right)
    \nonumber\\
    &&
    - 2 \chi^2 \lambda^2 \Lambda^2 x^2 \left(1 + \frac{x}{6} (\ln \lambda^2)'\right)
    \,,
    \label{eq:Ricci scalar - modified}
\end{eqnarray}
where $\lambda' = \partial \lambda / \partial x$.
Using equation~(\ref{eq:Minimum radius equation}) for the location $x_{\lambda}^{(i)}$ of
a reflection-symmetry hypersurface, we obtain the Ricci scalar
\begin{eqnarray}
    {\cal R} |_{x=x_\lambda^{(i)}} &=& \frac{2}{(x_\lambda^{(i)})^2}
    + \frac{2 \chi^2}{(x_\lambda^{(i)})^2} \lambda^2 \left(\frac{M}{x_\lambda^{(i)}}-\frac{\Lambda (x_\lambda^{(i)})^2}{3}\right) \left(\frac{3 M}{x_\lambda^{(i)}} + \Lambda (x_\lambda^{(i)})^2-2\right)
    \nonumber\\
    &&
    - \frac{2 \chi^2}{x_\lambda^{(i)}} \left(\frac{3 M}{x_\lambda^{(i)}} + \Lambda (x_\lambda^{(i)})^2-2\right) (\ln \lambda)' |_{x=x_\lambda^{(i)}} 
    \label{eq:Ricci scalar - reflection surface}
\end{eqnarray}
which is finite for any finite and differentiable $\lambda(x)$ as long as
$x_\lambda^{(i)}$ is finite and non-zero. Therefore, we need a minimum radius
$x_{\lambda}^{(-)}> 0$ or a maximum radius $x_\lambda^{(+)} \neq
\infty$. 

Finite negative values for $x_{\lambda}^{(-)}$ would imply finite
Ricci curvature at this location, but the classical singularity at $x=0$ would
then be reached by an infalling observer before it could be resolved by
holonomy-type effects.
Thus, a finite and positive minimum radius must exist for the interior geometry to be
regular at small scales.  
At large scales such that $\Lambda x^2 \gg 1$ and
$M/x\ll1$, the Ricci scalar is approximately given by
\begin{eqnarray}
    {\cal R} &\approx& 4 \chi^2 \Lambda
    + \frac{2}{x^2} \left(
    1 - \chi^2
    + \chi^2 \lambda^2 \left( 
    1
    + x \left(\ln \lambda^2\right)' \right)
    \right)
    \nonumber\\
    &&
    + 4 \chi^2 \Lambda \lambda^2 \left( \frac{3}{2}
    + \frac{5 x}{12} (\ln \lambda^2)'\right)
    - 2 \chi^2 \lambda^2 \Lambda^2 x^2 \left(1 + \frac{x}{6} (\ln \lambda^2)'\right)
    \,.
    \label{eq:Ricci scalar - reflection surface - large scales}
\end{eqnarray}
This is finite at $x \to \infty$ only if $\lambda(x)$ falls off at least as
$1/x=1/\sqrt{E^x}$, in which case no divergence occurs at large radii and no
maximum radius is needed to tame it, though the latter might still exist based on this argument alone.

Because the Ricci scalar is an invariant object, its value (\ref{eq:Ricci
  scalar - reflection surface})  at a reflection-symmetry hypersurface,
derived in static Schwarzschild coordinates, is the
same as the Ricci scalar of the homogeneous metric (\ref{eq:Spacetime metric
  homogeneous - modified - Internal time}) at the value
$t_\varphi = \pm \pi / (2 \tilde{\lambda})$ of our internal time.  The
advantage of the latter coordinate
is that it can be extended past the reflection-symmetry
hypersurface. There is therefore a robust sense in which our covariant holonomy
modifications resolve the singularity inside a static Schwarzschild black hole,
even without specifying the quantization ambiguity of the triangulation scheme
$\lambda(E^x)$ chosen to regularize the curvature. We shall explore
specific realizations of this reflection-symmetry hypersurface for a $\mu_0$ (constant $\lambda$) and a $\bar{\mu}$-type scheme in Sections~\ref{sec:mu0 scheme} and \ref{sec:mubar scheme}, respectively.


\subsection{Energy conditions in an emergent space-time theory}\label{sec:Energy conditions and effective Einstein equations}

The Einstein tensor of the emergent metric, defined as usual by
\begin{equation}\label{eq:Einstein tensor}
    G_{\mu\nu} = R_{\mu\nu} - \frac{1}{2} g_{\mu\nu} R
    \,,
\end{equation}
can give us useful information about energy conditions and suitable
interpretations related to singularity resolution.  We may then define an effective stress energy tensor $T_{\mu\nu}^{\rm eff} = (8\pi)^{-1}G_{\mu\nu}$ such that Einstein's equation $G_{\mu\nu}=8\pi T_{\mu\nu}^{\rm eff}$ formally holds. However, this equation does not play the role of an equation of motion because the space-time geometry is determined by emergence from the canonical equations. Moreover, there is no physical field with the effective stress-energy in our vacuum solutions, and in non-vacuum models the matter stress-energy in general differs from the effective stress energy as defined here. Nevertheless, the concept of $T_{\mu\nu}^{\rm eff}$ is useful because of its geometrical relationship with space-time curvature.
The usual energy conditions
can then be analyzed, which, for reference purposes, are given by
\begin{itemize}
    \item Null-energy condition (NEC): For any null $v^{\mu}_{(i)}$\\
    \textbf{Geometric:} $R_{\alpha\beta}v^{\alpha}_{(i)}v^{\beta}_{(i)}\geq 0$.\\
    \textbf{Physical:} $T^{\rm eff}_{\alpha\beta}v^{\alpha}_{(i)}v^{\beta}_{(i)}\geq 0$.
    \item Weak-energy condition (WEC): For any timelike $u^{\mu}_{(i)}$\\
    \textbf{Geometric:} $G_{\alpha\beta}u^{\alpha}_{(i)}u^{\beta}_{(i)}\geq 0$.\\
    \textbf{Physical:} $T^{\rm eff}_{\alpha\beta}u^{\alpha}_{(i)}u^{\beta}_{(i)}\geq 0$.
    \item Strong energy condition (SEC): For any timelike $u^{\mu}_{(i)}$\\
    \textbf{Geometric:} $R_{\alpha\beta}u^{\alpha}_{(i)}u^{\beta}_{(i)}\geq 0$.\\
    \textbf{Physical:}
    $\left(T^{\rm eff}_{\alpha\beta} - \frac{1}{2} T^{\rm eff} g_{\alpha\beta} \right)u^{\alpha}_{(i)}u^{\beta}_{(i)}\geq 0$.
\end{itemize}

For any null $v^{\mu}_{(i)}$, we have
$R_{\alpha\beta}v^{\alpha}_{(i)}v^{\beta}_{(i)} =
G_{\alpha\beta}v^{\alpha}_{(i)}v^{\beta}_{(i)} = 8\pi T^{\rm eff}_{\alpha\beta}v^{\alpha}_{(i)}v^{\beta}_{(i)}$ by definition of the Einstein tensor and the effective stress-energy tensor.
For any timelike $u^{\mu}_{(i)}$, whenever $G_{\alpha\beta}u^{\alpha}_{(i)}u^{\beta}_{(i)} \geq 0$ we also have $T^{\rm eff}_{\alpha\beta}u^{\alpha}_{(i)}u^{\beta}_{(i)} \geq 0$ by definition of the effective stress-energy tensor.
Finally, $8\pi(T^{\rm eff}_{\alpha\beta} - \frac{1}{2} T^{\rm eff}
g_{\alpha\beta}) = G_{\alpha\beta} + \frac{1}{2} R g_{\alpha\beta} =
R_{\alpha\beta}$ by definition of the effective stress-energy tensor. 
Therefore, the geometric and physical versions of the energy conditions are
equivalent to each other in any theory that can be formulated with effective Einstein equations.

For singularity resolution in a model, such as ours, in which solutions do not follow from effective Einstein equations, only the
geometric versions are meaningful. They are based on focusing or
decofusing properties of congruences of geodesics, geometrical properties captured by the Ricci
tensor. Physical matter fields that propagate stress-energy are not required for this purpose.
Importantly, such a theory still has to be fully covariant for the
Ricci tensor to be well-defined, which is not the case for many models
constructed in loop quantum gravity.

Taking these conclusions into account and for further clarity, we reduce the
number of conditions and rename them as follows:
\begin{itemize}
    \item Null geometric condition (NGC):
    \begin{eqnarray}\label{NEC}
    R_{\alpha\beta}v^{\alpha}_{(i)}v^{\beta}_{(i)}\geq 0\quad \text{for any null $v^{\mu}_{(i)}$}.
    \end{eqnarray}
This condition corresponds to the NEC in GR.

    \item Timelike geometric condition (TGC)
     \begin{eqnarray}\label{SEC}
     R_{\alpha\beta}u^{\alpha}_{(i)}u^{\beta}_{(i)}\geq 0\quad \text{for any timelike vector $u^{\mu}_{(i)}$}.
    \end{eqnarray}
This condition corresponds to the SEC in GR.
\end{itemize}
(The WEC has no known geometric interpretation because the
Einstein tensor itself does not have a direct geometric meaning; see \cite{Curiel} for a discussion.)


\subsection{Congruences near a minimum-radius hypersurface}\label{Repulsive}

It is a common scheme in effective approaches to loop quantum gravity that
black-hole singularities may be resolved by the appearance of a minimum
positive radius, determined by the phase-space variable $E^x$. In most cases,
this outcome is simply an extension of results from quantum cosmology to black
holes, using homogeneity of the Schwarzschild interior in order to transfer
cosmological models to this situation. Heuristically, the minimum radius can
then be interpreted as an effect of gravity becoming repulsive at large curvature
when a possibly discrete nature of space may change the dynamics.

However, the case of black holes is crucially different from cosmological
models because it is more sensitive to coordinate or slicing changes going beyond time-reparametrizations: the
homogeneous slicing of a Schwarzschild-type interior does not exist globally,
and therefore cannot be interpreted as a preferred slicing of co-moving
observers. Any physical statements such as a possibly repulsive behavior of
gravity at large curvature, as well as associated effects of singularity
resolution, are meaningful only if they can be shown to be covariant when different slicings are allowed. For instance, it is important to show that the position of the surface which replaces the classical singularity is invariant with respect to coordinate choices, something that is true for our reflection-symmetry hypersurface. Most models of loop quantum gravity have been
shown to fail this crucial condition, but emergent modified gravity is in a
position to present a detailed and consistent analysis.

Using this framework, we will demonstrate in this section that the geometry
near the reflection-symmetry hypersurface is such that the dynamics of a collection of
test particles is defocusing. We will carry out the analysis by studying the
expansion rate $\theta$, derived from the emergent space-time metric. As
usual, this parameter describes infinitesimal changes in the cross-sectional
area or volume for a congruence of null or timelike geodesics. Inside the horizon, both the ingoing and outgoing lightlike congruences
are converging, $\theta<0$, for any family of causally future-directed 
worldlines. Whether congruences are focusing or defocusing, depending on the
geometry of space-time, is determined by the sign of ${\rm d}\theta/{\rm d}\psi$,
where $\psi$ is the affine parameter along lightlike geodesics or proper time
along timelike geodesics.

\subsubsection{Lightlike congruences}

We start by studying null congruences. The relationship between components of ingoing and outgoing radial null worldlines has previously been used as (\ref{eq:Null components relation}) where $v_t$ is constant along null geodesics. Here, we need a more detailed version that includes all possible sign choices. Starting with the null condition
\begin{equation}
    -\frac{1}{N^2} v_t^2 +2\frac{N^x}{N^2} v_tv_x+(\tilde{q}^{xx}-(N^x)^2/N^2)v_x^2=0
\end{equation}
and setting $v_t=-e$ equal to a constant, we obtain the radial component
\begin{equation}
    v^{(s)}_x= e\frac{N^x+sN\sqrt{\tilde{q}^{xx}}}{N^2\tilde{q}^{xx}-(N^x)^2}
    = \frac{es}{N\sqrt{\tilde{q}^{xx}}-s(N^x)^2}
\end{equation}
where $s=\pm1$ is the sign choice obtained when solving the quadratic equation. The sign of $e$, which may be normalized to $e=\pm1$ for null geodesics, also remains free at this point. We therefore have four different combinations of the signs of $s$ and $e$, corresponding to the four directions of a light cone in two space-time dimensions.

The sign freedom should be partially fixed by restricting attention to future-pointing directions. This property is determined by the time component $v_s^t$ with a raised index, given by
\begin{equation}
    v_s^t=\frac{e}{N^2}+\frac{esN^x/N^2}{N\sqrt{\tilde{q}^{xx}}-s(N^x)^2}= \frac{e/N^2}{1-s\sqrt{\tilde{q}_{xx}}N^x/N}\,.
\end{equation}
The corresponding radial component with raised index equals
\begin{equation}
    v_s^x=-\frac{eN^x}{N^2}+\frac{es(\tilde{q}^{xx}-(N^x)^2/N^2)}{N\sqrt{\tilde{q}^{xx}}-s(N^x)^2}= es\frac{\sqrt{\tilde{q}^{xx}}}{N}\,.
\end{equation}
Our equations are based on the condition that $\tilde{q}^{xx}$ and $N$ are always positive.

The expansion parameter of a congruence of null geodesics starting at the
points of a sphere of some initial radius are obtained as
\begin{eqnarray}\label{theta}
    \theta_{s}&=&\frac{1}{\sqrt{-g}}\partial_{\mu}\left(\sqrt{-g}v^{\mu}_{(i)}\right)
    =\frac{es}{\sqrt{\tilde{q}_{xx}}N}\left(\ln q_{\theta\theta}\right)'
\end{eqnarray}
with a radial derivative denoted by the prime. The second equality assumes a stationary coordinate system with time-independent metric components. In this case, the signs of the expansion parameters are determined by the product $es$, or by the sign of $v_s^x$.

For further evaluation, we express this result entirely in terms of phase-space coordinates and the lapse function, independently of any gauge choice, by using $q_{\theta\theta}=E^x$ and Eq.~(\ref{eq:Modified structure function - non-periodic version - simple}), in which we ignore the cosmological constant for the present analysis:
\begin{eqnarray}
    \theta_{s}=es \frac{\chi \left(E^x\right)'}{N E^\varphi \sqrt{E^x}}\sqrt{1+\lambda\left(E^x\right)^2\left(1-\frac{2M}{\sqrt{E^x}}\right)}\,.
\end{eqnarray}
However, a discussion of relevant sign combinations of $e$ and $s$ requires a specific gauge.

It is convenient to evaluate this expression in a gauge, such as GP, which is non-degenerate across the horizon. In this case, using (\ref{eq:Spacetime metric - modified - GP simplified}), the sign of $v_s^t$ is determined by 
\begin{equation} \label{sign}
    1-s \sqrt{\tilde{q}_{xx}}N^x/N=1-s \frac{\chi_{\infty}}{\chi} \sqrt{\frac{2M}{x}+\frac{\chi^2}{\chi_{\infty}^2}-1}\,.
\end{equation}
This expression is always positive for $s=-1$, such that $e=1$ implies future-directed null geodesics. Since $v_s^x<0$ for any $x$, these are ingoing geodesics with negative expansion because $es<0$.

For $s=+1$, (\ref{sign}) changes sign at the Schwarzschild radius, $x=2M$, for any positive function $\chi(E^x)$. For $x>2M$, the expression is still positive and we have $e=+1$ for future-directed null geodesics. For $x<2M$, however, we need $e=-1$ for future direction, such that $es=-1$. Therefore, the expansion changes from positive values at $x>2M$ to negative values at $x<2M$. We conclude that the modification functions preserve the classical extent of the trapped region.

A complete expression for the expansion of interior null congruences in the GP gauge is given by
\begin{eqnarray}
    \theta_{s}=-                   \frac{2\chi_0}{x}\sqrt{1+\lambda(x)^2\left(1-\frac{2M}{x}\right)}\,,
\end{eqnarray}
using the solutions for $E^{\varphi}$ and $N$ and assuming constant $\chi=\chi_0$ for the sake of simplicity.
We can extract information about whether gravity focuses or defocuses these
congruences by deriving an evolution equation for this expansion parameter, obtained from
\begin{eqnarray}\label{evolution-expansion parameter1}
    \frac{{\rm d}\theta_{s}}{{\rm d}\psi}=v_{s}^{x}\frac{{\rm
  d}\theta_{s}}{{\rm d}x}
\end{eqnarray}
where $\psi$ is an affine parameter. The expression
(\ref{evolution-expansion parameter1}) is independent of the sign choice of $s$,
which appears in both factors on the right. For instance, in the classical
Schwarzschild space-time, still evaluated in a GP gauge, we have
${\rm d}\theta_{s}/{\rm d}\psi=-2/x^2<0$ which diverges as
$x\rightarrow0$. This means that both ingoing and outgoing congruences
converge to the center if they are within the horizon.

Quantum gravity is often expected to have some sort of defocusing effect near the
Planck scale, which can happen if the expression in
Eq.~(\ref{evolution-expansion parameter1}) turns positive somewhere in the
black-hole interior. (The defocusing effect may also be expressed as a repulsive behavior of modified gravity acting on test objects on our background vacuum solutions.)
 This outcome is indeed realized with our holonomy-modified
metric. Without loss of generality, we express this evolution in the GP gauge:
\begin{eqnarray}\label{evolution-expansion parameter}
    \frac{{\rm d}\theta_{s}}{{\rm d}\psi}=\frac{2\chi_{0}^{2}}{x^2}\left[-1+\lambda^2(x)\left(\frac{3M}{x}-1\right)-\lambda(x)\lambda'(x)\left(2M-x\right)\right]\,.
\end{eqnarray}
The last two terms in the parenthesis can contribute to defocusing effects at
sufficiently small $x$.  Rewriting them as
\begin{equation}
  \frac{M}{x}(3\lambda^2- x
  (\lambda^2)')- \frac{1}{2}(2\lambda^2- x(\lambda^2)')= -x^3\left(M
  \frac{{\rm d}(\lambda^2/x^3)}{{\rm d}x} -\frac{1}{2} \frac{{\rm
      d}(\lambda^2/x^2)}{{\rm d}x}\right)\,,
\end{equation}
we see that the contribution proportional to $M$ is most significant for small
$x$. It can overcome the classical focusing behavior implied by the constant $-1$ in
(\ref{evolution-expansion parameter}) provided $\lambda^2/x^3$ decreases at
the rate of $1/x^2$, or faster. Therefore, $\lambda(x)$ can increase at most
by $\lambda\propto \sqrt{x}$. The two cases of constant $\lambda$ and 
$\lambda(x)$ falling off  as $1/x$ are included in this range and will be
discussed in detail in the following sections.

Any congruence of null radial geodesics in spherically symmetric space-time will
maintain its shape, such that its shear vanishes, $\sigma_{\mu\nu}=0$.
It also crosses each hypersurface of constant radius in an orthogonal manner,
such that
$\omega_{\alpha\beta}=0$. Therefore, in  the Raychaudhuri equation, which
through the geodesic deviation equation holds in any Riemannian geometry
including an emergent one, only the self-coupling term
$\theta^2$ and curvature through the Ricci tensor contribute.
The Ricci tensor contributes by its contraction with the components of $v^{\mu}_{s}$,
\begin{eqnarray}\label{Ricci-null}
R_{\alpha\beta}v^{\alpha}_{s}v^{\beta}_{s}=-\frac{\chi_{0}^{2}\lambda^2}{x^3}\left(2M+x(\ln\lambda^2)'\left(x-2M\right)\right)\,,
\end{eqnarray}
and therefore we obtain
\begin{eqnarray}
    -\frac{1}{2}\theta_{s}^{2}-  R_{\alpha\beta}v^{\alpha}_{s}v^{\beta}_{s}=\frac{2\chi_{0}^{2}}{x^2}\left[-1+\lambda^2(x)\left(\frac{3M}{x}-1\right)+\lambda(x)\lambda'(x)\left(x-2M\right)\right]
\end{eqnarray}
in agreement with (\ref{evolution-expansion parameter}). This result confirms
that the Raychaudhuri equation indeed holds in unmodified form, emphasizing the importance of a reliable space-time geometry for solutions of a modified theory. Physically, there is an unambiguous conclusion that
emergent Ricci curvature is responsible for defocusing congruences of light rays.

\subsubsection{Timelike congruences and their implications}

A similar computation can be performed for timelike congruences, with slightly longer expressions and fewer cancellations owing to an extra term from the normalization condition. Moreover, we keep $e$ as a free real number in order to obtain a complete family of timelike geodesics rather than just two directions from
$e=\pm1$ used in the lightlike case.

Setting $u_t=-e$, normalization implies the radial component
\begin{equation}
    u_x^{(s)}= e\frac{N^x+s N\sqrt{(1-N^2/e^2)\tilde{q}^{xx}+(N^x)^2/e^2}}{N^2\tilde{q}^{xx}-(N^x)^2}\,.
\end{equation}
Velocity components with raised indices are then given by
\begin{eqnarray}
    u^{t}_{s}&=&\frac{e}{N^2-\tilde{q}_{xx}\left( N^x\right)^2}\left(1+s \frac{\sqrt{\tilde{q}_{xx}}N^x}{N}\sqrt{1-\frac{N^2-\tilde{q}_{xx}\left(N^x\right)^2}{e^2}}\right)\nonumber\\
    &=& \frac{1}{N^2} \frac{e^2+\tilde{q}_{xx}(N^x)^2}{e-s \sqrt{\tilde{q}_{xx}}N^xN^{-1}\sqrt{e^2-N^2+\tilde{q}_{xx}\left(N^x\right)^2}}
    \end{eqnarray}
    and
    \begin{equation}
    u^{x}_{s}=se\frac{\sqrt{\tilde{q}^{xx}}}{N}\sqrt{1-\frac{N^2-\tilde{q}_{xx}\left(N^x\right)^2}{e^2}}\,.
\end{equation}
For massive particles that reach infinity with zero initial
velocity, we have $e=\pm1$, where $e=1$ corresponds to a future-pointing worldline falling in from infinity. 

The sign of $u_s^t$ is determined by the sign of 
\begin{equation}
    \frac{e}{N}-s \sqrt{\tilde{q}_{xx}}N^xN^{-1}\sqrt{\frac{e^2}{N^2}-1+\frac{\tilde{q}_{xx}\left(N^x\right)^2}{N^2}} = X-sY\sqrt{X^2+Y^2-1}
\end{equation}
where we defined $X=e/N$ and $Y=\sqrt{\tilde{q}_{xx}}N^x/N>0$. It can easily
be seen that this expression can vanish only if $Y^2=1$. If $Y<1$, which
corresponds to $x>2M$ in the GP gauge, we have $X>Y\sqrt{X^2+Y^2-1}$ for any
$X>0$, and therefore $u_s^t>0$ if $e>0$ for both choices of $s=\pm1$. If
$Y>1$, or $x<2M$ in the GP gauge, we have $X<Y\sqrt{X^2+Y^2-1}$ for any $X$,
and therefore $u_s^t$ can be positive only for $e>0$ and $s=-1$. There are
only infalling timelike geodesics in this trapped region.

The definition (\ref{theta}) implies
\begin{eqnarray}
    \theta_{s}=\frac{se}{\sqrt{\tilde{q}_{xx}}Nq_{\theta\theta}}\partial_{x}\left(q_{\theta\theta}\sqrt{1+(\tilde{q}_{xx}\left(N^x\right)^2-N^2)/e^2}\right)
\end{eqnarray}
or
\begin{eqnarray}
    \theta_{s}=2 \ {\rm sgn}(e) \ s \ \chi_0 \frac{e^2-1+3M/(2x)}{x\sqrt{e^2-1+2M/x}} \sqrt{1+\lambda(x)^2\left(1-\frac{2M}{x}\right)} 
\end{eqnarray}
in the GP gauge.
The rate of change relative to proper time $\tau$ is given by
\begin{eqnarray}
   \frac{{\rm d}\theta_{s}}{{\rm d} \tau}&=&-\frac{\chi_{0}^{2}}{x^4 \left(2 M+\left(e^2-1\right) x\right)} \left(x \left(9 M^2+8 M \left(e^2-1\right) x+2 \left(e^2-1\right)^2 x^2\right)+\right.\nonumber \\ && \left.\lambda (x)^2 \left(-24 M^3+M^2 \left(32-23 e^2\right) x-2 M \left(3 e^4-10 e^2+7\right) x^2+2 \left(e^2-1\right)^2 x^3\right)\right. \nonumber \\ &&\left.-x (x-2 M) \lambda (x) \left(2 M+\left(e^2-1\right) x\right) \left(3 M+2 \left(e^2-1\right) x\right) \lambda '(x)\right)
\end{eqnarray}
The Ricci tensor contributes to the Raychaudhuri equation  for timelike congruences by
\begin{eqnarray}\label{Ricci-timelike}
    R_{\mu\nu}u^{\mu}_{s}u^{\nu}_{s}&=&-\chi_{0}^{2}\lambda^2\left(x\right)\left(2M+x\left(x-2M\right)(\ln\lambda^2)'\right)\frac{\left(6M^2+\left(4e^2-7\right)Mx-2\left(e^2-1\right)x^2\right)}{2\left(2M-x\right)x^4}\nonumber \\
\end{eqnarray}
which provides defocusing effects in a collection of massive particles. 

The long expressions above obscure the physical effects when considering a generic $e$. Hence, let us focus on the particular case where the collection of massive particles starts at rest from infinity ($e=1$), such  that the evolution of their geodesics takes the simpler form
\begin{eqnarray}\label{Raychaudurri}
    \frac{{\rm d}\theta_s}{{\rm d}\tau}=-\frac{9M \chi_{0}^{2}}{2x^3}\left(1+\lambda^2(x)\left(1-\frac{2M}{x}\right)\right)+\frac{3\chi_{0}^{2}M\lambda^2(x)}{2x^4}\left(x (x-2 M) \left(\ln\lambda^{2}\right)'+2M\right)
\end{eqnarray}
with
\begin{eqnarray}
    R_{\mu\nu}u^{\mu}_{s}u^{\nu}_{s}&=& - \frac{3M\chi_{0}^{2} \lambda^2 (x)}{2x^4} \left(x (x-2 M) \left(\ln\lambda^{2}\right)'+2M \right)
\end{eqnarray}
Unlike null geodesics, any cross-section of a timelike congruence is a 3-dimensional spatial hypersurface. Therefore,  moving from one hypersurface to another may imply non-vanishing shear. Non-zero effects of shear can be computed in the GP-gauge to provide
\begin{eqnarray}
   \sigma_{\mu\nu}\sigma^{\mu\nu}&=& \frac{6M \chi_{0}^{2}}{2x^3}\left(1+\lambda^2(x)\left(1-\frac{2M}{x}\right)\right)\,.
\end{eqnarray}
With this result, we can rewrite equation (\ref{Raychaudurri}) as Raychaudhuri's equation,
\begin{eqnarray}
     \frac{{\rm d}\theta_s}{{\rm d}\tau}=-\frac{1}{3}\theta_{s}^2-\sigma_{\mu\nu}\sigma^{\mu\nu}- R_{\mu\nu}u^{\mu}_{s}u^{\nu}_{s}
\end{eqnarray}

Both cases of geodesics show that the transition from focusing to defocusing is
captured by space-time curvature through the emergent Ricci
tensor. We can formalize this behavior by using our
purely geometric conditions (\ref{NEC}) and (\ref{SEC}):
\begin{itemize}
    \item Null geometric condition:\\
    Substituting the Ricci tensor contracted with tangent vectors to null
    geodesics, (\ref{Ricci-null}), into the NGC (\ref{NEC}), we obtain
    \begin{eqnarray}\label{eq:Null cond interior}
    -\frac{\lambda^2 \chi_0^2}{x^3}\left(2M-x(\ln\lambda^2)'\left(2M-x\right)\right) \ge 0
    \,.
    \end{eqnarray}
    Our vacuum model clearly violates the condition everywhere inside the
    horizon $x<2M$ for any constant or monotonically decreasing $\lambda$.

    \item Timelike geometric condition: \\
    Substituting the Ricci tensor contracted with tangent vectors to timelike
    geodesics falling in from some reference point, (\ref{eq:Null cond interior}), into the TGC  (\ref{SEC})
    we obtain 
    \begin{eqnarray}\label{eq:Timelike cond interior}
    -\chi_{0}^{2} \lambda^2 (x) \frac{\left(3 M+2 \left(e^2-1\right) x\right)}{2x^4} \left(2M-x(\ln\lambda^2)'\left(2M-x\right)\right) \geq 0
    \,.
    \end{eqnarray}
    This condition is violated everywhere in the interior for any constant or monotonically decreasing $\lambda$. 

\end{itemize}

A similar analysis can be done near the hypersurface of maximum radius.
However, a maximal radius does not appear for asymptotically vanishing
holonomy functions $h(x)=\lambda(E^x)/\tilde{\lambda}$. It may therefore be
interpreted as an undesired feature of the asymptotically constant (or increasing) case because it
implies strong modifications in low-curvature regimes.  The $\bar{\mu}$-type
scheme of $\lambda\propto 1/\sqrt{E^x}$, in particular, does not have a
hypersurface of
maximum radius.  We will therefore forego details of an analysis of repulsive
effects beyond a maximal radius.


\subsection{The net stress-energy tensor and gravitational energy}

We define the net stress-energy tensor
\begin{eqnarray}\label{eq:Net stress-energy tensor}
    \bar{T}_{\mu \nu} \equiv T_{\mu \nu} - \frac{G_{\mu \nu} + \Lambda g_{\mu \nu}}{8\pi}
    \,,
\end{eqnarray}
where $T_{\mu \nu}$ is the matter stress-energy tensor (if present, and
derived canonically from matter contributions to the constraints) and $G_{\mu
\nu}$ is the Einstein tensor (\ref{eq:Einstein tensor})  of the emergent
space-time metric.  
In the classical limit, where the Einstein equations hold, the net stress-energy is always zero.
But in general it is non-zero in a modified gravity theory (not necessarily
given by some effective version of LQG).
In particular, in vacuum such that $T_{\mu \nu}=0$, the net stress-energy
tensor has purely gravitational contributions. It inherits the important property
of a covariant conservation law,
$\nabla^\mu \bar{T}_{\mu \nu} = - \nabla^\mu ( G_{\mu \nu} +
\Lambda g_{\mu\nu})/(8\pi) = 0$, from the Bianchi identities, provided the
Einstein tensor is indeed well-defined and covariant.
We may refer to this contribution as local gravitational stress-energy, which has
only non-classical contributions. This new concept is only defined
  relative to general relativity and, to the best of our knowledge, does not
  have an independent physical motivation. Nevertheless, we will show that it
  leads to several statements about modified gravity and new interpretations
  that are intuitively meaningful.

One possible application of this new concept is a reformulation of the
energy conditions.  However, as we will see, these conditions do not play an
important role in singularity resolution when the classical Einstein equations
do not hold.  We will therefore not consider them as strict conditions on
possible modifications, but various signs in these expressions are still of
importance in interpretations related to energy. With this in mind, we make
the following definitions:
\begin{itemize}
    \item Null energy signature (NES):
    \begin{equation}\label{NEC net stress-energy}
        \sigma_{\rm NES}={\rm sgn}(\bar{T}_{\alpha\beta}v^{\alpha}_{(i)}v^{\beta}_{(i)})
        \quad \text{for any null $v^{\mu}_{(i)}$}
        \,.
    \end{equation}
    \item Timelike energy signature (TES):
    \begin{equation}\label{WEC net stress-energy}
    \sigma_{\rm TES}={\rm sgn}(\bar{T}_{\alpha\beta}u^{\alpha}_{(i)}u^{\beta}_{(i)})
        \quad \text{for any timelike $u^{\mu}_{(i)}$}
        \,.
    \end{equation}
  \end{itemize}
As mentioned earlier, the interpretation of the physical form of the SEC is
that of gravity being attractive for the sign imposed by this condition. As an
energy condition, this interpretation is available only
if Einstein equations hold, and therefore we will not consider
its analog as an energy condition in a modified theory.

The Ricci scalar in our models is given by
\begin{equation}\label{eq:Ricci scalar}
    R = \frac{2 \chi_0^2}{x^4} \left( \lambda_\infty^2 x^2
    - \lambda^2 \left(x^2-3 M^2\right)
    - \frac{x}{2} (\lambda^2)' (x-2 M) (2 x-3 M)\right)
\end{equation}
in an asymptotically flat space-time with a suitable value of the constant
$\lambda_{\infty}$ as discussed in Section~\ref{sec:Global structure}.  By
definition of the net stress-energy tensor, the NES (\ref{NEC net
  stress-energy}) has the opposite sign of the left-hand side of the NGC (\ref{NEC}) in
vacuum.  Using a family of null geodesics, we obtain
\begin{equation}\label{eq:Null cond interior - net energy}
  \sigma_{\rm NES}={\rm sgn}
  \left(M/x-(\ln\lambda)'\left(2M-x\right)\right)
\end{equation}
as the NES  by simply flipping the sign of (\ref{eq:Null cond interior}) and
removing positive factors.
Using a family of timelike geodesics at rest at infinity for the TES (\ref{WEC net stress-energy}) we obtain
\begin{equation} \label{TES}
  \sigma_{\rm TES}={\rm sgn}
  \left( 1- \lambda_\infty^2/\lambda^2
    - 2(\ln \lambda)' (2 M-x) \right)
\end{equation}
using (\ref{eq:Timelike cond interior}) and (\ref{eq:Ricci scalar}).
Both signatures are positive in the interior $x<2M$ for 
monotonically decreasing $\lambda$. For constant $\lambda$, (\ref{eq:Null cond
  interior - net energy}) remains positive while (\ref{TES}) vanishes.
Using the concept of a net stress-energy tensor, the NGC and TGC are violated,
while the NES and TES are non-negative  near the hypersurface of minimum
radius.

If we do not restrict these expressions to the interior, we can obtain
  varying signs. For constant $\lambda$, we have $\sigma_{\rm NES}=+1$ for any
  $x$, while $\sigma_{\rm TES}=0$. For $\lambda=\tilde{\lambda}/x$, we have
  $\sigma_{\rm NES}={\rm sgn}(3M/x-1)$ which is negative for $x>3M$, and
  $\sigma_{\rm TES}={\rm sgn}(4M/x-1)$ which is negative for $x>4M$. There is
  therefore negative net stress-energy in most of the exterior region in a
  $\bar{\mu}$-type scheme. We interpret this outcome as an additional
  gravitational binding energy compared with the classical behavior. This
  interpretation is compatible with the acceleration equations
  (\ref{evolution-expansion parameter}) and (\ref{Raychaudurri}) for null
  and timelike congruences, where the $\lambda$-contributions strengthen the classical
  negative term for large $x\gg M$. In the interior, however, the defocusing
  behavior is dominant, turning the signs of the acceleration equations as
  well as the signatures of net stress-energy.

\section{Exterior physics}
\label{sec:Exterior physics}

Different holonomy schemes qualitatively result in a similar
mechanism of singularity-resolution. A suitable regime  for
interesting physical effects that can distinguish between  different choices of $\lambda(x)$
is therefore primarily given by the exterior geometry, described by the emergent metric
(\ref{eq:Spacetime metric - modified - Schwarzschild - Asymptotic and zero
  mass - simp}).
For now we consider exterior effects around a black hole, but far from
any cosmological scales. We therefore continue to assume a vanishing cosmological constant.
We will focus on the asymptotically flat modified Schwarzschild metric with a
constant overall  factor $\chi$, given by (\ref{eq:Spacetime metric - modified
  - Schwarzschild - Asymptotic and zero mass}), such that
\begin{equation}
    {\rm d} s^2 =
    - \left(1 - \frac{2M}{x}\right) {\rm d} t^2
    +  \frac{{\rm d} x^2}{\chi_0^2( 1 + \lambda^2 \left( 1 - 2M/x \right)
    )(1 - 2M/x)}
    + x^2 {\rm d} \Omega^2
    \label{eq:Spacetime metric - modified - Schwarzschild - Asymptotic and zero mass - simp}
\end{equation}
with
\begin{equation} \label{Flat}
  \chi_0=\frac{1}{\sqrt{1+\lambda(\infty)^2}}\,.
\end{equation}

\subsection{Newtonian limit}

Using the coordinate transformation $x = r \left(1+ M/(2r)\right)^2$, which in
the case of the classical solution corresponds to the isotropic radial coordinate, the metric (\ref{eq:Spacetime metric - modified - Schwarzschild - Asymptotic and zero mass - simp}) becomes
\begin{eqnarray}
    {\rm d} s^2 = - \left(\frac{1-M/2r}{1+M/2r}\right)^2 {\rm d}t^2 + \left(1+\frac{M}{2r}\right)^4 \left( \left( 1 + \lambda(x)^2 \left(\frac{1-M/2r}{1+M/2r}\right)^2
    \right)^{-1} \frac{{\rm d} r^2}{\chi_0^2} +r^2 {\rm d}\Omega^2\right)\,.
\end{eqnarray}
In this form, it is easier to derive the weak-field regime defined by $M\ll r$ and $\lambda^2 \ll 1$,
\begin{eqnarray}
    {\rm d} s^2 \approx - \left(1-\frac{2M}{r}\right) {\rm d}t^2 + \left(1+\frac{2M}{r}\right) \left( \left( 1 - \lambda(x)^2 \left(1-\frac{2M}{r}\right)
    \right) \frac{{\rm d} r^2}{\chi_0^2} +r^2 {\rm d}\Omega^2\right)\,,
\end{eqnarray}
where $x \approx r  + M$ in the argument of $\lambda(x)$.
Defining the new radial coordinate
\begin{equation}\label{eq:Second isotropic coord transf}
    \mathfrak{r}(r) = r_0\exp\left( \int_{r_0}^r \frac{{\rm d} \tilde{r}}{\chi_0 \tilde{r}} \sqrt{ 1 - \lambda(\tilde{r}+M)^2 \left(1-\frac{2M}{\tilde{r}}\right)}\right)
  \end{equation}
  with a reference radius $r_0$ where $\mathfrak{r}(r_0)=r_0$,
the metric recovers its isotropic form in the weak-field regime,
\begin{eqnarray}
    {\rm d} s^2 \approx - \left(1-\frac{2M}{r(\mathfrak{r})}\right) {\rm d}t^2 + \left(1+\frac{2M}{r(\mathfrak{r})}\right) B(\mathfrak{r}) \left( {\rm d} \mathfrak{r}^2 + \mathfrak{r}^2 {\rm d}\Omega^2\right)\,.
\end{eqnarray}
The function $B(\mathfrak{r})$ results from the inversion of (\ref{eq:Second isotropic coord transf}) to $r^2 = B(\mathfrak{r}) \mathfrak{r}^2$.
An explicit integration of (\ref{eq:Second isotropic coord transf}) and hence
$B(\mathfrak{r})$ depends on the chosen holonomy function $\lambda(x)/\tilde{\lambda}$. For
our approximations of weak-field gravity and small holonomy modifications, it
can be written as
\begin{equation}
  \mathfrak{r} \approx r_0\exp \left(\frac{1}{\chi_0} \int_{r_0}^r
                         \frac{1-\frac{1}{2}\lambda(\tilde{r}+M)^2}{\tilde{r}}{\rm
                         d}\tilde{r}\right)   \approx r_0(r/r_0)^{1/\chi_0}
  \exp\left(-\frac{1}{2\chi_0}\int_{r_0}^r
    \frac{\lambda(\tilde{r})^2}{\tilde{r}}{\rm d}\tilde{r}\right)\,. 
\end{equation}
The shift by $M$ does not contribute because
$\lambda(r+M)^2\approx\lambda(r)^2+2M\lambda(r)\lambda'(r)$ has a subleading
contribution from $M$ for $M\ll r$ and $\lambda^2\ll1$.
For constant $\lambda=\tilde{\lambda}$, we have
\begin{equation}
  \mathfrak{r}(r)\approx
  r_0(r/r_0)^{(1-\frac{1}{2}\tilde{\lambda}^2)/\chi_0}\,,
\end{equation}
and for
$\lambda(r)=\tilde{\lambda}/r$ with constant $\tilde{\lambda}$ we obtain
\begin{equation} \label{rr2}
  \mathfrak{r}(r)\approx r_0(r/r_0)^{1/\chi_0}
  \exp\left(-\frac{\tilde{\lambda}^2}{4r_0^2\chi_0}\left(1-\frac{r_0^2}{r^2}\right)\right)\,.
\end{equation}
Condition (\ref{Flat}) for asymptotic flatness, applied for small $\lambda$,
implies that $\chi_0\approx 1-\frac{1}{2}\tilde{\lambda}^2$ for constant $\lambda$,
and $\chi_0=1$ for $\lambda(r)=\tilde{\lambda}/r$. Since the exponential
function in (\ref{rr2}) quickly approaches a constant for large $r$,
$\mathfrak{r}$ is close to a linear function, as in the classical solution.

\subsection{Geodesic properties}

The strict covariance conditions imposed on our effective space-time models
make it possible to derive reliable relativistic effects from properties of
timelike and lightlike geodesics, going beyond the Newtonian limit.

\subsubsection{Timelike geodesics and effective potential}

We first analyze worldlines of massive objects in the effective space-time.
Because the effective space-time is static and spherically symmetric, we have
corresponding Killing vectors and conserved quantities $\tilde{E}$ and
$\tilde{L}$ related to the energy and angular momentum (per mass) of the particle. 
Using these conserved quantities and the lapse function in the space-time
metric, we formulate the normalization condition
$g_{\mu \nu} u^\mu u^\nu = - \kappa$ for the particle's 4-velocity
$u^\mu$. For the sake of convenience, we
include both massive and massless options by the parameter $\kappa=0,1$:
\begin{eqnarray}
    0 &=&
    \frac{1}{2} \left(\frac{{\rm d} x}{{\rm d} \tau}\right)^2
    + \frac{\tilde{q}^{xx}}{2} \left( \kappa
    + \frac{\tilde{L}^2}{x^2}
    - \frac{\tilde{E}^2}{N^2} \right)
    \,.
    \label{eq:Particle motion simplified - Schwarzschild gauge - Static}
\end{eqnarray}
Here, $\tau$ is the particle's proper time (to be replaced by the affine
parameter in the massless case).

This equation can be interpreted as an energy-balance law between kinetic and
potential energy of classical 1-dimensional motion, resulting in the effective
potential (per unit mass)
\begin{eqnarray}
    \tilde{V}_{\rm eff}
    &=&
    \frac{\tilde{q}^{x x}}{2} \left( \kappa
    - \frac{\tilde{E}^2}{N^2}
    + \frac{\tilde{L}^2}{x^2} \right)
    \notag\\
    &=&
    \frac{\chi_0^2}{2} \left( 1 + \lambda^2 \left( 1 - \frac{2 M}{x} \right)
    \right) \left( \left( \kappa
    + \frac{\tilde{L}^2}{x^2} \right) \left(1 - \frac{2 M}{x}\right)
    - 
    \tilde{E}^2 \right)
    \,.
    \label{eq:Effective potential particle motion simplified - Schwarzschild gauge - Static}
\end{eqnarray}
The relevant metric components are given by (\ref{eq:Spacetime metric - modified - Schwarzschild}).
We will also use the first derivative of the potential,
\begin{eqnarray}
    \frac{\partial \tilde{V}_{\rm eff}}{\partial x} &=&
    \frac{\chi_0^2}{x^5}
    \Bigg[
    x \left(3 \tilde{L}^2 M-\tilde{L}^2 x+\kappa  M x^2\right)
    \nonumber\\
    &&
    - \lambda^2 \left(\tilde{L}^2 \left(8 M^2-6 M x+x^2\right)+M x^2 (4 \kappa  M+\tilde{E}^2 x-2 \kappa  x)\right)
    \nonumber\\
    &&
    + (\lambda^2)' \frac{x}{2} (2 M-x) \left(\tilde{L}^2 (2 M-x)+x^2 (2 \kappa  M+\tilde{E}^2 x-\kappa  x)\right)
    \Bigg]
    \label{eq:Effective force}
\end{eqnarray}
where $\lambda'={\rm d}\lambda/{\rm d} x$.

Circular orbits are characterized by
$\partial \tilde{V}_{\rm eff} / \partial x = 0$ and
${\rm d} x / {\rm d} \tau =0$. Inserting these conditions into equations
(\ref{eq:Particle motion simplified - Schwarzschild gauge - Static}) and
(\ref{eq:Effective force}), respectively, gives us two equations that relate
the energy and angular momentum:
\begin{equation} \label{E1}
    \tilde{E}^2 = \left(1-\frac{2 M}{x}\right) \left(\frac{\tilde{L}^2}{x^2}+\kappa \right)
  \end{equation}
  and
  \begin{eqnarray}
&&    \tilde{E}^2\lambda^2 \left(\frac{M}{x}  + \left(1-\frac{2M}{x}\right)
   \frac{{\rm d}\ln\lambda}{{\rm d}\ln x}\right)=
                                                    \frac{3M\tilde{L}^2}{x^3}-\frac{\tilde{L}^2}{x^2}
   +  \frac{\kappa M}{x}\label{E2}\\
&&  \qquad  + \lambda^2\left(1-\frac{2 M}{x}\right)  \left(\left(\frac{4\tilde{L}^2
    M}{x^3}-\frac{\tilde{L}^2}{x^2} + \frac{2\kappa M}{x^2} \right)
    +  \left(1-\frac{2 M}{x}\right)^2 \left(\frac{\tilde{L}^2}{x^2}+\kappa
    \right) \frac{{\rm d}\ln\lambda}{{\rm d}\ln x}\right)
    \,.\nonumber
  \end{eqnarray}
There is a qualitative difference in these equations compared with the
classical case, in which case the left-hand side of (\ref{E2}) vanishes. The
right-hand side then determines $\tilde{L}$, which then provides $\tilde{E}$ using
(\ref{E1}). For non-zero $\lambda$, (\ref{E2}) depends on both $\tilde{L}$ and
$\tilde{E}$, but $\tilde{E}$ can easily be eliminated by using (\ref{E1}). The
terms with derivatives of $\lambda$ then cancel out completely, and the
resulting equation for $\tilde{L}$ reads
\begin{eqnarray}
    \left(x + \lambda^2 (x-2 M)\right) \left(3 \tilde{L}^2 M-\tilde{L}^2 x+\kappa  M x^2\right) = 0
    \,.
\end{eqnarray}
The first parenthesis by definition vanishes at an extremal radius $x_{\lambda}^{(i)}$
(solving (\ref{eq:Minimum radius equation}) for $\Lambda=0$),
which never happens in the static exterior region as we have shown.
The second parenthesis must therefore vanish for circular orbits, which is
independent of $\lambda$ and
implies the classical stationary radius 
\begin{equation}\label{eq:Stationary radius}
    x_0^{\pm} = \frac{\tilde{L}^2}{2 M \kappa} \left(1\pm\sqrt{1- \kappa
        \frac{12 M^2}{\tilde{L}^2}}\right) 
    \,.
\end{equation}
For the null case, $\kappa=0$, only $x_0^{-}$ is finite, given by the value
$3M$ derived by taking the limit $\kappa\to0$.

At this radial coordinate, we have
\begin{eqnarray}
    \frac{\partial^2 \tilde{V}_{\rm eff}}{\partial x^2} \bigg|_{x_0^\pm} &=&
    \kappa^4 \chi_0^2 \frac{32 M^4}{\tilde{L}^{6}} \left(1\pm\sqrt{1-\kappa\frac{12 M^2}{\tilde{L}^2}}\right)^{-6}
    \Bigg[
    - 6 \kappa \frac{M^2}{\tilde{L}^2}
    + \kappa \lambda^2 \frac{6 M^2}{\tilde{L}^2} \left( \kappa \frac{4 M^2}{\tilde{L}^2}
    - 1\right)
    \nonumber\\
    &&\qquad
    + \left( 1
    - \kappa \frac{6 M^2}{\tilde{L}^2}
    + \lambda^2 \left( 1 - \kappa \frac{8 M^2}{\tilde{L}^2} \right) \right) \left( 1
    \pm \sqrt{1-\kappa\frac{12 M^2}{\tilde{L}^2}} \right)
    \Bigg]
    \,.
\end{eqnarray}
As in the classical case, only $x_0^+$ is a local minimum of the potential for
a massive particle, while $x_0^-$ is a local maximum for both massive and 
massless particles. 



\subsubsection{Nearly-circular orbits of massive objects}

For nearly-circular orbits we consider only the case of massive objects and hence set $\kappa=1$.
It will be useful to invert the relation (\ref{eq:Stationary radius}) for the angular momentum in terms of the equilibrium coordinate $x_0=x_0^+$:
\begin{equation}
    \tilde{L}^2 =
    \frac{M x_0^2}{x_0-3 M}\,.
\end{equation}
If the object is displaced slightly from its equilibrium radius, it 
oscillates around it with frequency 
\begin{eqnarray}
    \omega_r{}^2 &=& \frac{\partial^2 \tilde{V}_{\rm eff}}{\partial x^2} \bigg|_{x=x_0, \kappa=1}
    \nonumber\\
    &=&
    \frac{M}{x_0^3} \frac{x_0-6M}{x_0-3M} \chi_0^2 \left( 1 + \lambda^2 \left(1 - \frac{2 M}{x_0}\right)\right)
    \nonumber\\
    &=&
    \omega_\varphi{}^2 \left(1-\frac{6M}{x_0}\right) \chi_0^2 \left( 1 + \lambda^2 \left(1 - \frac{2 M}{x_0}\right)\right)
\end{eqnarray}
where $\omega_\varphi$ is the angular frequency of the orbit and given by
\begin{eqnarray}
    \omega_\varphi{}^2 =
    \frac{\tilde{L}^2}{x_0^4} =
    \frac{M}{x_0^2 (x_0-3 M)}
    \,.
\end{eqnarray}
Therefore, the precession rate for nearly-circular orbits equals
\begin{equation}
    \omega_p = \omega_\varphi - \omega_r
    = \left( 1 - \chi_0 \sqrt{1-\frac{6M}{x_0}}  \sqrt{1 + \lambda^2 \left(1 - \frac{2 M}{x_0}\right)} \right) \omega_\varphi
    \,.
\end{equation}

\subsubsection{Null rays: Deflection angle and redshift}

The effective potential (\ref{eq:Effective potential particle motion simplified - Schwarzschild gauge - Static}) for a massless particle is
\begin{eqnarray}
    \tilde{V}_{\rm null} &=&
    \frac{\chi_0^2}{2} \left( 1 + \lambda^2 \left( 1 - \frac{2 M}{x} \right)
    \right) \left( \frac{b^2}{x^2} \left(1 - \frac{2 M}{x}\right)
    - 1 \right) \tilde{E}^2
    \ ,
    \label{eq:Effective potential particle motion simplified - Schwarzschild gauge - Static - null}
\end{eqnarray}
where $b \equiv \tilde{L} / \tilde{E}$ is the impact parameter.
Using $\tilde{L} = x^2 \dot{\phi}$ we can obtain ${\rm d} \phi / {\rm d} x = \dot{\phi} / \dot{x}$ from (\ref{eq:Particle motion simplified - Schwarzschild gauge - Static}) with $\kappa=0$,
\begin{eqnarray}
    \frac{{\rm d} \phi}{{\rm d} x} &=&
    - \frac{1}{\chi_0} \frac{1}{x^2}
    \left( 1 + \lambda^2 \left( 1 - \frac{2 M}{x} \right)
    \right)^{-1/2}
    \left( \frac{1}{b^2} - \left(1 - \frac{2 M}{x}\right) \frac{1}{x^2} \right)^{-1/2}
    \ .
\end{eqnarray}
For a null ray that is not captured by the black hole, its turning point is
given by the largest solution $x_{\rm tp}$ to $V_{\text{null}} (x_{\rm tp}) =
0$:
\begin{eqnarray}
    \frac{x_{\rm tp}^2}{b^2} = 1 - \frac{2 M}{x_{\rm tp}}
    \ .
    \label{eq:Turning point radius equation - null ray - Schwarzschild gauge - Static}
\end{eqnarray}

The complete change in the angular coordinate can then be derived as
\begin{eqnarray}
    \Delta \phi &=& \frac{2}{\chi_0} \int^\infty_{x_{\rm tp}} \frac{{\rm d} x}{
    x^2\sqrt{
    \left( 1 + \lambda^2 \left( 1 - 2 M/x \right)
    \right)
    \left( 1/b^2 - \left(1 - 2 M/x\right)/x^2 \right)}}
    - \pi \label{eq:Total change in angular coordinate - null ray - Schwarzschild gauge - Static}
    \\
    &=&
    \frac{2}{\chi_0} \int_0^{1/x_{\rm tp}} \frac{{\rm d} u}{
    \sqrt{\left( 1 + \lambda^2 \left( 1 - 2 M u \right)
    \right)
    \left( \left(1 - 2 M x_{\rm tp}^{-1}\right) x_{\rm tp}^{-2} - \left(1 - 2 M u \right) u^2 \right)}}
    - \pi
    \,,\nonumber
\end{eqnarray}
where we used the substitution $u=1/x$ to obtain the second line. As
usual, the
subtraction of $\pi$ expresses the result relative to the change of angle in flat space.

This integral is complicated in analytical form, even in the classical case $\lambda\to0$.
We therefore consider the expression at leading order in $M/x_{\rm tp}$,
\begin{eqnarray}
    \Delta \phi
    &\approx&
    \pi \left( \chi_0^{-1} -1\right)
    - \frac{2 x_{\rm tp}}{\chi_0} \int_0^{1/x_{\rm tp}} {\rm d} u\ \frac{\lambda^2}{2\sqrt{1-u^2 x_{\rm tp}^2}}
    \nonumber\\
    &&
    + \frac{2 M}{\chi_0} \int_0^{1/x_{\rm tp}} {\rm d} u\ \frac{1 + u x_{\rm tp} (1+u x_{\rm tp}) + \lambda^2 \left(1 + 2 u x_{\rm tp} (1+u x_{\rm tp})\right)}
    {\left(1+u x_{\rm tp}\right) \left(1-u^2 x_{\rm tp}^2\right)^{1/2} \left(1+\lambda^2\right)^{3/2}}
    \,.
    \label{eq:Deflection angle - simp}
\end{eqnarray}

The deflection angle, as well as the effective potential and the force it implies on
massive objects,  can be
used for standard derivations of physical scenarios once we have chosen a
holonomy function $h=\lambda/\tilde{\lambda}$. 
In the next two sections, we will do this in the cases of constant $h$ and a decreasing
$h(x)\propto 1/x$, respectively.
In addition, one could also consider redshift effects, but we now show that
there are no $\lambda$-modifications in this effect for radial null rays.

A stationary observer is described by a 4-velocity $u^\mu \partial_\mu = u^t \partial_t$.
Timelike normalization implies
\begin{eqnarray}
    u^\mu \partial_\mu &=&
    \frac{\partial_t}{\sqrt{- g_{t t}}}
    =
    \left(1 - \frac{2 M}{x}\right)^{-1/2} \partial_t
    \,.
    \label{eq:Static observer 4-velocity}
\end{eqnarray}
Such an observer, located at a radial position $x$,
observes the frequency
\begin{eqnarray}
    \omega (x) &=&
    - g_{\mu \nu} k^\mu u^\nu \big|_{x}
    = \sqrt{- g_{t t}} k^t
    = \sqrt{1 - \frac{2 M}{x}} k^t\,,
\end{eqnarray}
of a light ray with wave 4-vector $k^{\mu}$.
Completing $u^{\nu}$ to a local inertial frame in the radial manifold by
including the normalized spacelike vector
\begin{eqnarray}
    \zeta^\mu \partial_\mu &=&
    \frac{\partial_x}{\sqrt{\tilde{q}_{x x}}}
    \ ,
\end{eqnarray}
the observed radial wave-number is
\begin{eqnarray}
    k (x) = g_{\mu \nu} k^\mu \zeta^\nu \big|_{x}
    = \chi_0^{-1} \left( 1 + \lambda(x)^2 \left( 1 - \frac{2 M}{x} \right)
    \right)^{-1/2} \left(1 - \frac{2 M}{x}\right)^{-1/2} k^x
    \ .
\end{eqnarray}
The lightlike condition $g_{\mu \nu} k^\mu k^\nu = 0$ implies
\begin{eqnarray}
    k (x) = \pm \omega (x)
\end{eqnarray}
in the orthonormal frame, 
where the positive sign determines properties of outgoing light rays.

If the light ray was emitted near the horizon at $x = x_{\rm e} \gtrsim 2 M$
with frequency $\omega_{\rm e}$, as measured by a stationary observer at $x_{\rm
  e}$, and follows a null geodesic,  conservation of
$k_t = g_{t t} k^t$ implies that it is simply subject to the classical redshift 
\begin{eqnarray}
    \frac{\omega_{\rm o}}{\omega_{\rm e}}
    = \sqrt{\frac{1 - 2 M / x_{\rm e}}{1 - 2 M / x_{\rm o}}}
    \ ,
\end{eqnarray}
where $\omega_{\rm o}$ is the frequency observed by a stationary observer at $x = x_{\rm o}$.
We therefore see no holonomy modifications to gravitational redshift, for arbitrary $\lambda$.

The underlying reason for this result is that the component $g_{tt}$ of the
exterior metric \eqref{eq:Spacetime metric - modified - Schwarzschild -
  Asymptotic and zero mass - simp} has the same form as the classical
component in Schwarzschild space-time. Only the spatial metric of the emergent
line element is directly changed for a modified structure function, and the
static gauge implies that $\lambda$-modifications do not affect conditions on
the lapse function.  For the same reason, gravitational time-dilation, for
instance for an observer in a geostationary orbit at some distance $R$ from
the Earth with radius $r_{\rm E}$, retains the classical expression
\begin{eqnarray}\label{time-dilation}
    \Delta \tau = \Delta t \, \sqrt{\frac{1-2M/r_{\rm E}}{1-2M/(r_{\rm E}+R)}}\,,
\end{eqnarray}
devoid of any holonomy corrections for arbitrary $\lambda$. 

\subsection{Energy and thermodynamics}

Physical properties of our space-time, rather than of objects moving on this
background, are determined by curvature tensors. We will first derive relevant
ones and then apply them to various considerations related to energy and
thermodynamics.

\subsubsection{Curvature tensors}

The metric (\ref{eq:Spacetime metric - modified - Schwarzschild - Asymptotic
  and zero mass - simp}) is static and therefore has the Killing vector
$\xi^\mu \partial_\mu = \partial_t$.
The unit normal vector of the foliation given by constant $t$ equals
\begin{equation}
    n^\mu \partial_\mu = \left(1 - \frac{2 M}{x}\right)^{-1/2} \partial_t
    = \left(1 - \frac{2 M}{x}\right)^{-1/2} \xi^\mu \partial_\mu
    \ .
\end{equation}

The Einstein tensor for an emergent metric of the form (\ref{eq:Spacetime metric - modified - Schwarzschild - Asymptotic and zero mass - simp}) with arbitrary $\lambda$ takes the form
\begin{eqnarray}
    G_{\mu\nu} {\rm d} x^\mu \otimes {\rm d} x^\nu
    &=&
    - \frac{\chi_0^2}{x^2} \left(1 - \frac{2M}{x}\right) \left( 1 - \frac{\lambda_\infty^2}{\lambda^2} -\frac{4 M^2}{x^2} + x (\ln\lambda^2)' \left(1-\frac{2 M}{x}\right)^2 \right) {\rm d} t^2
    \nonumber\\
    &&
    + \frac{\lambda^2}{x^2} \left(1-\frac{2 M}{x}\right)^{-1} \left(1+\lambda^2 \left(1-\frac{2 M}{x}\right)\right)^{-1} \left( 1 - \frac{\lambda_\infty^2}{\lambda^2} - \frac{2M}{x} \right) {\rm d} x^2
    \nonumber\\
    &&
    + \frac{\chi_0^2 \lambda^2}{2} \left(1-\frac{M}{x}\right) \left(\frac{2 M}{x} + x (\ln\lambda^2)' \left(1-\frac{2 M}{x}\right)\right) {\rm d} \Omega^2
    \ ,
    \label{eq:Einstein tensor - modified}
\end{eqnarray}
where $\lambda'=\partial\lambda/\partial x$.

Each symmetric 2-sphere has a normal vector
\begin{equation}
    r^\mu \partial_\mu = \sqrt{\tilde{q}^{xx}} \partial_x
    \ ,
\end{equation}
in space, such that $\tilde{g}_{\mu \nu} r^\mu r^\nu=1$.
With this information, we can compute the extrinsic-curvature tensor of the spheres to find
\begin{eqnarray}
    \mathcal{K}^{(S)}_{\mu \nu} {\rm d} x^\mu {\rm d} x^\nu &\equiv& \left(\frac{1}{2} \mathcal{L}_r \tilde{q}_{\mu \nu}\right) {\rm d} x^\mu {\rm d} x^\nu
    = \sqrt{\tilde{q}^{xx}} x {\rm d} \Omega^2
    \nonumber\\
    &=&
    \chi_0 x \ \sqrt{1 + \lambda^2 \left( 1 - \frac{2 M}{x} \right)} \ \ \sqrt{1 - \frac{2 M}{x}} {\rm d} \Omega^2
    \ ,
\end{eqnarray}
with trace given by
\begin{equation}
    \mathcal{K}^{(S)} = \frac{2 \chi_0}{x} \sqrt{1 + \lambda^2 \left( 1 - \frac{2 M}{x} \right)} \  \ \sqrt{1 - \frac{2 M}{x}}
    \ . \label{eq:Trace of 2-sphere extrinsic curvature - modified}
\end{equation}

\subsubsection{Geometric conditions, net stress-energy tensor, and gravitational energy}

The null and timelike conditions, (\ref{eq:Null cond interior}) and
(\ref{eq:Timelike cond interior}), are satisfied at sufficiently large
distances for any $\lambda(x)$ that falls off at least as $1/\sqrt{x}$. In
general, for any $\lambda(x)= a x^{-n}$, the NGC is satisfied for
$x\geq 2 M (n+1/2)/n$, which lies outside the horizon and gets closer to it
for large $n$.  Violations of the geometric conditions therefore generically
start somewhere in the exterior. The corresponding coordinate value can be interpreted as the
location where quantum effects become significant.  The special
case of constant $\lambda$  may be included in this
parameterization as the limit $n\to 0$, in which case the NGC is violated
everywhere.

In Schwarzschild coordinates, the net stress-energy tensor, defined by
(\ref{eq:Net stress-energy tensor}) and using the Einstein tensor defined by
(\ref{eq:Einstein tensor}) for the emergent space-time metric, takes the form
\begin{eqnarray}
    8\pi\bar{T}_{\mu\nu} {\rm d} x^\mu \otimes {\rm d} x^\nu
    &=&
    \frac{\chi_0^2\lambda^2}{x^2} \left(1 - \frac{2M}{x}\right) \left( 1 - \frac{\lambda_\infty^2}{\lambda^2} -\frac{4 M^2}{x^2} + x (\ln\lambda^2)' \left(1-\frac{2 M}{x}\right)^2 \right) {\rm d} t^2
    \nonumber\\
    &&
    - \frac{\lambda^2}{x^2} \left(1-\frac{2 M}{x}\right)^{-1} \left(1+\lambda^2 \left(1-\frac{2 M}{x}\right)\right)^{-1} \left( 1 - \frac{\lambda_\infty^2}{\lambda^2} - \frac{2M}{x} \right) {\rm d} x^2
    \nonumber\\
    &&
    - \frac{\chi_0^2 \lambda^2}{2} \left(1-\frac{M}{x}\right) \left(\frac{2 M}{x} + x (\ln\lambda^2)' \left(1-\frac{2 M}{x}\right)\right) {\rm d} \Omega^2
    \,,
    \label{eq:Einstein tensor - Schwarzschild no cosmo}
\end{eqnarray}
using (\ref{eq:Einstein tensor - modified}).
In the exterior, $x>2M$, and for monotonically decreasing $\lambda$, we obtain $\bar{T}_{xx}<0$ everywhere, while $\bar{T}_{tt}$ and $\bar{T}_{\varphi\varphi}$ potentially change sign at some point, leading to violations of the energy conditions.

%

The energy density measured by a static observer with 4-velocity given by
(\ref{eq:Static observer 4-velocity}) equals
\begin{eqnarray}
    \rho \equiv \bar{T}_{\mu\nu}u^\mu u^\nu
    &=&
    \frac{\chi_0^2\lambda^2}{x^2} \left( 1 - \frac{\lambda_\infty^2}{\lambda^2} -\frac{4 M^2}{x^2} + x (\ln\lambda^2)' \left(1-\frac{2 M}{x}\right)^2 \right)
    \,.
\end{eqnarray}
For monotonically decreasing $\lambda$, this expression is negative at
sufficiently large values of $x$ where the last term dominates.  This shows
that, unlike in the interior where the gravitational energy density is
positive, in the exterior at large scales it does act as a binding energy
for these observers. (The sign of the energy depends on the
observers. In particular, for comoving observers the interior energy density
is proportional to, and of the same sign as, the component $\bar{T}_{xx}$ of
(\ref{eq:Einstein tensor - Schwarzschild no cosmo}) which is negative.)

\subsection{Black-hole thermodynamics}

A useful energy quantity in general relativity is the (normalized) Brown-York quasilocal energy as seen by observers characterized by a vector $\hat{t}^\mu=\hat{N}n^\mu+\hat{N}^xs^\mu_x$,
\begin{equation}
    E = - \frac{1}{8 \pi} \int {\rm d}^{2} z\ \hat{N} \left( \sqrt{\det \sigma} \mathcal{K}^{(S)} - \sqrt{\det \bar{\sigma}} \bar{\mathcal{K}}^{(S)} \right)
    \,.
    \label{eq:Brown-York quasilocal energy}
\end{equation}
The integration is over a 2-dimensional surface with coordinates $z$ and
the induced 2-metric $\sigma$, and $\mathcal{K}^{(S)}$ is the trace of
extrinsic curvature on the 2-sphere. The barred quantities are evaluated in
Minkowski space-time

The first steps of the derivation can be reproduced in a canonical setting,
which we show here. However, it turns out that a crucial ingredient requires
an extension of canonical variables into a timelike direction, which is not
available in emergent modified gravity without using a specific
solution. Brown-York quasilocal energy therefore cannot be defined by an
independent derivation within
emergent modified gravity. Instead, we will consider the classical geometrical
expression (\ref{eq:Brown-York quasilocal energy}) as a net quasilocal energy
in an emergent space-time, just as we used the classical Einstein tensor in
our definition of net stress-energy.

As for the derivation of the Brown-York expression, consider the classical
Hamiltonian action over a space-time region ${\cal R}$, 
\begin{equation}
    S [N,\vec{N},q,p] = \int_{\cal R} {\rm d}^4x\ \left(p^{ab}\dot{q}_{ab} - H N - H_aN^a\right)
\end{equation}
where $p^{ab}=(16\pi )^{-1}\sqrt{\det q}\left(K^{ab}-Kq^{ab}\right)$ is the
momentum conjugate to the 3-metric $q_{ab}$, and $K_{ab}$ is the extrinsic
curvature with trace $K=K_{ab}q^{ab}$.
The boundary contribution associated to the first term in the action is
\begin{eqnarray}
    S_{\partial {\cal R}} = \int_{\partial {\cal R}} {\rm d}^3x\ p^{ab}q_{ab}
    = - \frac{1}{8\pi}\int_{\partial {\cal R}} {\rm d}^3x\ \sqrt{|\det q^{(\partial{\cal R})}|} K
\end{eqnarray}
where $q^{(\partial{\cal R})}$ is the 3-metric of the boundary $\partial {\cal
  R}$, which in general may be spacelike or timelike.
We obtain the Brown-York term by restricting $S_{\partial {\cal R}}$ to the
boundary of the (non-smooth) boundary $\partial\partial {\cal R}$, 
\begin{eqnarray} \label{Sdd}
    S_{\partial\partial {\cal R}}
    = - \frac{1}{8\pi}\int_{\partial\partial {\cal R}} {\rm d}^2z\ \hat{N} \sqrt{|\det \sigma|} K\,.
\end{eqnarray}
Here, $\sigma$ is the 2-metric and $z$ are coordinates of the 2-dimensional
surface $\partial\partial {\cal R}$, while $\hat{N}$ is the lapse function of the chosen
observer which is normal to $\partial {\cal R}$. 
In triad variables we obtain the same boundary term since
\begin{eqnarray}
    S_{\partial {\cal R}} = \int_{\partial {\cal R}} {\rm d}^3x\ \left(E^\varphi K_\varphi + E^x K_x\right) = - \frac{1}{8\pi}\int_{\partial {\cal R}} {\rm d}^3x\ \sqrt{|\det q^{(\partial{\cal R})}|} K\,.
\end{eqnarray}

This derivation shows that $K$ is directly derived from a linear combination
of canonical variables. In emergent modified gravity, $K$ is therefore not
given by extrinsic curvature of the emergent space-time. The canonical
variables $K_x$ and $K_{\varphi}$ are readily available on spacelike parts of
the boundary $\partial {\cal R}$, but not on the timelike components if there
is no fundamental space-time theory with a corresponding action
principle. However, the timelike extension is precisely what is required for a
derivation of the Brown--York quasilocal energy, making it impossible to
derive a strict analog of (\ref{Sdd}). The final restriction to
$\partial\partial{\cal R}$ is spacelike, and may be expressed in the same way
as (\ref{Sdd}) in emergent modified gravity with the trace $K$ of the
canonical variables. However, the presence of a crucial gap in the derivation
suggests that such a definition may be of limited value.

In what follows, we therefore proceed by using the classical result
(\ref{eq:Brown-York quasilocal energy}) as the net quasilocal energy, defined
geometrically through extrinsic curvature in the emergent space-time.  The
barred functions are the corresponding quantities in the ground state which in
classical general relativity is given by Minkowski space-time. But, as
discussed in the previous subsection, this may no longer be the case in a
modified theory.  We therefore choose our ground state as the space-time
obtained in the zero-mass limit.

We are interested in the net quasilocal energy defined by 2-spheres enclosing the black hole.
Thus, the net quasilocal energy in the holonomy-modified model for observers
with $\hat{N}=1$ and $\hat{N}^x=0$ is given by
\begin{eqnarray}
    E (x) &=& x \chi_0 \left( \sqrt{1 + \lambda^2} - \sqrt{1 - \frac{2 M}{x}}
              \sqrt{1 + \lambda^2 \left( 1 - \frac{2 M}{x} \right)} \right)
    \,,
    \label{eq:Brown-York quasilocal energy - modified}
  \end{eqnarray}
using (\ref{eq:Trace of 2-sphere extrinsic curvature - modified}), and $\chi_0=1/\sqrt{1+\lambda_\infty^2}$ for asymptotically flat solutions.
This expression is everywhere positive and monotonically decreasing for
$\lambda'\leq 0$. Since we
are considering vacuum solutions, we conclude that
the gravitational field itself has an energy contribution.
In the asymptotic limit of $x \to \infty$, the energy acquires a correction to its classical
value, given by the mass, unless the holonomy parameter is asymptotically vanishing,
\begin{equation}
      \lim_{x\to\infty} E(x) = M \left(1+\frac{\lambda_\infty^2}{1+\lambda_\infty^2}\right)
      \,.
\end{equation}

A static observer has the 4-acceleration
\begin{eqnarray}
    a^\mu \partial_\mu
    &=& \chi_0^2 \frac{2 M}{x^2} \left( 1 + \lambda^2 \left( 1 - \frac{2 M}{x} \right) \right) \partial_x
    \,,
\end{eqnarray}
with norm
\begin{eqnarray}
    a &=& \sqrt{\tilde{g}_{\mu \nu} a^\mu a^\nu}
    = \sqrt{\tilde{q}_{x x} (a^x)^2}
    \nonumber\\
    &=&
    \chi_0^2 \frac{2 M}{x^2} \left(1 - \frac{2 M}{x}\right)^{-1/2}
    \left( 1 + \lambda^2 \left( 1 - \frac{2 M}{x} \right)
    \right)^{1/2}
    \,,
    \label{eq:Proper acceleration - holonomy modification}
\end{eqnarray}
subject to holonomy modifications.
We perform a near-horizon expansion at $x = 2 M + \rho^2 / 8 M = 2 M (1 +
\Bar{\rho}^2)$ for small $\Bar{\rho}$:
\begin{eqnarray}
    a &=&
    \frac{\chi_0^2}{2 M \Bar{\rho}}
    \left(1 - \frac{3}{2} \left(1 - \frac{\lambda_{\rm H}^2}{3} \right) \Bar{\rho}^2
    + O(\bar{\rho}^4) \right)
    \label{eq:Proper acceleration - Schwarzschild - Near horizon expansion}
\end{eqnarray}
where $\lambda_{\rm H} = \lambda(2 M)$.

The modified Schwarzschild metric (\ref{eq:Spacetime metric - modified - Schwarzschild}) to leading order in $\Bar{\rho}$ becomes
\begin{eqnarray}
    {\rm d} s^2
    &=&
    - \Bar{\rho}^2 \left(1-\Bar{\rho}^2\right) {\rm d} t^2
    + \frac{1 - \lambda_{\rm H}^2 \Bar{\rho}^2 + \lambda_{\rm H}^4 \Bar{\rho}^4}{1-\Bar{\rho}^2} (4 M)^2 \frac{{\rm d} \Bar{\rho}^2}{\chi_0^2}
    + x^2 {\rm d} \Omega^2
    \nonumber\\
    &=&
    - \Bar{\rho}^2 \left(1-\Bar{\rho}^2\right) {\rm d} t^2
    + \frac{1 - \lambda_{\rm H}^2 \Bar{\rho}^2 + \lambda_{\rm H}^4 \Bar{\rho}^4}{1-\Bar{\rho}^2} \frac{(4 M)^2}{\chi_0^2} {\rm d} \Bar{\rho}^2
    + x^2 {\rm d} \Omega^2
    \nonumber\\
    &=&
    \frac{(4 M)^2}{\chi_0^2} \left( - \Bar{\rho}^2 {\rm d} \Bar{\tau}^2
    + {\rm d} \Bar{\rho}^2 \right)
    + {\rm d} X_\perp^2
    \ ,
    \label{eq:Spacetime metric - Schwarzschild - Near horizon expansion}
\end{eqnarray}
where $\Bar{\tau} = \chi_0 t / (4 M)$, and $X_{\perp}$ are 2-dimensional local coordinates on the spheres.
This is simply the Rindler metric up to a constant conformal scaling.
Thus, we do not see any major local modifications to leading order in $\bar{\rho}$, for arbitrary $\lambda$.
The black-hole temperature is therefore classical near the horizon,
\begin{equation}
    T = \frac{1}{4 \pi \sqrt{2 M x \left(1-2 M/x\right)}}
    \ .
\end{equation}
Since the space-time is static, it must be in thermal equilibrium, and the
temperature away from the horizon is related to the horizon temperature
through redshift. Thus, for some $\bar{x} > x$,
\begin{equation}
    T (\bar{x}) = \frac{1}{4 \pi \sqrt{2 M x \left(1-2 M/\bar{x}\right)}}
    \approx
    \frac{1}{8 \pi M} \left(1-\frac{2 M}{\bar{x}}\right)^{-1/2}
    \ .
    \label{eq:Temperature of observer at x'}
\end{equation}
Taking $\bar{x} \to \infty$, we recover the Hawking temperature $T_{\rm H} \equiv T (\infty) = 1 / (8\pi M)$.

Let us now consider an observer at a fixed coordinate $\bar{x}$.
A change $\delta M$ in the mass will then change the net quasilocal energy (\ref{eq:Brown-York quasilocal energy - modified}) by 
\begin{eqnarray}
    \delta E (\bar{x}) &=& \chi_0 \left(1 - \frac{2 M}{\bar{x}}\right)^{-1/2} \left( 1 + 2 \lambda^2 \left( 1 - \frac{2 M}{\bar{x}} \right) \right) \left(1 + \lambda^2 \left( 1 - \frac{2 M}{\bar{x}} \right)\right)^{-1/2} \delta M
    \nonumber\\
    &=&
    T (\bar{x}) 8 \pi M \chi_0 \left( 1 + 2 \lambda^2 \left( 1 - \frac{2 M}{\bar{x}} \right) \right) \left(1 + \lambda^2 \left( 1 - \frac{2 M}{\bar{x}} \right)\right)^{-1/2} \delta M
    \ ,
    \label{eq:Variation Brown-York quasilocal energy - mu-scheme}
\end{eqnarray}
where we used the temperature (\ref{eq:Temperature of observer at x'}) and have assumed that $\lambda$ is independent of the mass.
Using the thermodynamic analog $\delta E = T \delta S$, we find that the entropy seen by a static observer at $\bar{x}$ is given by
\begin{eqnarray}
    S (\bar{x}) &=& \int_0^M 8 \pi M' \chi_0 \left( 1 + 2 \lambda^2 \left( 1 - \frac{2 M'}{\bar{x}} \right) \right) \left(1 + \lambda^2 \left( 1 - \frac{2 M'}{\bar{x}} \right)\right)^{-1/2} {\rm d} M'
    \nonumber\\
    &=&
    \chi_0 \frac{8 \pi \bar{x}^2}{15 \lambda^4} \Bigg( \sqrt{1 + \lambda^2 \left(1-\frac{2M}{\bar{x}}\right)} \left( 3 + \lambda^2 \left( 1 + \frac{3 M}{\bar{x}} \right) - 2 \lambda^4 \left( 1 + \frac{M}{\bar{x}} - \frac{6 M^2}{\bar{x}^2} \right) \right)
    \nonumber\\
    &&\qquad
    - \sqrt{1 + \lambda^2} \left( 3 + \lambda^2 - 2 \lambda^4 \right)
    \Bigg)
    \,,
    \label{eq:Quasilocal entropy - modified}
\end{eqnarray}
for arbitrary $\lambda$ independent of the mass.
Notice that the entropy is independent of the reference ground state used in the definition of the quasilocal energy (\ref{eq:Brown-York quasilocal energy - modified}) because this ground state does not depend on the mass (as long as $\lambda$ is mass independent) and hence does not enter the thermodynamic relation $\delta S = T \delta E$.
In the zero-mass limit, the entropy is vanishing $S|_{M\to0}=0$ and is a minimum $\delta S / \delta M|_{M\to0}=0$. Therefore, no work can be extracted from the gravitational ground-state energy.

The entropy (\ref{eq:Quasilocal entropy - modified}) has the correct classical limit,
\begin{equation}
    S (\bar{x}) \xrightarrow[\lambda\to0]{\chi_0 \to 1} \pi (2 M)^2
    = \frac{A_{\rm H}}{4}
    = S_{\rm BH}
    \ ,
\end{equation}
which is constant outside the event horizon. Therefore, all the information is
contained inside the black hole.
The general expression of the entropy, however, is not homogeneous.
The asymptotic limit of the entropy in general depends on the asymptotic behavior of $\lambda$,
\begin{eqnarray}
    S (\infty)
    = \left(1 + \lambda_\infty^2 + O \left(\lambda_\infty^4\right) \right) S_{\rm BH}
    \,,
    \label{eq:Quasilocal entropy - modified - asymptotic}
\end{eqnarray}
and hence may not correspond to the Bekenstein-Hawking entropy unless $\lambda$ has a vanishing asymptotic limit.
Finally, entropy in general  does not acquire the classical expression at the horizon,
\begin{eqnarray}
    S (2M)
    &=&
    \frac{8 \chi_0}{15 \lambda_{\rm H}^4} \left( 3 + \frac{5}{2} \lambda_{\rm H}^2
    - \sqrt{1 + \lambda_{\rm H}^2} \left( 3 + \lambda_{\rm H}^2 - 2 \lambda_{\rm H}^4 \right)
    \right) S_{\rm BH}
    \nonumber\\
    &=&
    \left( 1 + \frac{\lambda_{\rm H}^2-\lambda_\infty^2}{2} + O \left( \lambda_{\rm H}^2 \lambda_\infty^2 , \lambda_{\rm H}^4 , \lambda_\infty^4\right)
    \right) S_{\rm BH}
    \,,
    \label{eq:Quasilocal entropy - modified - horizon}
\end{eqnarray}
where $\lambda_{\rm H} = \lambda (2M)$. In fact $S(2M)>S(\infty)$ if $\lambda_{\rm H} \ge \lambda_\infty$.
This is a sign that the gravitational field itself contains some information
even in the vacuum exterior, a feature that is supported by the concept of net
gravitational energy.  The fact that the thermodynamic quasilocal entropy of the black hole is not a constant outside the horizon, in the presence of holonomy modifications, is a signal that the constant classical Bekenstein-Hawking entropy  \textit{does not} count the quantum degrees of freedom of the black hole. Using general arguments, it has previously been suggested that once black to white hole transitions are allowed, it can be shown that the Bekenstein-Hawking entropy is not a measure of the black-hole Hilbert space but rather of the degrees of freedom of the horizon through which the black hole interacts with its surroundings \cite{Bardeen1, Bardeen2}.
In the presence of holonomy terms, we find that this expectation is further extended in the sense that the gravitational field itself also contains information and hence the entropy expressions computed by different external observers need not be the same. It will be interesting to check  in the future whether information lost inside the black hole can indeed be `recovered' through the white hole  once matter fields are added to the mix and the entropy stored in the gravitational field is taken into account.

\section{Constant holonomy function}
\label{sec:mu0 scheme}

In this section we revisit the simplest, but nontrivial, case of constant
$\lambda=\tilde{\lambda}$, such that $h(x)=1$, which had been previously solved in
\cite{alonso2022nonsingular}.  In the LQG literature, a constant holonomy
parameter is often referred to as the $\mu_0$-scheme. Since the angular lengths
of holonomies on the symmetry spheres are constant, irrespective of their area,
the geometrical length $\tilde{\lambda} \sqrt{E^x}$ or the area of a plaquette
increase with the radius and is unbounded asymptotically.  In the
Schwarzschild gauge, for instance, the geometrical length equals
$\tilde{\lambda} x$ and grows linearly with the radial coordinate. This increase has
no effect on holonomy modifications in the static
region of the Schwarzschild geometry because the increasing length is
multiplied by $K_{\varphi}=0$. However, it could be a problem in non-static
gauges or when the static region is extended to a dynamical homogeneous region
in the presence of a positive cosmological constant. For instance, if
$E^x=t_{\rm h}^2$, the geometrical holonomy length increases with time and is
multiplied by $K_{\varphi}$ which, according to its classical behavior, could
be expected to be constant and non-zero:
$K_{\varphi}\propto \dot{E}^x/\sqrt{E^x}$. However,  the crucial point is that the dynamical behavior of
$K_{\varphi}$ is changed by holonomy modifications. The modified equation of
motion (\ref{Kphi}) then shows that, for constant $\lambda=\tilde{\lambda}$
and $E^x=t_{\rm h}^2$, holonomy modifications, quantified as the ratio
$(2\tilde{\lambda})^{-1}\sin(2\tilde{\lambda}K_{\varphi})/(2\dot{E}^2/\sqrt{E^x})$
of the modified and classical terms representing $K_{\varphi}$, are in fact
constant (for constant $c_f$) and do not show any potential implications of an increasing
geometrical length of the holonomy.

In a covariant formulation of holonomy modifications based on emergent
modified gravity, holonomy functions have implications not only for the
Hamiltonian constraint and the canonical dynamics but also for the geometrical
structure of space-time.  Unlike the Hamiltonian constraint, the emergent
metric contains a $K_{\varphi}$-independent $\lambda$-term that remains
non-zero even in the static region of a Schwarzschild gauge.  Imposing
asymptotic flatness on the metric (\ref{eq:Spacetime metric - modified -
  Schwarzschild}) with $\Lambda=0$, which requires $\alpha=1/\chi_0$ and
$\chi_0 = 1/\sqrt{1+\tilde{\lambda}^2}$, we obtain the line element
\begin{eqnarray}
    {\rm d} s^2 =
    - \left(1 - \frac{2 M}{x} \right) {\rm d} t^2
    + \left( 1 - \frac{\tilde{\lambda}^2}{1+\tilde{\lambda}^2} \frac{2M}{x}
    \right)^{-1} \left(1 - \frac{2 M}{x} \right)^{-1} {\rm d} x^2
    + x^2 \ {\rm d} \Omega^2
    \,.
    \label{eq:Spacetime metric - mu0-scheme - exterior}
\end{eqnarray}
The heuristic relationship between $\tilde{\lambda}$ and a holonomy length is lost
in this expression, and $\tilde{\lambda}$ is not explicitly multiplied by factors of
$x$. Nevertheless, this line element has been derived from a holonomy-modified
Hamiltonian constraint before the gauge was fixed.  

In spite of these cautionary remarks about customary statements concerning
holonomy modifications with constant length $\lambda=\tilde{\lambda}$, we will demonstrate in this
section that there are several unexpected features in this case, including:
\begin{enumerate}
    \item The deflection angle of a light ray moving past a massive object
      receives a holonomy modification for arbitrarily large impact parameters,
      even in the zero-mass limit. 
    \item The black-hole entropy monotonically increases as the observer moves
      farther away from the horizon. The maximum, given by the asymptotic
      value does not correspond to the Bekenstein-Hawking entropy. 
    \item In the presence of a positive cosmological constant, the areal
      radius is bounded not only from below but also from above. 
\end{enumerate}

\subsection{Minimum radius and geometric conditions}

The minimum radius of space-times for constant $\lambda=\tilde{\lambda}$ is
obtained by solving equation~(\ref{eq:Minimum radius equation}):
\begin{equation}
    x_{\tilde{\lambda}} = \frac{2 M \tilde{\lambda}^2}{1+\tilde{\lambda}^2}
    \,.
\end{equation}
At this minimum radius, the Ricci scalar (\ref{eq:Ricci scalar - reflection surface}) takes the value
\begin{eqnarray}
    {\cal R} |_{x=x_{\tilde{\lambda}}} = \frac{3 M}{x_{\tilde{\lambda}}^3}
    = \frac{3}{8 M^2} \left(\frac{1+\tilde{\lambda}^2}{\tilde{\lambda}^2}\right)^3
    \,,
    \label{eq:Ricci scalar - reflection surface - mu0}
\end{eqnarray}
and the Kretschmann scalar equals
\begin{eqnarray}
    K |_{x=x_{\tilde{\lambda}}} \equiv R_{\mu\nu\alpha\beta} R^{\mu\nu\alpha\beta} |_{x=x_{\tilde{\lambda}}}
    = 
    \frac{1}{x_{\tilde{\lambda}}^4 \tilde{\lambda}^4} \left( \frac{9}{4} + \frac{\tilde{\lambda}^2}{2} + 4 \tilde{\lambda}^4\right)
    \,.
\end{eqnarray}
Both expressions are finite but diverge in the classical limit,  $\tilde{\lambda}
\rightarrow 0$, in which $x_{\tilde{\lambda}}\to0$.
A comoving observer in the homogeneous  interior passes through this
region in a finite amount of proper time, given by
\begin{eqnarray}
    \tau_{\rm cross} = \pi \left(2 M + x_{\tilde{\lambda}}\right)
    \,.
\end{eqnarray}

The expansion parameter for null geodesic congruences is
\begin{eqnarray}
    \theta_{\pm}=\pm\frac{2}{x}\sqrt{1-\frac{x_{\tilde{\lambda}}}{x}}
\end{eqnarray}
and changes relative to the affine parameter $\psi$ according to
\begin{eqnarray}\label{mu0-theta}
    \frac{{\rm d}\theta_{\pm}}{{\rm d}\psi}=-\frac{2}{x^2} \left(1-\frac{3 x_{\tilde{\lambda}}}{2 x}\right)\,.
\end{eqnarray}
In the classical limit, $\tilde{\lambda}\rightarrow0$, we have
\begin{eqnarray}
    \lim_{\tilde{\lambda}\rightarrow0}\frac{{\rm d}\theta_{\pm}}{{\rm d}\psi}&=&-\frac{2}{x^2}<0
\end{eqnarray}
which means that both ingoing and outgoing congruences converge towards the
singularity, and therefore the geometry focuses null rays.  For non-zero
$\tilde{\lambda}$, however, light rays are defocused near the reflection symmetry surface in
the region $x<3x_{\tilde{\lambda}}/2$.

For timelike geodesics at rest at infinity, we have
\begin{eqnarray}
\theta_{\pm}=-\frac{3}{2 x}\sqrt{\frac{2M}{x}}\sqrt{1-\frac{x_{\tilde\lambda}}{x}} 
\end{eqnarray}
with proper-time rate of change
\begin{eqnarray}
    \frac{{\rm d}\theta_{\pm}}{{\rm d}\tau}=-\frac{9M}{2x^3}\left(1-\frac{4 x_{\tilde\lambda}}{3 x}\right)\,.
\end{eqnarray}
These geodesics are defocused in the region $x<4x_{\tilde{\lambda}}/3$. Thus, photons start feeling the repulsive effects of the emergent space-time before massive particles do.

The geometric conditions, discussed in Subsection~\ref{Repulsive} and now
evaluated  with $\lambda(x)=\tilde{\lambda}$, imply
\begin{itemize}
    \item Null geometric condition:\\
    Using (\ref{Ricci-null}),
    \begin{eqnarray}
        R_{\alpha\beta}v^{\alpha}_{\pm}v^{\beta}_{\pm}=-\frac{x_{\tilde\lambda}}{x^3}<0
    \end{eqnarray}
    and therefore the condition is violated everywhere in space-time, even at large scales in the exterior.
    \item Timelike geometric condition:\\
    Using (\ref{Ricci-timelike}),
    \begin{eqnarray}
    R_{\alpha\beta}u^{\alpha}_{\pm}u^{\beta}_{\pm}=-\frac{3Mx_{\tilde\lambda}}{2x^4}<0    
    \end{eqnarray}
    and therefore the condition as well is violated everywhere in space-time.
\end{itemize}
The violation of geometric conditions everywhere for constant
$\lambda=\tilde{\lambda}$ may be interpreted as an indication of large-scale
problems since quantum effects such as defocusing are not expected to be
dominant unless we are at large curvature. A quantitative assessment requires
a detailed analysis of observational features.

\subsection{Observational features}

As in our general discussion, the covariant nature of our emergent space-time
metric makes it possible to apply standard methods in order to derive
modifications to weak-field gravity, geodesic motion, and thermodynamics.

\subsubsection{Newtonian and zero-mass limits}

The metric (\ref{eq:Spacetime metric - mu0-scheme - exterior}) can be written in terms of the isotropic coordinate
\begin{equation}
    \mathfrak{r}(x) = \frac{M}{2 k} \frac{-2 \sqrt{x-2 M} \sqrt{x-x_{\tilde\lambda}}+2 M-2 x+x_{\tilde\lambda}}{2 M-x_{\tilde\lambda}}
    \,,
\end{equation}
where $k$ is a free constant. Inverting this function,
\begin{equation}
    x = k \mathfrak{r} \left( \left(1+\frac{M}{2 k \mathfrak{r}}\right)^2
    + \frac{x_{\tilde\lambda}}{2 k \mathfrak{r}} \left(1-\frac{M}{4 k \mathfrak{r}}-\frac{k \mathfrak{r}}{M}\right) \right)
    \,,
\end{equation}
leads to the line element
\begin{eqnarray}
    {\rm d} s^2 &=&
    - \left(1 - \frac{2 M}{k \mathfrak{r}} \left( \left(1+\frac{M}{2 k \mathfrak{r}}\right)^2
    + \frac{x_{\tilde\lambda}}{2 k \mathfrak{r}} \left(1-\frac{M}{4 k \mathfrak{r}}-\frac{k \mathfrak{r}}{M}\right) \right)^{-1} \right) {\rm d} t^2
    \nonumber\\
    &&\qquad
    + \left( \left(1+\frac{M}{2 k \mathfrak{r}}\right)^2
    + \frac{x_{\tilde\lambda}}{2 k \mathfrak{r}} \left(1-\frac{M}{4 k \mathfrak{r}}-\frac{k \mathfrak{r}}{M}\right) \right)^2 k^2 \left( {\rm d} \mathfrak{r}^2
    + \mathfrak{r}^2 \ {\rm d} \Omega^2 \right)
    \,.
    \label{eq:Spacetime metric - mu0-scheme - exterior - isotropic}
\end{eqnarray}

In the weak-field regime, $M\ll \mathfrak{r}$, the line element can be approximated by
\begin{eqnarray}
    {\rm d} s^2 &\approx&
    - \left(1 - \frac{2 M}{k \mathfrak{r}} \left(1 +\tilde{\lambda}^2
    \right) \right) {\rm d} t^2
    \nonumber\\
    &&\qquad
    + \left(1 +\tilde{\lambda}^2
    \right)^{-1} \left( \left(1 +\tilde{\lambda}^2
    \right)^{-1}
    + \frac{2M + x_{\tilde\lambda}}{k \mathfrak{r}} \right) k^2 \left( {\rm d} \mathfrak{r}^2
    + \mathfrak{r}^2 \ {\rm d} \Omega^2 \right)
    \,.
    \label{eq:Spacetime metric - mu0-scheme - exterior - isotropic - weak field}
\end{eqnarray}
Both $k$ and $1+\tilde{\lambda}^2$ can be eliminated by absorbing a factor of
  $k/(1+\tilde{\lambda}^2)$ in the coordinate $\mathfrak{r}$, or by simply choosing
  the free constant $k$ to equal $1+\tilde{\lambda}^2$.
Asymptotically, $\mathfrak{r}\to \infty$, we then obtain Minkowski space-time
\begin{eqnarray}
    {\rm d} s^2 &\approx&
    - {\rm d} t^2
    + {\rm d} \mathfrak{r}^2
    + \mathfrak{r}^2 \ {\rm d} \Omega^2
    \,.
    \label{eq:Spacetime metric - mu0-scheme - exterior - isotropic - asymptotic}
\end{eqnarray}

Minkowski space-time is also recovered in the zero-mass limit at any value of
$x$ or $\mathfrak{r}$, as can be seen by simply inserting $M=0$ in the
Schwarzschild coordinates (\ref{eq:Spacetime metric - mu0-scheme - exterior})
or in isotropic coordinates (\ref{eq:Spacetime metric - mu0-scheme - exterior
  - isotropic - weak field}), and using
$x_{\tilde\lambda}=2M\tilde{\lambda}^2/(1+\tilde{\lambda}^2)$ in the latter case.  As we saw previously,
a constant holonomy function $h(x)=1$, or
$\lambda=\tilde{\lambda}$, is the only choice that
recovers a Minkowski space-time in the zero-mass limit $M\to0$  for the case $\Lambda=0$.

\subsubsection{Nearly-circular orbits of massive objects}

Massive objects in nearly circular orbits for constant $\lambda=\tilde{\lambda}$ oscillate
about the equilibrium radius $x_0$ with frequency 
\begin{equation}
    \omega_r = \omega_\varphi \; \sqrt{1-\frac{6M}{x_0}} \; \sqrt{1 - \frac{x_{\tilde{\lambda}}}{x_0}}
    \,.
\end{equation}
The orbit has a precession rate
\begin{equation}
    \omega_p = \left( 1 - \sqrt{1-\frac{6M}{x_0}} \sqrt{1 - \frac{x_{\tilde{\lambda}}}{x_0}} \right) \omega_\varphi \ .
\end{equation}

\subsubsection{Deflection angle}

The deflection angle (\ref{eq:Deflection angle - simp}) for constant
$\lambda=\tilde{\lambda}$, expanded to leading order in $M/x_{\rm tp}$ with the turning-point
radius $x_{\rm tp}$,  is given by
\begin{eqnarray}
    \Delta \phi
    = \pi\left( \left(1-\frac{\tilde{\lambda}^2}{2}\right) \sqrt{1+\tilde{\lambda}^2} - 1\right)
    + \frac{4 M}{x_{\rm tp}} \frac{1 + 3 \tilde{\lambda}^2/2}{1+\tilde{\lambda}^2}
    + O(M^2/x_{\rm tp}^2)
    \label{eq:Deflection angle - m0}
\end{eqnarray}
and has the correct classical limit as $\tilde{\lambda}\to 0$.
However, for non-zero $\tilde{\lambda}$, there is a correction  independent of the
turning-point radius of the light ray and of the mass $M$.
The limit for $x_{\rm tp}\to\infty$ or $M\to0$ of $\Delta\phi$ therefore
has a non-zero remnant deflection angle, even at infinite distance from a
massive object or in the absence of any mass. In particular, even though the
zero-mass limit of the line element is equal to Minkowski space-time, the
deflection angle can deviate from the Minkowski result if it is computed
before the limit $M\to0$ is taken.

This result is surprising, but it is not necessarily problematic in an
observational context. The traditional, Eddington-type measurement of a
deflection angle does not determine a single $\Delta\phi$ but rather the
difference of two $\Delta\phi$, one with a heavy mass close to the light ray
($M\not=0$) and one with a light ray far from any heavy mass ($M\to0$). Since
the unexpected $\tilde{\lambda}$-term in $\Delta\phi$ is mass-independent, it
cancels out in any difference of two deflection angles.

\subsubsection{Net stress-energy tensor}

For constant $\lambda=\tilde{\lambda}$, the Ricci scalar and the net stress-energy
tensor, defined as the negative Einstein tensor, 
are given by
\begin{eqnarray}
    R^{(\tilde{\lambda})} &=& \frac{3 M x_{\tilde{\lambda}}}{x^4}
    \ ,
    \label{eq:Ricci scalar - constant holonomy}
    \\
    \bar{T}_{\mu\nu}^{(\tilde{\lambda})} {\rm d} x^\mu \otimes {\rm d} x^\nu
    &=&
    - \frac{2 M x_{\tilde{\lambda}}}{x^4} \left(1 - \frac{2 M}{x}\right) {\rm d} t^2
    + \frac{x_{\tilde{\lambda}}}{x^3} \left(1 - \frac{x_{\tilde{\lambda}}}{x} \right)^{-1} \left(1 - \frac{2 M}{x}\right)^{-1} {\rm d} x^2
    \nonumber\\
    &&
    - \frac{x_{\tilde{\lambda}}}{2 x} \left(1-\frac{M}{x}\right) {\rm d} \Omega^2
    \ ,
    \label{eq:Einstein tensor - constant holonomy}
\end{eqnarray}
using the metric (\ref{eq:Spacetime metric - mu0-scheme - exterior}).  We have
included the superscript $(\tilde{\lambda})$ in order to distinguish these
quantities from the corresponding versions in a $\bar{\mu}$-type scheme
studied in the next section. The energy densities according to a static
observer in the exterior and a comoving observer in the homogeneous interior
are given by
\begin{eqnarray}
    \rho^{(\tilde{\lambda})} &\equiv& 
    \bar{T}_{\mu\nu}^{(\tilde{\lambda})} \hat{t}^\mu \hat{t}^\nu
    = - \frac{2 M x_{\tilde{\lambda}}}{x^4}
    \,,
    \label{eq:Energy density - constant holonomy}
    \\
    \rho_{\rm h}^{(\tilde{\lambda})} &\equiv& 
    \bar{T}_{\mu\nu}^{(\tilde{\lambda})} \hat{t}_{\rm h}^\mu \hat{t}_{\rm h}^\nu
    = - \frac{x_{\tilde{\lambda}}}{t_{\rm h}^3} \left(1 - \frac{x_{\tilde{\lambda}}}{t_{\rm h}} \right)^{-2} \left(\frac{2 M}{t_{\rm h}}-1\right)^{-2}
    \,,
    \label{eq:Energy density - interior - constant holonomy}
\end{eqnarray}
respectively. Here, $\hat{t}^\mu \partial_\mu = N^{-1} \partial_t$ is
the normalized time-like Killing vector field, and
$\hat{t}_{\rm h}^\mu \partial_\mu = N^{-1} \partial_{t_{\rm h}}$ is
the normalized comoving velocity associated to the metric (\ref{eq:Spacetime
  metric - mu0-scheme - exterior}) under the appropriate coordinate swap.

Let us list four important observations.
\begin{enumerate}
    \item All the components of the Einstein tensor are asymptotically vanishing, with the angular pressure being the slowest to decay at a rate of $x^{-1}$ and the energy density being the fastest at a rate of $x^{-4}$.
    \item The Ricci scalar is positive everywhere,
      $R^{(\tilde{\lambda})}>0$, even in vacuum.
    \item Restricting to the exterior, the energy density is
      negative everywhere, $\rho^{(\tilde{\lambda})}<0$, the radial pressure is
      positive everywhere, $\bar{T}_{xx}^{(\tilde{\lambda})}>0$, and the angular
      pressure is negative everywhere, $\bar{T}_{\varphi\varphi}^{(\tilde{\lambda})}<0$.
    \item Restricting to the interior, the energy density for comoving observers is negative everywhere, the radial pressure is positive everywhere, while the angular pressure is negative for $t_{\rm h}>M$ and positive for $t_{\rm h}<M$.
\end{enumerate}

\subsubsection{Black-hole thermodynamics}

The net quasilocal energy (\ref{eq:Brown-York quasilocal energy - modified}) for
constant $\lambda=\tilde{\lambda}$ and with $\hat{N}=1$ is given by
\begin{equation}
    E_{\tilde{\lambda}} (x) = x \left(1-\sqrt{1 - \frac{2 M}{x}} \sqrt{1 - \frac{x_{\tilde{\lambda}}}{x}}  \right)
    \ ,
    \label{eq:Brown-York quasilocal energy - mu0-scheme}
\end{equation}
implying the net quasilocal entropy
\begin{equation}
    S_{\tilde{\lambda}} (x)
    =
    \frac{8 \pi x^2}{15 \tilde{\lambda}^4} \Bigg( \sqrt{1 -\frac{x_{\tilde{\lambda}}}{x}} \left( 3 + \tilde{\lambda}^2 \left( 1 + \frac{3 M}{x} \right) - 2 \tilde{\lambda}^4 \left( 1 + \frac{M}{x} - \frac{6 M^2}{x^2} \right) \right)
    - \left( 3 + \tilde{\lambda}^2 - 2 \tilde{\lambda}^4 \right)
    \Bigg)
    \label{eq:Quasilocal entropy - constant l}
\end{equation}
from  (\ref{eq:Quasilocal entropy - modified}).
This expression does not recover the classical value asymptotically:
\begin{equation}
   \lim_{x\to\infty} S_{\tilde{\lambda}} (x)
    =
    S_{\rm BH} \frac{1+2 \tilde{\lambda}^2}{1+\tilde{\lambda}^2}
    > S_{\rm BH}
    \,.
\end{equation}
At the horizon, it takes the value
\begin{equation}
    S_{\tilde{\lambda}} (2 M) = 
    S_{\rm BH} \left(1 + \frac{\tilde{\lambda}^4}{48}
    + \mathcal{O}(\tilde{\lambda}^6)
    \right)
    \,,
\end{equation}
and therefore $\lim_{x\to\infty}S_{\tilde{\lambda}} (x)>S_{\tilde{\lambda}} (2 M)$. (The
corrections to the horizon value are of order $\tilde{\lambda}^4$, while the
asymptotic value is corrected to order $\tilde{\lambda}^2$.)

In fact, the expression (\ref{eq:Quasilocal entropy - constant l}) is a
monotonically increasing function: the entropy increases as the observer moves
farther from the horizon.  One possible interpretation is that the
gravitational field in the vacuum exterior contains entropy and reduces the
information.  The overall entropy is then maximal for asymptotic observers
that have access to the whole gravitational field.  This is contrary to what
one would expect, namely, that if the gravitational field contains some
information, then asymptotic observers, who unlike near-horizon observers have
access to this information, would measure an entropy that would be the minimum
of the above expression. Since the increase of entropy is
  implied by holonomy modifications, heuristically related to discrete spatial
  structures, one might argue that our result for $S_{\tilde{\lambda}}$ can only
  take into account information accessible to measurements in a continuum
  model. Information related to spatial discreteness cannot be observed in
  this way, and
  therefore increases the entropy because it indirectly affects the geometry.

Whether this behavior is considered
problematic depends on the physical significance attached to the net energy
and entropy of a modified vacuum space-time.

\subsection{Maximal radius in the presence of a cosmological constant}

The presence of a cosmological constant can significantly alter the asymptotic
and global structure of the modified space-time.
Including a cosmological constant, the space-time line element for constant
$\lambda=\tilde{\lambda}$  is given by
\begin{eqnarray}
    {\rm d} s^2 &=&
    - \left(1 - \frac{2 M}{x} - \frac{\Lambda x^2}{3}\right) {\rm d} t^2
    \\
    &&
    + \left( 1 - \frac{\tilde{\lambda}^2}{1+\tilde{\lambda}^2}\left(\frac{2M}{x} + \frac{\Lambda x^2}{3} \right)
    \right)^{-1} \left(1 - \frac{2 M}{x}-\frac{\Lambda x^2}{3}\right)^{-1} {\rm d} x^2
    + x^2 {\rm d} \Omega^2
    \,.\nonumber
    \label{eq:Spacetime metric - mu0-scheme - exterior - cosmo}
\end{eqnarray}
In the classical case, $\tilde{\lambda}\to0$, a nonvanishing, positive $\Lambda$ implies that a second horizon exists at
\begin{equation}
    x_\Lambda \approx \sqrt{\frac{3}{\Lambda}}
\end{equation}
(neglecting $M/x$ for small $\Lambda$), beyond which the space-time is homogeneous.

With $\tilde{\lambda} \neq0$, the coordinate singularity determined by equation
(\ref{eq:Minimum radius equation}) is a third-order polynomial and hence has
three rather complicated solutions some of which may be complex in general.  For small
$x \ll 2 M$, the Newtonian term dominates and the solution is approximately
that of the $\Lambda=0$ case,
\begin{equation}
    x_{\tilde{\lambda}}^{(-)} \approx \frac{2 M \tilde{\lambda}^2}{1+\tilde{\lambda}^2}
    \,.
\end{equation}
For large $x\gg 2M$, the cosmological constant dominates and the solution is approximately
\begin{equation}
    x_{\tilde{\lambda}}^{(+)} \approx \sqrt{\frac{3}{\Lambda} \frac{1+\tilde{\lambda}^2}{\tilde{\lambda}^2}}
\end{equation}
if $\Lambda>0$, which is the case we will focus on since it agrees with
observations.  This value is always outside the cosmological horizon and is a
maximum-radius surface, beyond which the space-time starts collapsing;  see
Fig.~\ref{fig:Holonomy_KS_Vacuum_Wormhole-Periodic-Cosmo-MaxRadius}.

The existence of a maximum radius due to the presence of a quantum parameter
is unexpected since it occurs at macroscopic scales where quantum gravity is
not expected to play a significant role. This outcome of holonomy
  modifications in the traditional treatment is usually related to a
  growing geometrical length $\tilde{\lambda}t_{\rm h}$ in a homogeneous region,
  which, unlike in the static region, does contribute to holonomy
  modifications in the Hamiltonian constraint because
  $K_{\varphi}\not=0$. However, as already discussed, covariant holonomy modifications in the
  Hamiltonian constraint do not show an increasing holonomy length but rather
  imply a constant magnitude of holonomy modifications related to the
  classical form of extrinsic curvature via (\ref{Kphi}). The reason for
  unexpected effects on large scales is therefore not a growing holonomy
  length but rather the fact that holonomy modifications do not decrease
  quickly enough as $t_{\rm h}$ increases.

  Although non-classical behavior on large scales is unexpected as an effect
  of quantum gravity, it is not impossible. While local quantum-gravity
  effects should be small at low curvature, global properties in large regions
  are subject to a combination of a huge number of tiny quantum-gravity
  corrections, which could conceivably add up to sizeable
  contributions. Moreover, non-classical behavior on large scales may well
  happen in the general context of modified gravity. Large-scale modifications
  of the classical behavior are not necessarily incompatible with observations
  if they happen sufficiently far outside of the cosmological horizon. Making
  $\tilde{\lambda}$ sufficiently small, all the unexpected effects found in this
  section may therefore be acceptable in an individual model.
    Nevertheless, the physical viability of a fundamental theory, such as the
    concept of holonomy
    modifications in general or of LQG, can easily be endangered if
    microscopic features have significant implications on macroscopic
    scales. Corrections to the classical theory may then be hard to control
    when one goes beyond relatively simple models such as spherically
    symmetric ones. The cosmological constant itself, in its traditional
    explanation as vacuum energy, is an example of such
    UV-IR mixing. Its macroscopic problems seem to be exacerbated by holonomy
    modifications with constant length parameter. However, while there is
    potential observational evidence for the presence of a cosmological
    constant, there is no indication for the existence of a maximal
    radius. From this perspective, it is therefore best to avoid models that
    place an upper bound on the range of the radial coordinate.

Models with constant holonomy length may still be relevant if they are used in
a finite range of radii. For instance, the complete function $\lambda(x)$ may not follow a
  strict power law, but be nearly constant in a bounded region of
  space-time outside of which it decreases toward zero asymptotically. It is then important to
  show that non-constant $\lambda(E^x)$ can indeed suppress holonomy-type
  effects in classical regimes and restrict implications of holonomy
  modifications to the expected quantum-gravity regime, as we will do in the
  next section. We also reiterate the importance of a covariant formulation
  based on an emergent space-time metric, which makes it possible to have
  consistent effects in the static gauge, in which traditional holonomy
  modifications would disappear, and in non-static homogeneous regions of
  space-time.

\section{Decreasing holonomy function}
\label{sec:mubar scheme}

In this section, we use a holonomy function (\ref{eq:mu-scheme holonomy
  function}) $h(E^x)=\lambda(E^x)/\tilde{\lambda}= \hat{r}/\sqrt{E^x}$
analogous to a $\bar{\mu}$-type
scheme, and derive the particular physical effects of this choice. As in
equation~(\ref{mur}), $\hat{r}$ is a reference radius where $h(\hat{r})=1$.  In models
of loop quantum gravity, a specific coefficient is often chosen in the parameterization
\begin{equation}
    \lambda^2 = \tilde{\lambda}^2\hat{r}^2/E^x
     \,,\label{eq:mu-scheme holonomy function - repeated}
\end{equation}
where the constant $\tilde{\lambda}=\lambda\sqrt{E^x}/\hat{r}$
can be interpreted as the side length of 
plaquettes in a uniform triangulation of spheres.  If one uses regular
polygons with $n$ sides and assigns the area
$\Delta=8 \pi \gamma \ell_{\rm Pl}^2$ to each of them, suggested for instance
by the area gap of LQG, it follows that
\begin{equation}
  \tilde{\lambda}^2\hat{r}^2=4 n^{-1}\tan (\pi/n) \Delta=:\tilde{\Delta}\,.
\end{equation}
(See Sec.~\ref{sec:Effective loop quantum gravity} for the full details of
these relations.) Note, however, that holonomies of a gauge connection do not
depend on the metric, and therefore it is not required to have a strict
relationship between the constant $\tilde{\lambda}$ in (\ref{eq:mu-scheme
  holonomy function - repeated}) and the area spectrum, even in a
fundamentally discrete theory such as loop quantum gravity.

Using this parameterization of $\lambda(E^x)$, we have $\lambda_\infty = 0$
and, for $\Lambda=0$, the asymptotically flat line element
(\ref{eq:Spacetime metric - modified - Schwarzschild - Asymptotic and zero
  mass - simp}) for the black hole exterior is simplified to
\begin{eqnarray}
    {\rm d} s^2 =
    - \left(1 - \frac{2 M}{x}\right) {\rm d} t^2
    + \left( 1 + \frac{\tilde \Delta}{x^2} \left( 1 - \frac{2 M}{x} \right)
    \right)^{-1} \left(1 - \frac{2 M}{x}\right)^{-1} {\rm d} x^2
    + x^2 {\rm d} \Omega^2
    \,.
    \label{eq:Spacetime metric - mu-scheme - exterior}
\end{eqnarray}
We will reincorporate the cosmological constant in the final subsection and
revisit the status of a maximum radius that appeared for constant $\lambda=\tilde{\lambda}$.
Our detailed derivations show that the decreasing parameterization of
$\lambda(E^x)$ removes all the unexpected features of constant $\lambda$ in
low-curvature regimes.

\subsection{Minimum radius and geometric conditions}

The minimum radius of space-times for $\lambda\propto1/\sqrt{E^x}$,
obtained by solving equation (\ref{eq:Minimum radius equation}), is given by
\begin{eqnarray}
    x_{\tilde\Delta} &=&
    (\tilde{\Delta} M)^{1/3} \frac{\left( 1 + \sqrt{1+\tilde{\Delta}/(27 M^{2})}\right)^{2/3} - \left(\tilde{\Delta}/(27 M^{2})\right)^{1/3}}{\left(1 + \sqrt{1+\tilde{\Delta}/(27 M^{2})} \right)^{1/3}}
    \ .
    \label{eq:Minimum radius - mu-scheme}
\end{eqnarray}
To analyze the effect for small masses, we define $\delta \equiv 4 M^2 / \tilde{\Delta}$ and expand around $\delta=0$:
\begin{eqnarray}
    \frac{x_{\tilde\Delta}}{2 M} &=&
    1 - \delta + 3 \delta^2
    + O \left( \delta^3 \right)
    \,.
    \label{eq:Minimum radius - mu-scheme - small masses}
\end{eqnarray}
Therefore, to leading order, the horizon of a small mass corresponds to the minimum
radius such that the wormhole interior will be very short, determined as a
timelike distance
since $x$ is a time coordinate in the interior.
This cannot happen in the case of constant $\lambda=\tilde{\lambda}$, where
$x_{\tilde{\lambda}} = 2 M \tilde{\lambda}^2 / (1+\tilde{\lambda}^2)$ and the
 horizon and minimal radius do not match to leading order unless
$M=0$. On the other hand, for large $M$ black-holes, we define $\tilde{\delta}=\tilde{\Delta}/(27M^2)$, and by expanding around $\tilde{\delta}=0$, we obtain
\begin{eqnarray}
    \frac{x_{\tilde\Delta}}{2 M} &=&\frac{3}{4^{1/3}}\left(\tilde{\delta}^{1/3}-\frac{1}{4^{1/3}}\tilde{\delta}^{2/3}+O \left( \tilde{\delta}^{4/3} \right)
    \,.\right)
\end{eqnarray}
The results indicate that the minimum radius is relatively small compared to the horizon, suggesting that the wormhole interior will be significantly large, similar to the constant $\lambda = \tilde{\lambda}$ case.

The expansion parameter for null geodesic congruences is
\begin{eqnarray}
    \theta_{\pm}=-\frac{2}{x}\sqrt{1+\frac{\tilde{\Delta}}{x^2}\left(1-\frac{2M}{x}\right)}
\end{eqnarray}
and changes according to
\begin{eqnarray}
    \frac{{\rm d}\theta_{\pm}}{{\rm d}\psi}=-\frac{2}{x^2}\left[1+\frac{\tilde{\Delta}}{x^2}\left(2-\frac{5M}{x}\right)\right]
\end{eqnarray}
The correction term is dominant near the surface of reflection symmetry, and the sign changes at some surface $\bar{x}=f(M,\tilde{\Delta})x_{\tilde{\Delta}}$ with $f(M,\tilde{\Delta})>1$. Similarly, for timelike congruences, we have
\begin{eqnarray}
    \theta_{\pm}=-\frac{3}{2x}\sqrt{\frac{2M}{x}}\sqrt{1+\frac{\tilde{\Delta}}{x^2}\left(1-\frac{2M}{x}\right)}
\end{eqnarray}
with the rate of change
\begin{eqnarray}
    \frac{{\rm d}\theta_{\pm}}{{\rm d}\tau}=-\frac{9M}{2x^3}\left[1+\left(\frac{5}{3}-\frac{4M}{x}\right) \frac{\tilde{\Delta}}{x^2}\right].
\end{eqnarray}
The correction terms are dominant near the hypersurface of reflection symmetry, changing
sign at $\tilde{x}=g(M,\tilde{\Delta})x_{\tilde\Delta}$, where once more
$g(M,\tilde{\Delta})>1$. It can be demonstrated that
$g(M,\tilde{\Delta})<f(M,\tilde{\Delta})$. Thus, as with constant
$\lambda=\tilde{\Delta}$, free-falling photons start being defocused at larger
scales compared with massive particles. 

The geometric conditions for $\lambda\propto1/\sqrt{E^x}$ are:
\begin{itemize}
    \item Null geometric condition:
    \begin{eqnarray}
        R_{\mu\nu}v^{\mu}_{\pm}v^{\nu}_{\pm}=\frac{2}{x^2} \frac{\tilde{\Delta}}{x^2}\left(1-\frac{3M}{x}\right)\geq 0\,,
    \end{eqnarray}
    This condition is violated only in the region $x<3M$.
    \item Timelike geometric condition:
    \begin{eqnarray}
        R_{\mu\nu}u^{\mu}_{\pm}u^{\nu}_{\pm}=\frac{3M}{x^3}\frac{\tilde{\Delta}}{x^2}\left(1-\frac{3M}{x}\right)\geq 0\,,
    \end{eqnarray}
    Also this condition is violated only in the region $x<3M$.
\end{itemize}
Unlike for constant $\lambda=\tilde{\lambda}$, here the geometric conditions are
satisfied at large scales and therefore the modified gravitational force focuses 
congruences far from the black hole, similar to its classical effect.

\subsection{Observational features}

As before, the weak-field behavior and properties of geodesic motion can help
to distinguish physical implications of different modification functions.

\subsubsection{Newtonian limit}

For a decreasing holonomy function $h(E^x)=\hat{r}/\sqrt{E^x}$ and using the coordinate transformation $x = r \left(1+ M/(2r)\right)^2$, the emergent space-time line element is given by
\begin{eqnarray}
    {\rm d} s^2 &=& - \left(\frac{1-\frac{M}{2r}}{1+\frac{M}{2r}}\right)^2 {\rm d}t^2 + \left(1+\frac{M}{2r}\right)^4 \left( \left( 1 + \frac{\tilde{\Delta}}{r^2} \frac{\left(1-\frac{M}{2r}\right)^2}{\left(1+\frac{M}{2r}\right)^4}
    \right)^{-1} \frac{{\rm d} r^2}{\chi_0^2} +r^2 {\rm d}\Omega^2\right)
    \nonumber\\
    &\approx&
    - \left(1-\frac{2M}{r}\right) {\rm d}t^2 + \left(1+\frac{2M}{r}\right) \left( \left( 1 + \frac{\tilde{\Delta}}{r^2} \left(1-\frac{3 M}{r}\right)
    \right)^{-1} \frac{{\rm d} r^2}{\chi_0^2} +r^2 {\rm d}\Omega^2\right)
    \,.
\end{eqnarray}
where we assumed $M/r\ll 1$ in the second line.
The isotropic coordinate (\ref{eq:Second isotropic coord transf}) is now given by
\begin{equation}\label{eq:Second isotropic - mubar}
    \mathfrak{r} = k \exp \int \frac{{\rm d} r}{r} \left( 1 - \frac{\tilde{\Delta}}{r^2} \left(1-\frac{3M}{r}\right)
    \right)^{1/2}\approx k \exp \int\frac{{\rm d}r}{r}
    \sqrt{1-\tilde{\Delta}/r^2} \left(1+\frac{3\tilde{\Delta}M}{2r(r^2-\tilde{\Delta})}\right)\,.
  \end{equation}
The integration can be done in closed form, but the resulting expressions are
lengthy.

In most cases of interest, the weak-field limit $M\ll r$ also implies
$\tilde{\Delta}\ll r^2$, which can be used to simplify (\ref{eq:Second
  isotropic - mubar}). To leading order in $\tilde{\Delta}/r^2$ and $M/r$, the
integral is independent of $M$ and therefore agrees with its value in the zero-mass limit.

\subsubsection{Zero-mass limit}

The decreasing holonomy function implies a metric (\ref{eq:Spacetime metric -
  modified - Schwarzschild}) that is not flat in the zero mass limit $M\to0$
in the case of $\Lambda=0$, and instead approaches the line element
\begin{eqnarray}
    {\rm d} s^2 &=&
    - {\rm d} t^2
    + \left( 1 + \frac{\tilde \Delta}{x^2}
    \right)^{-1} {\rm d} x^2
    + x^2 {\rm d} \Omega^2
    \,.
    \label{eq:Spacetime metric - mu-scheme - exterior - zero mass limit}
\end{eqnarray}
We interpret this result as suggesting  that the spatial geometry is not smooth at the Planck scale $x \sim \sqrt{\tilde \Delta}$.
The 3-dimensional volume in a slice of constant $t$ and enclosed by $x<r$ is given by
\begin{eqnarray}
    V_{\Delta}(r) &=& 4 \pi \int_0^r {\rm d} x\ x^2 \left( 1 + \frac{\tilde{\Delta}}{x^2}
    \right)^{-1/2}
    \nonumber\\
    &=&
    \frac{4 \pi r^3}{3} \left( \left(1-2 \frac{\tilde{\Delta}}{r^2} \right) \sqrt{1+\frac{\tilde{\Delta}}{r^2}} + 2 \left(\frac{\tilde{\Delta}}{r^2}\right)^{3/2}\right)
    \nonumber\\
    &=&
    \frac{4 \pi r^3}{3} \left( 1-\frac{3}{2} \frac{\tilde{\Delta}}{r^2}
    + 2 \frac{\tilde{\Delta}^{3/2}}{r^3}
    + O \left(\frac{\tilde{\Delta}^2}{r^4}\right)\right)
    \,.
\end{eqnarray}
The volume of a given region is therefore always smaller compared with flat space.

The zero-mass metric (\ref{eq:Spacetime metric - mu-scheme - exterior - zero mass limit}) can be expressed in the isotropic coordinate
\begin{equation}
    \mathfrak{r} = \sqrt{\frac{\tilde{\Delta}}{4}} \sqrt{\frac{\sqrt{1+\tilde{\Delta}/x^2} + 1}{\sqrt{1+\tilde{\Delta}/x^2} - 1}}
    \,,
\end{equation}
such that
\begin{eqnarray}
    {\rm d} s^2 &=&
    - {\rm d} t^2
    + \left(1 - \frac{\tilde{\Delta}}{4\mathfrak{r}^2} \right)^2 \left( {\rm d} \mathfrak{r}^2
    + \mathfrak{r}^2 {\rm d} \Omega^2 \right)
    \,.
    \label{eq:Spacetime metric - mu-scheme - exterior - zero mass limit - isotropic}
\end{eqnarray}

\subsubsection{Nearly-circular orbits of massive objects}

Massive objects in nearly circular orbits, using a decreasing holonomy function
$h(E^x)=\hat{r}/\sqrt{E^x}$, oscillate around the equilibrium radius $x_0$ with frequency
\begin{equation}
    \omega_r = \omega_\varphi \sqrt{1-\frac{6M}{x_0}} \sqrt{1 + \frac{\tilde{\Delta}}{x^2} \left(1 - \frac{2 M}{x_0}\right)}
    \,.
\end{equation}
Their orbits have the precession rate
\begin{equation}
    \omega_p = \left( 1 - \sqrt{1-\frac{6M}{x_0}} \sqrt{1 + \frac{\tilde{\Delta}}{x^2} \left(1 - \frac{2 M}{x_0}\right)} \right) \omega_\varphi
    \,.
\end{equation}

\subsubsection{Deflection angle}

The deflection angle (\ref{eq:Deflection angle - simp}), to leading order in $M/x_{\rm tp}$, is given by
\begin{eqnarray}
    \Delta \phi
    &=&
    - \frac{\pi}{4} \frac{\tilde{\Delta}}{x_{\rm tp}^2}
    + \frac{4 M}{x_{\rm tp}} \left(1 + \frac{3}{2} \frac{\tilde{\Delta}}{x^2}\right) \left(1 + \frac{\tilde{\Delta}}{x^2}\right)^{-3/2}
    + O \left(M^2/x_{\rm tp}^2\right)
    \nonumber\\
    &=&
    - \frac{\pi}{4} \frac{\tilde{\Delta}}{x_{\rm tp}^2}
    + \frac{4 M}{x_{\rm tp}} \left(1-\frac{3}{8} \left(\frac{\tilde{\Delta}}{x^2}\right)^2 + O \left(\frac{\tilde{\Delta}^3}{x^6}\right)\right)
    + O \left(M^2/x_{\rm tp}^2\right)
    \,.
    \label{eq:Deflection angle - mbar}
\end{eqnarray}
While a repulsive correction survives the zero-mass limit, it decreases
quickly for $x_{\rm tp}^2\gg\tilde{\Delta}$, unlike in the case of a constant
holonomy function.

\subsubsection{Net stress-energy tensor}

The Ricci scalar for $h(x)=\hat{r}/x$ is given by
\begin{eqnarray}
    R &=&
    \frac{2 \tilde{\Delta}}{x^4} \left( 1 - \frac{7 M}{x} + \frac{9 M^2}{x^2} \right)
    \ ,
    \label{eq:Ricci scalar - mu-scheme}
\end{eqnarray}
and, using (\ref{eq:Einstein tensor - modified}), the net stress-energy tensor equals
\begin{eqnarray}
    \bar{T}_{\mu\nu}^{(\tilde\Delta)} {\rm d} x^\mu \otimes {\rm d} x^\nu
    &=&
    - \frac{\tilde{\Delta}}{x^4} \left(1 - \frac{6 M}{x}\right) \left(1 - \frac{2 M}{x}\right)^2 {\rm d} t^2
    - \frac{\tilde{\Delta}}{x^4} \left(1 + \frac{\tilde{\Delta}}{x^2} \left( 1 - \frac{2 M}{x} \right) \right)^{-1} {\rm d} x^2
    \nonumber\\
    &&
    + \frac{\tilde{\Delta}}{x^4} \left( 1 - \frac{3 M}{x} \right) \left( 1 - \frac{M}{x} \right) x^2 {\rm d} \Omega^2
    \,.
    \label{eq:Einstein tensor - mu-scheme}
\end{eqnarray}
The gravitational energy densities in the exterior and interior are given by
\begin{eqnarray}
    \rho &\equiv& 
    \bar{T}^{(\tilde\Delta)}_{\mu \nu} \hat{t}^\mu \hat{t}^\nu
    = - \frac{\tilde{\Delta}}{x^4} \left(1 - \frac{6 M}{x}\right) \left(1 - \frac{2 M}{x}\right)
    \,,
    \label{eq:Energy density - mu-scheme}
    \\
    \rho_{\rm h} &\equiv& 
    \bar{T}^{(\tilde\Delta)}_{\mu \nu} \hat{t}_{\rm h}^\mu \hat{t}_{\rm h}^\nu
    = - \frac{\tilde{\Delta}}{t_{\rm h}^4} \left( 1 - \frac{\tilde{\Delta}}{t_{\rm h}^2} \left(\frac{2 M}{t_{\rm h}} -1\right)
    \right)^{-2} \left(\frac{2 M}{t_{\rm h}}-1\right)^{-1}
    \,,
    \label{eq:Energy density - interior - mu-scheme}
\end{eqnarray}
respectively, where $\hat{t}^\mu \partial_\mu = N^{-1} \partial_t$ is the
normalized stationary velocity, while $\hat{t}_{\rm h}^\mu \partial_\mu = N^{-1} \partial_{t_{\rm h}}$ is the normalized comoving velocity associated to the metric (\ref{eq:Spacetime metric - mu-scheme - exterior}) under the appropriate coordinate swap.

The four main differences between a decreasing holonomy function $h(x)=\hat{r}/x$
and a constant $h(x)=1$ are:
\begin{enumerate}
    \item All the components of the Einstein tensor decay faster with increasing $x$ for a
      decreasing holonomy function, with the exception of $G_{tt}$ which has a
      similar decay in both schemes. For $h(x)=\hat{r}/x$, all components decrease
      at the same rate to leading order in $1/x$. 
    \item For $h(x)=\hat{r}/x$, the Ricci curvature (\ref{eq:Ricci scalar -
        mu-scheme}) vanishes at the coordinates 
    \begin{eqnarray}
        \frac{x_0^{\pm}}{2 M} &=& \frac{7 \pm \sqrt{13}}{4}
        \ ,
    \end{eqnarray}
    where $x_0^+ \approx 5.3 M >2M$ and $x_0^- \approx 1.69 M < 2M$.
    Therefore, unlike for a constant holonomy function, the Ricci scalar
    is not positive everywhere, becoming negative outside of but near
    the horizon, 
    \begin{eqnarray}
        R &\xrightarrow[x\to 2M (1 + \delta)]{\empty}&
        - \frac{\tilde{\Delta}}{32 M^4}
        + O\left(\delta^2/(2M)^2\right)
        \,.
    \end{eqnarray}
    \item Restricted to the exterior, the energy density (\ref{eq:Energy
        density - mu-scheme}) has different signs in different regions,
      $\rho<0$ for $x>6M$ and $\rho>0$ for $2M<x<6M$; the radial pressure is
      negative everywhere, $\bar{T}_{xx}<0$, and the angular pressure changes
      sign, such that $\bar{T}_{\varphi\varphi}>0$ for $x>3M$ and
      $\bar{T}_{\varphi\varphi}<0$ for $2M<x<3M$. 
    \item Restricted to comoving observers in the interior, the energy density
      is negative everywhere, the radial pressure is positive everywhere,
      $\bar{T}_{tt}>0$, and the angular pressure changes sign at $t_{\rm h}=M$
      such that it is negative for $t_{\rm h}>M$ and positive for $t_{\rm
        h}<M$. 
\end{enumerate}

Finally, we note that in the zero-mass limit, $M\to0$, the net stress-energy tensor (\ref{eq:Einstein tensor - mu-scheme}) does not vanish but reduces to
\begin{equation}
    \bar{T}_{\mu\nu}^{(\tilde\Delta)} {\rm d} x^\mu \otimes {\rm d} x^\nu \big|_{M\to 0}
    =
    - \frac{\tilde{\Delta}}{x^4} {\rm d} t^2
    - \frac{\tilde{\Delta}}{x^4} \left(1 + \frac{\tilde{\Delta}}{x^2} \right)^{-1} {\rm d} x^2
    + \frac{\tilde{\Delta}}{x^4} x^2 {\rm d} \Omega^2
    \,,
    \label{eq:Einstein tensor - mu-scheme - zero mass}
\end{equation}
while the energy density is simply given by $\rho|_{M\to 0} = \bar{T}_{tt}|_{M\to 0}$.
Some gravitational stress-energy therefore survives in the vacuum but quickly decays as $\tilde{\Delta}/x^4$ and may therefore be observable only at the Planck scale.

\subsubsection{Black-hole thermodynamics}

The net quasilocal energy (\ref{eq:Brown-York quasilocal energy - modified}) for
$h(x)=\hat{r}/x$ with $\hat{N}=1$ is given by 
\begin{equation}
    E_{\tilde\Delta} (x) = - x \left( \sqrt{1 - \frac{2 M}{x}} \sqrt{1 + \frac{\tilde{\Delta}}{x^2} \left( 1 - \frac{2 M}{x} \right)} - \sqrt{1 + \frac{\tilde{\Delta}}{x^2}} \right)
    \ ,
    \label{eq:Brown-York quasilocal energy - mu-scheme}
\end{equation}
and the net quasilocal entropy (\ref{eq:Quasilocal entropy - modified}) by
\begin{eqnarray}
    S_{\tilde\Delta} (\bar{x})
    &=&
    \frac{8 \pi \bar{x}^6}{15 \tilde{\Delta}^2} \Bigg( \sqrt{1 + \frac{\tilde{\Delta}}{\bar{x}^2} \left(1-\frac{2M}{\bar{x}}\right)} \left( 3 + \frac{\tilde{\Delta}}{\bar{x}^2} \left( 1 + \frac{3 M}{\bar{x}} \right) - 2 \frac{\tilde{\Delta}^2}{\bar{x}^2} \left( 1 + \frac{M}{\bar{x}} - \frac{6 M^2}{\bar{x}^2} \right) \right)
    \nonumber\\
    &&\qquad
    - \sqrt{1 + \frac{\tilde{\Delta}}{\bar{x}^2}} \left( 3 + \frac{\tilde{\Delta}}{\bar{x}^2} - 2 \frac{\tilde{\Delta}^2}{\bar{x}^4} \right)
    \Bigg)
    \ .
    \label{eq:Quasilocal entropy - mu-scheme}
\end{eqnarray}
This entropy recovers its classical value asymptotically, $\lim_{x\to\infty}S_{\tilde\Delta} (x) = S_{\rm BH}$.
At the horizon, however, it has the non-classical value
\begin{eqnarray}
    S_{\tilde\Delta} (2 M) &=& \frac{2 (4 \pi)^4}{15} \frac{A_{\rm H}^3}{\tilde{\Delta}^2} \left( 3 + 10 \pi \frac{\tilde{\Delta}}{A_{\rm H}}
    - \sqrt{1 + 4\pi\frac{\tilde{\Delta}}{A_{\rm H}}} \left( 3 + 4 \pi \frac{\tilde{\Delta}}{A_{\rm H}} \left( 1 - 8 \pi \frac{\tilde{\Delta}}{A_{\rm H}} \right) \right)
    \right)
    \nonumber\\
    &=& S_{\rm BH} \left(
    1 + 2 \pi \frac{\tilde{\Delta}}{A_{\rm H}}
    + O \left( \frac{\tilde{\Delta}^2}{A_{\rm H}^2}\right) \right)
    \ ,
    \label{eq:Quasilocal entropy at horizon - mu-scheme}
\end{eqnarray}
such that $S_{\tilde\Delta} (2 M)> S_{\rm BH}$.
In contrast to constant $h(x)=1$, the entropy $S_{\tilde\Delta}$ is monotonically decreasing.
This means that the gravitational field in the exterior vacuum has some information and hence the entropy is minimal for asymptotic observers who have access to the whole gravitational field.

\subsection{Presence of a cosmological constant}

After reincorporating the cosmological constant, the metric (\ref{eq:Spacetime
  metric - mu-scheme - exterior}) for $h(x)=\hat{r}/x$ such that
$\lambda(x)=\sqrt{\tilde{\Delta}}/x$  reads 
\begin{eqnarray}
    {\rm d} s^2 &=&
    - \left(1 - \frac{2 M}{x} - \frac{\Lambda x^2}{3}\right) {\rm d} t^2
    \\
    &&
    + \left( 1 + \frac{\tilde{\Delta}}{x^2} \left( 1 - \frac{2 M}{x} - \frac{\Lambda x^2}{3} \right)
    \right)^{-1} \left(1 - \frac{2 M}{x} - \frac{\Lambda x^2}{3}\right)^{-1} {\rm d} x^2
    + x^2 {\rm d} \Omega^2
    \,.\nonumber
    \label{eq:Spacetime metric - mu-scheme - exterior - Cosmo}
\end{eqnarray}
The presence of a cosmological constant can significantly alter the global
structure of the modified space-time.  Even in the classical case,
$\tilde{\Delta}\to0$, a nonvanishing, positive $\Lambda$ implies that a second
horizon exists at
\begin{equation}
    x_\Lambda \approx \sqrt{\frac{3}{\Lambda}}
    \,,
\end{equation}
beyond which the space-time is homogeneous.

If we consider a decreasing holonomy function, $h(x)=\hat{r}/x$,
equation~(\ref{eq:Minimum radius equation}) for coordinate singularities,
evaluated at small $x\ll 2M$, has the approximate solution (\ref{eq:Minimum radius - mu-scheme}) for a minimum radius similar to constant $\lambda$.
However, for large $x\gg 2M$, the Newtonian term can be neglected, and the approximate solution
\begin{equation}
    x_{\tilde\Delta}^{(+)} \approx \sqrt{\frac{3 \tilde{\Delta}}{\tilde{\Delta} \Lambda - 3}}
\end{equation}
is imaginary for $\tilde{\Delta} \Lambda < 3$. Hence, the radius is not
bounded from above for a decreasing holonomy function $h(x)=\hat{r}/x$ in such a case.  Based on
the observed value of the cosmological constant,
$\tilde\Delta \Lambda \sim 10^{-122}$ is negligible for any Planck-sized area
$\tilde\Delta$, and we do not need to
consider the case of $\tilde{\Delta} \Lambda >3$ in which
$x^{(+)}_{\Delta}$ is a real solution. This simple solution is exact only if $M=0$.

With this in mind,
we now write the only real, exact solution to the minimal radius for
$h(x)=\hat{r}/x$, including the cosmological constant:
\begin{eqnarray}
    x_{\tilde\Delta} &=&
    \left( - \tilde{\Delta} (3-\tilde{\Delta}\Lambda) + (3 \tilde{\Delta}  M)^{2/3} \left((3 -\tilde{\Delta} \Lambda)^2 + \sqrt{(3-\tilde{\Delta}  \Lambda)^3 \left(3-\tilde{\Delta} \Lambda+\tilde{\Delta}/(9 M^2) \right)} \right)^{2/3} \right)
    \nonumber\\
    &&\times
    (3-\tilde{\Delta}\Lambda)^{-1} (3 \tilde{\Delta}  M)^{-1/3} \left((3 -\tilde{\Delta} \Lambda)^2 + \sqrt{(3-\tilde{\Delta}  \Lambda)^3 \left(3-\tilde{\Delta} \Lambda+\tilde{\Delta}/(9 M^2) \right)} \right)^{-1/3}
    \,.
\end{eqnarray}
The expression (\ref{eq:Minimum radius equation}) is recovered in the limit
$\Lambda\to0$, and appears as the leading zeroth order in a $\tilde{\Delta} \Lambda$ expansion.
For massive black holes where $\tilde{\Delta}/M^2\ll1$ we have, to leading order,
\begin{equation}
    x_{\tilde\Delta} = \left(\frac{2 M \tilde{\Delta}}{1-\tilde{\Delta} \Lambda/3}\right)^{1/3}
    \,.
\end{equation}
For small masses, defining $\delta \equiv 4 M^2 / \tilde{\Delta}$, we have
\begin{eqnarray}
    \frac{x_{\tilde\Delta}}{2 M} &=&
    1 - (1-\tilde{\Delta} \Lambda/3) \delta
    \label{eq:Minimum radius - mu-scheme - small masses}
\end{eqnarray}
to leading order in $\delta$. For $h(x)=\hat{r}/x$, the conformal diagram of the maximal extension
of the vacuum solution in a de Sitter background is shown in
Fig.~\ref{fig:Holonomy_KS_Vacuum_Wormhole-Periodic-Cosmo-NO MaxRadius} with
$x_\lambda=x_{\tilde\Delta}$. 

\section{Conclusions}

Emergent modified gravity provides a classification of modified canonical
theories compatible with a covariant space-time geometry for each of its
solutions. So far, explicit realizations have been derived in spherically
symmetric models \cite{Higher,HigherCov} and for polarized Gowdy models
\cite{EmergentGowdy}, with up to second-order in derivatives. Even though this
is the same as the classical derivative order, a large number of modifications
are possible that are not of higher-curvature form. The direct canonical
formulation makes this formalism an ideal candidate for an analysis of
holonomy or other modifications proposed in models of loop quantum gravity.

Our main result is the first covariant formulation of an effective theory of
holonomy-modified gravity. The theory provides a consistent set of constraints
that resemble what has been analyzed before in this context, going back to
\cite{JR}. But an important new ingredient is the appearance of an emergent
space-time metric that differs from the classical expression in terms of
canonical fields. This metric is emergent rather than effective because, in
this setting, it is the only metric object that obeys the tensor
transformation law. An effective metric would instead be a correction to a
classical metric, such that there would be at least two different metric
objects within the same theory. Our explicit examples of space-time solutions
in various gauges revealed the importance of the emergent space-time metric
for a consistent interpretation of holonomy modifications. For instance, the
old problem of how to reconcile, on one hand, zero holonomy modifications in a
Schwarzschild gauge where extrinsic curvature vanishes, and non-trivial
holonomy modifications in a non-static gauge such as Gullstrand--Painlev\'e on
the other hand, is resolved by the observation that the emergent metric does
receive corrections at the kinematical level. These corrections survive even
in the Schwarzschild gauge.

Our formulation of holonomy modifications is covariant in space-time because
its solutions are compatible with a coordinate and slicing invariant
space-time structure. It is also covariant in phase-space in that it provides
an unambiguous interpretation of holonomy modifications and their specific
ingredients, taking into account the option to apply canonical
transformations. Our detailed discussion showed that the traditional
distinction between different triangulation schemes that determine the behavior of holonomies in homogeneous reduced models, such as $\mu_0$ or $\bar{\mu}$, cannot be maintained at the covariant level, in particular when canonical transformations are taken into account. Instead, the requirement that holonomy terms have a certain form, in particular periodicity in the relevant phase-space variable, distinguish a specific set of canonical variables in
which the modification function $\lambda$ that initially introduced holonomy modifications is split into two separate parts. This function, which may
depend on the spherical area $E^x$ and at first
resembles the traditional holonomy length $\mu$ in models of loop quantum
gravity, must be split into two contributions,
$\lambda(E^x)=\tilde{\lambda}h(E^x)$ with a constant $\tilde{\lambda}$ and a
function $h(E^x)$ that remains finite and non-zero for
$\tilde{\lambda}\to0$. Only the constant $\tilde{\lambda}$ then appears in
strictly periodic holonomy modifications and therefore represents the holonomy
length, while the holonomy function $h(E^x)$ provides non-trivial
modifications of the triad-dependent coefficients of holonomy terms. In this
way, $\tilde{\lambda}$ resembles a traditional constant holonomy parameter
$\mu_0$, and $h(E^x)$ would usually be interpreted as an inverse-triad
correction.  For covariant holonomy corrections, however, both expressions
originate in the same modification function $\lambda$ of a generic modified
theory. A non-constant holonomy function $h(E^x)$, rather than a non-constant
holonomy length $\tilde{\lambda}$, can then be used to control the strength of
holonomy modifications in various regimes.  Such an approach also clearly elucidates how holonomy modifications can come from fundamental holonomies of a gauge connection that do not refer to a metric for their definition, thereby preserving the holonomy-flux algebra. In particular, it is possible to
have suppressed holonomy effects at low curvature provided $h(E^x)$ decreases
sufficiently quickly for large $E^x$,  even if the holonomy length
$\tilde{\lambda}$ remains constant.

Another significant advantage of a covariant formulation is that it allows a
well-defined application of standard space-time concepts, such as the
definition of black holes through horizons, an analysis of curvature
singularities, the description of motion in curved space-time by geodesics,
and the introduction of various physical quantities related to thermodynamical
behavior.  For the latter, we introduced several new concepts of net
  energy expressions, including a net stress-energy tensor and a net
  quasilocal energy, that quantify deviations of emergent space-time from
  classical space-time in general relativity. Even in vacuum, these quantities
  are in general non-zero and may be positive or negative, allowing a
  generalization of energy conditions and conclusions about the focusing or
  defocusing behavior of space-time on different scales. Our covariant analysis fills a lacuna in the LQG literature by exhibiting an effective repulsive behaviour of the emergent space-time,  underlying the singularity resolution in this model, that is independent of any gauge or coordinate choice.

We have presented detailed derivations of such examples for two cases, a
constant holonomy function $h(E^x)=1$ and a decreasing one,
$h(E^x)\propto 1/x$, respectively, confirming the generic resolution of
black-hole singularities in spherically symmetric models. By and large, we did
not encounter significant problems in these models, even for constant $h(E^x)$
which in the usual interpretation as a $\mu_0$-scheme is often considered
problematic. However, there were several unexpected features in the latter
case which are not as severe as often claimed in traditional formulations but
still lead to modifications on small curvature scales. In particular, as
already observed in \cite{alonso2023charged}, the presence of a positive
cosmological constant implies the existence of a maximal radius for constant
holonomy functions. This outcome represents a global nonclassical effect that
may be interpreted as resulting from the accumulation of a large number of
small quantum corrections. Such instances of UV/IR mixing need to be
interpreted carefully, keeping in mind that they may be artefacts of
symmetry-reduced models in which a large number of identical small corrections
distributed over a strictly homogeneous sphere can only add up.  The symmetry
assumption implies that they do not average out to small values as may be
expected in an inhomogeneous context.

Heuristic holonomy-modification schemes, motivated by phenomenological
considerations, typically rule out such possibilities from the get
go. However, reliably evaluating a fundamental theory requires careful
considerations of all possible outcomes. Our constructions present a major
step in this direction by working, for the most part, with general holonomy
functions $h(E^x)$.  A constant holonomy function may then be realized in
bounded regions of space-time, provided it merges into a suitable fall-off
behavior for large $E^x$.  Our discussion of decreasing holonomy functions
show that the appearance of a maximal radius can easily be avoided. But they
also reveal additional subtleties, such as remnants of quantum corrections
that make the zero-mass limit of black-hole solutions differ from classical
Minkowski space-time. Nevertheless, our derivation of various potentially
observable properties demonstrated the overall consistency of this framework.

\section*{Acknowledgements:}

IHB is supported by the Indonesia Endowment Fund for Education (LPDP) grant from the Ministry of Indonesia. The work of MB and ED was supported in part by NSF grant PHY-2206591. SB is supported in part by the Higgs Fellowship and by the STFC Consolidated Grant “Particle Physics at the Higgs Centre”. 

For the purpose of open access, the author has applied a Creative Commons Attribution (CC BY) licence to any Author Accepted Manuscript version arising from this submission.

\begin{appendix}

  \section{Equations of Motion}\label{Appendix A}
  
  In this appendix, we provide the equation of motion for both non-periodic
  and periodic phase-space variables, taking the classical values for all the
  modification functions except $\lambda$ and $\chi_0$. It is worth mentioning
  the structure of the basic canonical bracket,
\begin{eqnarray}
    \left\{K_\varphi\left(x\right),E^\varphi\left(y\right)\right\}&=&\delta\left(x-y\right)\nonumber \\
    \left\{K_x\left(x\right),E^x\left(y\right)\right\}&=&\delta\left(x-y\right)\,,
\end{eqnarray}
and the fact that the diffeomorphism constraint remains classical:
\begin{eqnarray}
     H_x =& E^\varphi K_\varphi'
    - K_x (E^x)'
    \ .
\end{eqnarray}
The time evolution of any space function $f$ depending on $E^x$, $E^\varphi$,
$K_x$, $K_\varphi$, and their spatial derivatives can be written as 
\begin{eqnarray}
    \dot{f}=\{f,\tilde{H}[N]+H_x[N^x]\}
\end{eqnarray}
where we implicitly absorb the global factor $\chi$ in the lapse function
\begin{eqnarray}
    \bar{N}\equiv\chi N\,.
\end{eqnarray}

We provide the explicit equation of motion for $E^x$, $E^\varphi$, and
$K_\varphi$. The dynamics of $K_x$ is not included for two reasons: first, it
is quite lengthy compared to the other phase-space variables. Second, we
can always use the vanishing of the constraints, either $\tilde{H}=0$ or $H_x=0$,
to solve for $K_x$.

\subsection{Non-periodic variables}

For non-periodic phase-space variables, the Hamiltonian constraint is given by
(\ref{eq:Modified constraint - non-periodic version - simple}). Therefore, the
equations of motion are
\begin{equation}
\dot{E}^x=N^x\left(E^x\right)'+\bar{N}\frac{\sin\left(2\lambda
    K_\varphi\right)}{\lambda}\sqrt{E^x}\left(1+\left(\frac{\lambda
      \left(E^x\right)'}{2E^\varphi}\right)^2\right)
\end{equation}
and
\begin{eqnarray}
 \dot{E}^\varphi&=&\left(N^x E^\varphi\right)'+\bar{N}\left[\frac{E_\varphi}{2\sqrt{E^x}}\frac{\sin\left(2\lambda K_\varphi\right)}{\lambda}+2\sqrt{E^x}K_x\cos\left(2\lambda K_\varphi\right)\left(1+\left(\frac{\lambda \left(E^x\right)'}{2E^\varphi}\right)^2\right)\right.\nonumber \\&&\left.+\lambda \sin\left(2\lambda K_\varphi\right)\frac{\sqrt{E^x}}{2}\left(\frac{E^\varphi}{E^x}\left(\frac{\left(E^x\right)'}{2E^\varphi}\right)^2+\left(\frac{\left(E^x\right)'}{2E^\varphi}\right)'\right)\right.\nonumber \\&&
\left.+2\sqrt{E^x}E^\varphi\frac{\partial \ln\lambda}{\partial
                                                                                                                                                                                                                                                                                                                                                                                                                                                                                                                                 E^x}\left(K_\varphi\cos\left(2\lambda K_\varphi\right)-\frac{\sin\left(2\lambda K_\varphi\right)}{2\lambda}\right.\right.\nonumber \\ && \left.\left.+\left(\frac{\lambda \left(E^x\right)'}{2E^\varphi}\right)^2\left(K_\varphi\cos(2\lambda K_\varphi)+\frac{\sin\left(2\lambda K_\varphi\right)}{2\lambda}\right)\right)\right]\,,
\end{eqnarray}
as well as
\begin{eqnarray}
\dot{K}_\varphi&=&N^x K_{\varphi}'+\bar{N}\left[\frac{\sqrt{E^x}}{2}\left(
\Lambda-\frac{\lambda^2+\sin^2\left(\lambda K_\varphi\right)}{\lambda^2 E^x}\right)+\frac{1}{2\sqrt{E^x}}\left(\frac{\left(E^x\right)'}{2E^\varphi}\right)^2\cos^2\left(\lambda K_\varphi\right)\right.\nonumber \\ &&\left. +\frac{\sqrt{E^x}}{2}\frac{\partial \ln\lambda}{\partial E^x}\left(4\frac{\sin^2\left(\lambda K_\varphi\right)}{\lambda^2}-4K_\varphi \frac{\sin\left(2\lambda K_\varphi\right)}{2\lambda}-2\lambda K_\varphi \left(\frac{\left(E^x\right)'}{2E^\varphi}\right)^2 \sin\left(2\lambda K_\varphi\right)\right)\right]\nonumber \\ &&+\bar{N}'\frac{\sqrt{E^x}\left(E^x\right)'}{2\left(E^\varphi\right)^2}\cos^2\left(\lambda K_\varphi\right) \label{Kphidot}
\end{eqnarray}

\subsection{Periodic variables}

For periodic phase-space variables (constant $\tilde{\lambda}$), the Hamiltonian constraint is given by
(\ref{eq:Hamiltonian constraint - modified - periodic}). The equations of
motion is
\begin{equation}
\dot{E}^x=N^x\left(E^x\right)'+\bar{N}\frac{\tilde{\lambda}\sin\left(2\lambda
    K_\varphi\right)}{\lambda}\sqrt{E^x}\left(1+\left(\frac{\tilde{\lambda}
      \left(E^x\right)'}{2E^\varphi}\right)^2\right)
\end{equation}
and
\begin{eqnarray}
\dot{E}^\varphi&=&\left(N^x E^\varphi\right)'+\frac{\bar{N}\tilde{\lambda}\sqrt{E^x}}{2\lambda}\left[4K_x\left(1+\left(\frac{\tilde{\lambda} \left(E^x\right)'}{2E^\varphi}\right)^2\right)\cos\left(2\tilde{\lambda}K_\varphi\right)\right. \nonumber \\
&&\left.+\left(\left(\frac{4E^\varphi}{\tilde{\lambda}^2}+\frac{\left(\left(E^x\right)'\right)^2}{E^\varphi}\right)\left(\frac{1}{4E^x}-\frac{\partial\ln\lambda}{\partial E^x}\right)+\left(\frac{\left(E^x\right)'}{E^\varphi}\right)'\right)
   \tilde{\lambda}\sin\left(2\tilde{\lambda}K_\varphi\right)\right]\,,
\end{eqnarray}
as well as
\begin{eqnarray}
\dot{K}_\varphi&=&K'_{\varphi}N^x+\frac{\tilde{\lambda}}{2\lambda}\frac{\bar{N}'\sqrt{E^x}\left(E^x\right)'\cos\left(\tilde{\lambda}K_\varphi\right)}{\left(E^\varphi\right)^2}+\frac{\tilde{\lambda}\bar{N}\left(\left(E^x\right)'\right)^2}{8\lambda \sqrt{E^x}\left(E^\varphi\right)^2}\cos\left(\tilde{\lambda}K_\varphi\right)^2 \nonumber \\
&&-\bar{N}\frac{\tilde{\lambda}}{\lambda}\frac{\sqrt{E^x}}{2}\left[\frac{\lambda^2}{\tilde{\lambda}^2}\left(\frac{1}{E^x}-\Lambda\right)+4\left(\frac{1}{4E^x}-\frac{\partial \ln\lambda}{\partial E^x}\right)\frac{\sin^2\left(\tilde{\lambda}K_\varphi\right)}{\tilde{\lambda}^2}\right]
\end{eqnarray}

\end{appendix}



\begin{thebibliography}{10}

\bibitem{PhysRevLett.14.57}
R. Penrose, Phys.\ Rev.\ Lett {\bf 14}, 57 (1965).

\bibitem{rovelli2004quantum}
C. Rovelli, \textit{Quantum Gravity}, (Cambridge University Press, 2004).

\bibitem{thiemann2008modern}
T. Thiemann, \textit{Introduction to Modern Canonical Quantum General Relativity}, (Cambridge University Press, 2008).
   
\bibitem{bojowald2018effective}
M. ~Bojowald, S. ~Brahma, and D. H. Yeom. \textit{Effective Line elements and black hole models in canonical loop quantum gravity}
Phys. Rev. D \textbf{98}, 046015 (2018).
doi:10.1103/PhysRevD.7.2333

\bibitem{EMG}
M. Bojowald, and E.~I. Duque, Class.\ Quant.\ Grav. \textbf{42}, 095008 (2024), arXiv:2404.06375.

\bibitem{EMGCov}
M. Bojowald, and E.~I. Duque, Phys.\ Rev.\ D 108, 084066, (2023), 	arXiv:2310.06798.

\bibitem{alonso2022nonsingular}
A. A. Bardaji, D. Brizuela, and R. Vera. \textit{Nonsingular spherically symmetric black-hole model with holonomy corrections}
Phys.\ Rev.\ D \textbf{106}, 024035 (2022).

\bibitem{BBVeffBH}
A. A. Bardaji, D. Brizuela, and R. Vera. \textit{An effective model for the quantum Schwarzschild black hole}
Phys.\ Lett.\ B \textbf{829}, 137075 (2022).

\bibitem{APSII}
A.\ Ashtekar, T.\ Pawlowski, and P.\ Singh,
\newblock Quantum Nature of the Big Bang: Improved dynamics,
\newblock {\em Phys.\ Rev.\ D} 74 (2006) 084003, [gr-qc/0607039]

\bibitem{SchwarzN}
M.\ Bojowald, D.\ Cartin, and G.\ Khanna,
\newblock Lattice refining loop quantum cosmology, anisotropic models and
  stability,
\newblock {\em Phys.\ Rev.\ D} 76 (2007) 064018, [arXiv:0704.1137]

\bibitem{rovelli1995discreteness}
C. Rovelli, L. Smolin, Nuclear Physics B \textbf{442}, 593 (1995).

\bibitem{Kelly_2020}
J.~G. Kelly, R. Santacruz, and E.~W.Ewing, \textit{Effective loop quantum gravity framework for vacuum spherically symmetric spacetimes}, Phys. Rev. D. \textbf{102}, 106024 (2020).

\bibitem{ADM}
R. Arnowitt, S. Deser, and C.~W. Misner,  in {\em Gravitation: An Introduction to Current Research}, edited by L. Witten (Wiley, New York, 1962), reprinted in \cite{arnowitt2008republication}.

\bibitem{arnowitt2008republication}
R. Arnowitt, S. Deser, and C.~W. Misner, Gen.\ Rel.\ Grav. {\bf 40}, 1997 (2008).

\bibitem{hojman1976geometrodynamics}
S.~A. Hojman, K. Kucha\v{r}, and C. Teitelboim, Ann.\ Phys.\ (New York) {\bf 96},  88  (1976).

\bibitem{kuchar1974geometrodynamics}
K.~V. Kucha\v{r}, J.\ Math.\ Phys. {\bf 15}, 708 (1974).

\bibitem{pons1997gauge}
J.~M. Pons, D.~C. Salisbury, and L.~C. Shepley, Phys.\ Rev.\ D {55}, 658 (1997), arXiv:gr-qc/9612037.

\bibitem{salisbury1983realization}
D.~C. Salisbury, and K. Sundermeyer, Phys.\ Rev.\ D {27}, 740 (1983).

\bibitem{bojowald2000symmetry}
M. Bojowald, and H. Kastrup, Class.\ Quant.\ Grav. {\bf 17}, 3009 (2000), arXiv:hep-th/9907042.

\bibitem{bojowald2004spherically}
M. Bojowald, Class.\ Quant.\ Grav. {\bf 21}, 3733 (2004), arXiv:gr-qc/0407017.

\bibitem{alonso2021anomaly}
A. A. Bardaji and D. Brizuela.
\textit{Anomaly-free deformations of spherical general relativity coupled to matter}.
Phys. Rev. D. \textbf{104}, 084064 (2021).

\bibitem{Gambini2022}
R. Gambini, F. Benítez, and J. Pullin, Universe \textbf{8}, 526, (2022), arXiv:2102.09501.

\bibitem{EMGscalar}
M. ~Bojowald and E. I. ~Duque Phys.\ Rev.\ D \textbf{109}, 084006, (2024) arxiv:2311.10693.

\bibitem{EMGPF}
E.~I. Duque, Phys.\ Rev.\ D {\bf 109}, 044014, (2024), arXiv:2311.08616.

\bibitem{Axion}
I.~H.\ Belfaqih, M.\ Bojowald, S.\ Brahma, and E.~I.\ Duque, [to appear].

\bibitem{alonso2023charged}
A. A. Bardaji, D. Brizuela, and R. Vera. \textit{Singularity resolution by holonomy corrections: Spherical charged black holes in cosmological backgrounds}
Phys.\ Rev.\ D \textbf{107}, 064067 (2023).

\bibitem{Higher}
M.\ Bojowald and E.~I.\ Duque,
\newblock Emergent modified gravity,
\newblock {\em Class.\ Quantum Grav.} 41 (2024) 095008, arXiv:2404.06375.

\bibitem{HigherCov}
M.\ Bojowald and E.~I.\ Duque,
\newblock Emergent modified gravity: Covariance regained,
\newblock {\em Phys.\ Rev.\ D} 108 (2023) 084066, arXiv:2310.06798.

\bibitem{EmergentGowdy}
M.\ Bojowald and E.~I.\ Duque,
\newblock Emergent modified gravity: Polarized Gowdy model on a torus, [to
  appear]

\bibitem{JR}
J.~D.\ Reyes,
\newblock {\em Spherically Symmetric Loop Quantum Gravity: Connections to
  2-Dimensional Models and Applications to Gravitational Collapse},
\newblock PhD thesis, The Pennsylvania State University, 2009.

\bibitem{AOS}
A. Ashtekar, J. Olmedo, and P. Singh, Phys.\ Rev.\ D {\bf 98}, 126003, (2018)
arXiv:1806.02406.

\bibitem{Curiel}
E. Curiel, \textit{ A primer on energy conditions}, arXiv:1405.0403 [physics.hist-ph].


\bibitem{Bardeen1}
J.~M.~Bardeen, \textit{Black holes to white holes I. A complete quasi-classical model},
arXiv:2006.16804.

\bibitem{Bardeen2}
J.~M.~Bardeen, \textit{Black holes to white holes II: quasi-classical scenarios for white hole evolution},
arXiv:2007.00190.
\end{thebibliography}
\end{document}